\documentclass[twocolumn,
numberedappendix,
twocolappendix,
appendixfloats]{openjournal_shmr}

\usepackage{graphicx}
\usepackage{amsmath,amssymb}
\usepackage{float}
\usepackage{mathtools}
\usepackage{tabularx}
\usepackage{bbm}
\usepackage{outlines}
\usepackage{multirow}
\usepackage{rotating}
\usepackage{afterpage}
\usepackage{hyperref}
\hypersetup{colorlinks=true,
    urlcolor=blue,
    anchorcolor=blue,
    citecolor=blue,
    filecolor=blue,
    linkcolor=blue,
    menucolor=blue,
    linktocpage=true,
    pdfproducer=medialab,
    pdfa=true}

\usepackage[normalem]{ulem}

\newcommand{\OrcidID}[1]{ \href[urlcolor = red]{https://orcid.org/#1}{\textcolor{lightgray}{\faOrcid}}}
\newcommand{\OrcidIDName}[2]{\href{https://orcid.org/#1}{#2}}

\usepackage{eso-pic}

\AddToShipoutPictureBG*{%
  \AtPageUpperLeft{%
    \hspace{0.75\paperwidth}%
    \raisebox{-10.0\baselineskip}{%
      \makebox[0pt][l]{\textnormal{DES-2023-0808}}
}}}%

\AddToShipoutPictureBG*{%
  \AtPageUpperLeft{%
    \hspace{0.75\paperwidth}%
    \raisebox{-11.0\baselineskip}{%
      \makebox[0pt][l]{\textnormal{FERMILAB-PUB-25-0419-PPD}}
}}}%

\begin{document}


\title{Constraining the Stellar-to-Halo Mass Relation with Galaxy Clustering and Weak Lensing from DES Year 3 Data}

\shorttitle{Stellar-mass HOD}



\shortauthors{Zacharegkas \& Chang et al.}

\author{\OrcidIDName{0000-0002-2890-6758}{G.~Zacharegkas}$^\dagger$}
\affiliation{HEP Division, Argonne National Laboratory, 9700 South Cass Avenue, Lemont, IL 60439, USA}
\affiliation{Kavli Institute for Cosmological Physics, University of Chicago, Chicago, IL 60637, USA}

\author{\OrcidIDName{0000-0002-7887-0896}{C.~Chang}$^\star$}
\affiliation{Department of Astronomy and Astrophysics, University of Chicago, Chicago, IL 60637, USA}
\affiliation{Kavli Institute for Cosmological Physics, University of Chicago, Chicago, IL 60637, USA}

\author{\OrcidIDName{0000-0002-5933-5150}{J.~Prat}}
\affiliation{Nordita, KTH Royal Institute of Technology and Stockholm University, SE-106 91 Stockholm.}

\author{\OrcidIDName{0000-0001-9994-1115}{W.~G.~Hartley}}
\affiliation{Department of Astronomy, University of Geneva, ch. d’Ecogia 16, 1290 Versoix, Switzerland}

\author{\OrcidIDName{0000-0001-7725-7590}{S.~Mucesh}}
\affiliation{Department of Physics \& Astronomy, University College London, Gower Street, London, WC1E 6BT, UK}

\author{\OrcidIDName{0000-0001-8505-1269}{A.~Alarcon}}
\affiliation{Institute of Space Sciences (ICE, CSIC),  Campus UAB, Carrer de Can Magrans, s/n,  08193 Barcelona, Spain}

\author{\OrcidIDName{0000-0002-7394-9466}{O.~Alves}}
\affiliation{Instituto de Fısica Teorica, Universidade Estadual Paulista, Sao Paulo, 01140-070, Brazil}
\affiliation{Department of Physics, University of Michigan, Ann Arbor, MI 48109, USA}

\author{\OrcidIDName{0000-0002-6445-0559}{A.~Amon}}
\affiliation{Department of Astrophysical Sciences, Princeton University, Peyton Hall, Princeton, NJ 08544, USA}

\author{\OrcidIDName{0000-0001-8156-0429}{K.~Bechtol}}
\affiliation{Physics Department, 2320 Chamberlin Hall, University of Wisconsin-Madison, 1150 University Avenue, Madison, WI 53706-1390, USA}

\author{\OrcidIDName{0000-0001-7774-2246}{M.~R.~Becker}}
\affiliation{Argonne National Laboratory, 9700 South Cass Avenue, Lemont, IL 60439, USA}

% ...

\collaboration{DES Collaboration}

\email{$^{\dagger}$gzacharegkas@gmail.com}

\email{$^{\star}$chihway@kicp.uchicago.edu}



\begin{abstract}
We develop a framework to study the relation between the stellar mass of a galaxy and the total mass of its host dark matter halo using galaxy clustering and galaxy-galaxy lensing measurements. We model a wide range of scales, roughly from $\sim 100 \; {\rm kpc}$ to $\sim 100 \; {\rm Mpc}$, using a theoretical framework based on the Halo Occupation Distribution and data from Year 3 of the Dark Energy Survey (DES) dataset. The new advances of this work include: 1) the generation and validation of a new stellar mass-selected galaxy sample in the range of $\log M_\star/M_\odot \sim 9.6$ to $\sim 11.5$; 2) the joint-modeling framework of galaxy clustering and galaxy-galaxy lensing that is able to describe our stellar mass-selected sample deep into the 1-halo regime; and 3) stellar-to-halo mass relation (SHMR) constraints from this dataset. In general, our SHMR constraints agree well with existing literature with various weak lensing measurements. We constrain the free parameters in the SHMR functional form $\log M_\star (M_h) = \log(\epsilon M_1) + f\left[ \log\left( M_h / M_1 \right) \right] - f(0)$, with $f(x) \equiv -\log(10^{\alpha x}+1) + \delta [\log(1+\exp(x))]^\gamma / [1+\exp(10^{-x})]$, to be $\log M_1 = 11.506^{+0.325}_{-0.404}$, $\log \epsilon = -1.632^{+0.306}_{-0.181}$, $\alpha = -1.638^{+0.108}_{-0.099}$, $\gamma = 0.596^{+0.251}_{-0.210}$ and $\delta = 3.810^{+2.045}_{-1.811}$. The inferred average satellite fraction is within $\sim 5-35\%$ for our fiducial results and we do not see any clear trends with redshift or stellar mass. Furthermore, we find that the inferred average galaxy bias values follow the generally expected trends with stellar mass and redshift. Our study is the first SHMR in DES in this mass range, and we expect the stellar mass sample to be of general interest for other science cases. \\
\end{abstract}

\section{Introduction}\label{sec:Introduction}

Understanding the connection between galaxies and dark matter is one of the essential ingredients needed to form a comprehensive interpretation of the observed Universe. Analyses of Large-Scale Structure (LSS) from galaxy surveys can be biased if details of the galaxy-halo connection are ignored \citep{McDonald2009,Baldauf2012,Krause2017}. 
In modern analyses, one common approach to side-step this complication is to remove the data points on the small scales and only use data in the linear to quasilinear regime, where a simple prescription of the galaxy-halo connection such as \textit{linear galaxy bias} is sufficient \citep[see e.g.][]{y3-3x2ptkp}. In some cases, more complicated galaxy bias models on small scales \citep[such as][]{Heymans2020} are used and marginalize over the model parameters. For either approach, a data-driven model of the galaxy-halo connection on scales below a few Mpc could allow us to significantly improve the cosmological constraints achievable by a given dataset \cite[e.g.][]{Kwan2016,Mandelbaum2013,Cacciato2013,Park2016,Krause2017,Singh2019}. On the other hand, understanding the connection between different galaxy samples and their host halos also has implications for galaxy evolution \citep[see][for a review of studies for galaxy-halo connection]{Wechsler2018}.

In this work we study the combination of two probes for the galaxy-halo connection: galaxy-galaxy lensing and galaxy clustering. Galaxy-galaxy lensing \citep{tyson1984,McKay2001,hoekstra2004,Sheldon_2004,mandelbaum2004,Mandelbaum2006,seljak2005,Johnston2007,Heymans2006,leauthaud2011} refers to the measurement of the cross-correlation between the positions of foreground galaxies and shapes of background galaxies. Due to gravitational lensing, the images of background galaxies appear distorted due to the deflection of light as it passes by foreground galaxies and the dark matter halos they reside in. As a result, this measurement maps the average mass profile of the dark matter halos hosting the foreground galaxy sample -- a direct measurement of the galaxy-halo connection. 

Galaxy clustering \citep{Kaiser1984,Fry1993,leauthaud2011,Abbott2018} refers to the measurement of spatial distribution of galaxies. If we know how galaxies and dark matter halos are connected, the galaxy distribution can be straightforwardly translated to the distribution of dark matter, which makes them a direct probe of LSS. However, without knowledge of the galaxy-halo connection, galaxy clustering on its own suffers from the degeneracy between galaxy bias and the amplitude of the matter power spectrum. As a result, to extract the information in the galaxy clustering measurements, one has to combine it with other probes such as galaxy-galaxy lensing, in order to yield meaningful constraints on the galaxy-halo connection and cosmology \citep{vandenBosch2003,tinker2005,seljak2005}. Combining the two has been proved \citep{yoo2006,Cacciato2009} to be sensitive to cosmological information, and particularly to $\sigma_8$ and $\Omega_m$, since this combination probes the shape and amplitude of the halo mass function on small scales and the overall matter density and the bias of the galaxy sample at large scales.

A common approach to modeling both of these measurements is via the {\it Halo Model} \citep{PeacockSmith2000,Seljak2000,Ma2000,cooray2002} and the \textit{Halo Occupation Distribution} (HOD) framework \citep{PeacockSmith2000,Scoccimarro2001,Berlind2002,Tinker07,zheng2007,zehavi2011,white2011}, where most of the complexities lie within the HOD framework. The foundational assumption in the basic HOD framework is that a {\it central galaxy} is considered to exist in the centers of the host halos and the {\it satellite galaxies} are distributed within the halo \citep{KravtsovHOD2004}. The particular way that the galaxies are distributed depends on the mass of the host halo.

Various alternatives of the HOD framework have been proposed in the literature. For example, the \textit{conditional luminosity function} $\Phi(L|M_h)$ provides the average number of galaxies of luminosity $L \pm dL$ that reside in a halo of mass $M_h$ \citep{Tasitsiomi2004,Yang2004,vandenBosch2004,Vale2006,Cooray2006}; the \textit{conditional stellar mass function} $\Phi(M_\star|M_h)dM_\star$ describes the average number density of galaxies in a halo of mass $M_h$ with stellar masses in the range of $M_\star \pm dM_\star$ \citep{Yang2007,Moster2010,Behroozi2013}; and \textit{abundance matching} relies on simulations and assumes that there is there is a monotonic correspondence between halo mass (or circular velocity) and galaxy stellar mass (or luminosity) \cite[e.g.,][]{Tasitsiomi2004,Conroy06,KravtsovHOD2004,Vale2006,Conroy2009,Drory2009,Moster2010,Behroozi2013,Guo2010}.

In this paper we present a joint analysis of galaxy-galaxy lensing and galaxy clustering using a new stellar mass-selected galaxy sample from the first three years (Y3) of the Dark Energy Survey (DES) data. We model scales deep in the 1-halo regime using an HOD modeling framework based on our previous work in \cite{Zacharegkas2022}.

In \cite{Zacharegkas2022}, we analyzed the galaxy-galaxy lensing signal from two lens galaxy samples that were used by DES Y3 for the main cosmological studies \citep[e.g.][]{y3-3x2pt,Prat2022}. Our HOD model was widely based on \cite{zheng2007,zehavi2011} and the halo occupation was inferred as a function of the halo mass. The HOD constraints in that work were important in various aspects of understanding the DES Y3 cosmology results, but are not suited for understanding the general galaxy population given that the lens samples had various selections that were imposed to optimize their photometric redshifts. In this work, we expand upon \cite{Zacharegkas2022} in three main ways, which we discuss in what follows. First, we incorporate theory predictions for galaxy clustering on small scales in addition to galaxy-galaxy lensing. In addition, our new model accounts for the stellar mass-information of the lens galaxies. This is done by direct parametrization of the HOD, but is equivalent to using, for example, the conditional stellar mass function mentioned before. Finally, we construct and validate a new, more generic stellar-mass-selected lens sample compared to what was used in \cite{Zacharegkas2022}. This allows us to constrain the stellar-to-halo-mass relation of DES galaxies. We 
also note here that, the new stellar mass-selected galaxy sample can be used in future analyses of other science cases.

The structure of the paper is as follows. In Section~\ref{sec:LensSample} we introduce the new lens sample by describing in detail the machine learning algorithm used (training, validation and its products). We also present the final selection (in redshift and stellar mass) of our sample. In Section~\ref{sec:SourceSample} we discuss the source galaxy sample that we use in the galaxy-galaxy lensing part of the data vector. Section~\ref{sec:smMeasurements} describes the measurements of the data vector, the covariance matrix, and systematic tests. We then proceed to describe in depth our model in Section~\ref{sec:HODmodel} and Section~\ref{sec:modelobservables}. Equipped with the data vector and model, in Section~\ref{sec:Fits} we present the likelihood analysis we follow to fit the measurements and present the results in Section~\ref{sec:smResults}. We conclude in Section~\ref{sec:Conclusions}. 

Throughout this study we fix the $\Omega_m$ and $\sigma_8$ cosmological parameters to the best-fit for the \texttt{MagLim} sample from DES Y3 \citep{y3-3x2pt}, while we use the best-fit cosmology from {\it Planck 2018} \citep{Planck2018VI} for the rest of them: $\Omega_{\rm m}=0.339$, $\sigma_{8}=0.733$, $\Omega_{\rm b}=0.0486$, $H_0=67.37 \; {\rm km/s/Mpc}$, $n_{s}=0.9649$, $\Omega_\nu h^2=0.0006$, where $h$ is the Hubble constant in units of $100 \; {\rm km/s/Mpc}$. Since we consider the $\Lambda$-Cold Dark Matter ($\Lambda$CDM) cosmological model, we set $w=-1$ for the dark energy equation of state parameter. We test in Section~\ref{sec:sensitivity} the sensitivity of our results to the assumed cosmological model. In addition, all the halo masses use the definition of $M_{\rm 200c}$, based on the mass enclosed by radius $R_{\rm 200c}$ so that the mean density of a halo is 200 times the critical density at the redshift of the halo. However, our results are valid for any common choice of mass definition.

\section{The Halo Occupation Distribution model}\label{sec:HODmodel}

In this section we discuss how we model the Halo Occupation Distribution (HOD) for this work. This model is based on \cite{Zacharegkas2022} and it is modified so that the stellar-mass information of the galaxies is accounted for. This is done by incorporating a stellar-to-halo-mass relation (SHMR) into the HOD framework. Below we introduce the SHMR used in this study (Section~\ref{sec:SHMR}) and we then proceed to describe our HOD model (Section~\ref{sec:HODtophat}).

\subsection{The stellar-to-halo-mass relation}\label{sec:SHMR}

Henceforth, we will be using the notation $f_{\rm SHMR}(M_h) \equiv M_\star (M_h)$ for the SHMR. There are a large number of variants of this relation used in the literature and, for the most part, they are in good agreement. However, especially at large masses, there can exist significant differences. In this work we follow the SHMR form in \cite{Kravtsov2018}, which is the same as in \cite{Behroozi2013}. In that paper the authors carefully calibrated the SHMR of high-mass halos and found that the relation in that regime is shallower than the HOD constraints by \cite{Leauthaud2012b} and the abundance matching analyses by \cite{Moster2013,Behroozi2013}. The authors suggest that the main reason is that in these previous studies the measured stellar-mass of the central galaxies was underestimated significantly.

We parametrize the SHMR as follows:
\begin{align}\label{eq:SHMRdef}
    \log f_{\rm SHMR} (M_h) &\equiv \log M_\star (M_h) \nonumber \\
    &= \log(\epsilon M_1) + f\left[ \log\left( \frac{M_h}{M_1} \right) \right] - f(0) \; ,
\end{align}
with
$$f(x) \equiv -\log(10^{\alpha x}+1) + \delta \frac{[\log(1+\exp(x))]^\gamma}{1+\exp(10^{-x})} \; .$$
In the above expressions, we allow $\{ M_1, \epsilon, \alpha, \gamma, \delta \}$ to vary during our fits to the data. This functional form is a power-law with slope $-\alpha$ at the low-mass limit, $M_h \ll M_1$, and breaks into a different power-law with slope $\gamma$ for high-mass halos, $M_h \gg M_1$, at the characteristic halo-mass scale $M_1$ with a characteristic stellar-mass-to-halo-mass ratio $\epsilon$. Finally, $\delta$ describes the strength of the power-law in the massive end of the relation.

\subsection{The HOD parametrization}\label{sec:HODtophat}
For a galaxy sample selected by a stellar-mass threshold $M_{\star}^t$, i.e. $M_\star \geq M_\star^t$, we can write down the HOD for the central galaxy similar to the form presented in \cite{zheng2007}: 
\begin{equation}\label{eqHOD:CentralThresholded}
    \langle N_c(M_h|M_\star^t) \rangle = \frac{f_{\rm cen}}{2} \left[ 1 + {\rm erf} \left( \frac{\log f_{\rm SHMR}(M_h) - \log M_\star^{t}}{\sigma_{\log M_\star}} \right) \right] \; ,
\end{equation}
where the parameter $\sigma_{\log M_\star}$ describes the log-normal scatter in stellar mass at a given halo mass. The parameters that characterize the centrals are then: (1) $f_{\rm cen}$ which represents the fraction of central galaxies that made it into the sample after selection; (2) the parameter $\sigma_{\log M_\star}$ which quantifies the scatter in $\log M_\star$ and is related to the scatter in $M_h$ via the relation $\sigma_{\log M_\star}/\sigma_{\log M_h}=d\log M_\star/d\log M_h$; (3) the parameters that enter the SHMR as mentioned in Section~\ref{sec:SHMR}. 

To derive the HOD for a central occupation \textit{conditioned on a range of stellar masses} $[M_{\star,1},M_{\star,2}]$, we simply subtract two thresholded HOD's from Equation~\eqref{eqHOD:CentralThresholded} at $M_{\star,1}$ and $M_{\star,2}$: 
\begin{equation}\label{eqHOD:CentralRangeFromThresholded}
    \langle N_c(M_h) \rangle = \langle N_c(M_h| M_\star^{t1}) \rangle - \langle N_c(M_h| M_\star^{t2}) \rangle \; .
\end{equation}

For the satellite galaxies, a similar approach can be used when the same range of stellar masses is considered. We can thus write:
\begin{equation}\label{eqHOD:SatelliteRangeFromThresholded}
    \langle N_s(M_h) \rangle = \langle N_s(M_h| M_\star^{t1}) \rangle - \langle N_s(M_h| M_\star^{t2}) \rangle \; ,
\end{equation}
where we model the satellite occupation for a $M_\star$-thresholded sample as a re-scaled version of the central HOD multiplied by a power-law:
\begin{align}\label{eqHOD:SatelliteThresholdedInit}
    \langle N_s(M_h|M_\star^{t}) \rangle &= f_{\rm sat} \lambda_s(M_h) \; ; \nonumber \\
    \lambda_s (M_h)&\equiv \left( \frac{M_h- \kappa M_{\min}}{M_{\rm sat}} \right)^{\alpha_{\rm sat}} \exp \left( - \frac{M_{\rm cut}}{M_h} \right) \; ,
\end{align}
where $f_{\rm sat}$, $M_{\rm sat}$, $M_{\rm cut}$, $\alpha_{\rm sat}$ and $\kappa$ are free parameters, and we have defined $M_{\min} \equiv f_{\rm SHMR}^{-1}(M_\star^t)$. 
Specifically, $f_{\rm sat}$ is the equivalent of $f_{\rm cen}$ in the central HOD of Equation~\eqref{eqHOD:CentralThresholded} for the satellite occupation, $\alpha_{\rm sat}$ controls the power-law behaviour of the satellite occupation at large halo masses, and the stellar-mass dependence enters into the satellite modeling via the parameters $M_{\rm sat}$ and $M_{\rm cut}$ \citep[see, e.g.][]{leauthaud2011,zu2015}, that we model as:
\begin{align}\label{eq:SatelliteSHMRinM1}
    \frac{M_{\rm sat}}{10^{12} \; M_\odot} = B_{\rm sat} \left( \frac{f_{\rm SHMR}^{-1}(M_\star^{t})}{10^{12} \; M_\odot} \right)^{\beta_{\rm sat}} = B_{\rm sat} \left( \frac{M_{\min}}{10^{12} \; M_\odot} \right)^{\beta_{\rm sat}} \; ,
\end{align}
\begin{align}\label{eq:SatelliteSHMRinMcut}
    \frac{M_{\rm cut}}{10^{12} \; M_\odot} = B_{\rm cut} \left( \frac{f_{\rm SHMR}^{-1}(M_\star^{t})}{10^{12} \; M_\odot} \right)^{\beta_{\rm cut}} = B_{\rm cut} \left( \frac{M_{\min}}{10^{12} \; M_\odot} \right)^{\beta_{\rm cut}} \; ,
\end{align}
where the new free parameters ($B_{\rm sat}$, $B_{\rm cut}$, $\beta_{\rm sat}$, $\beta_{\rm cut}$) have been introduced to allow the satellites to have a different form of HOD from the centrals. As an additional note, as opposed to \cite{Zacharegkas2022}, in this current paper we decouple the HOD of the satellites in Equation~\eqref{eqHOD:SatelliteThresholdedInit} from the central one to give the model additional freedom in order to fit the more complex data we analyze here.

The total HOD prediction in the thresholded case is then given by:
\begin{align}\label{eqHOD:TotalThresholded}
    \langle N(M_h| M_\star^t) \rangle =& \langle N_c(M_h| M_\star^t) \rangle + \langle N_s(M_h| M_\star^t) \rangle \; .
\end{align}
The above approach effectively assumes a tophat distribution of the stellar mass in the range, and has been shown to be a good model for certain datasets \citep[see e.g.][]{leauthaud2011}. This is also what we will use as our fiducial model in this work. We describe a more general case in Appendix~\ref{app:HODexact}, where the stellar mass distribution takes on a more flexible form. 

\begin{figure}
\centering
\includegraphics[width=\columnwidth]{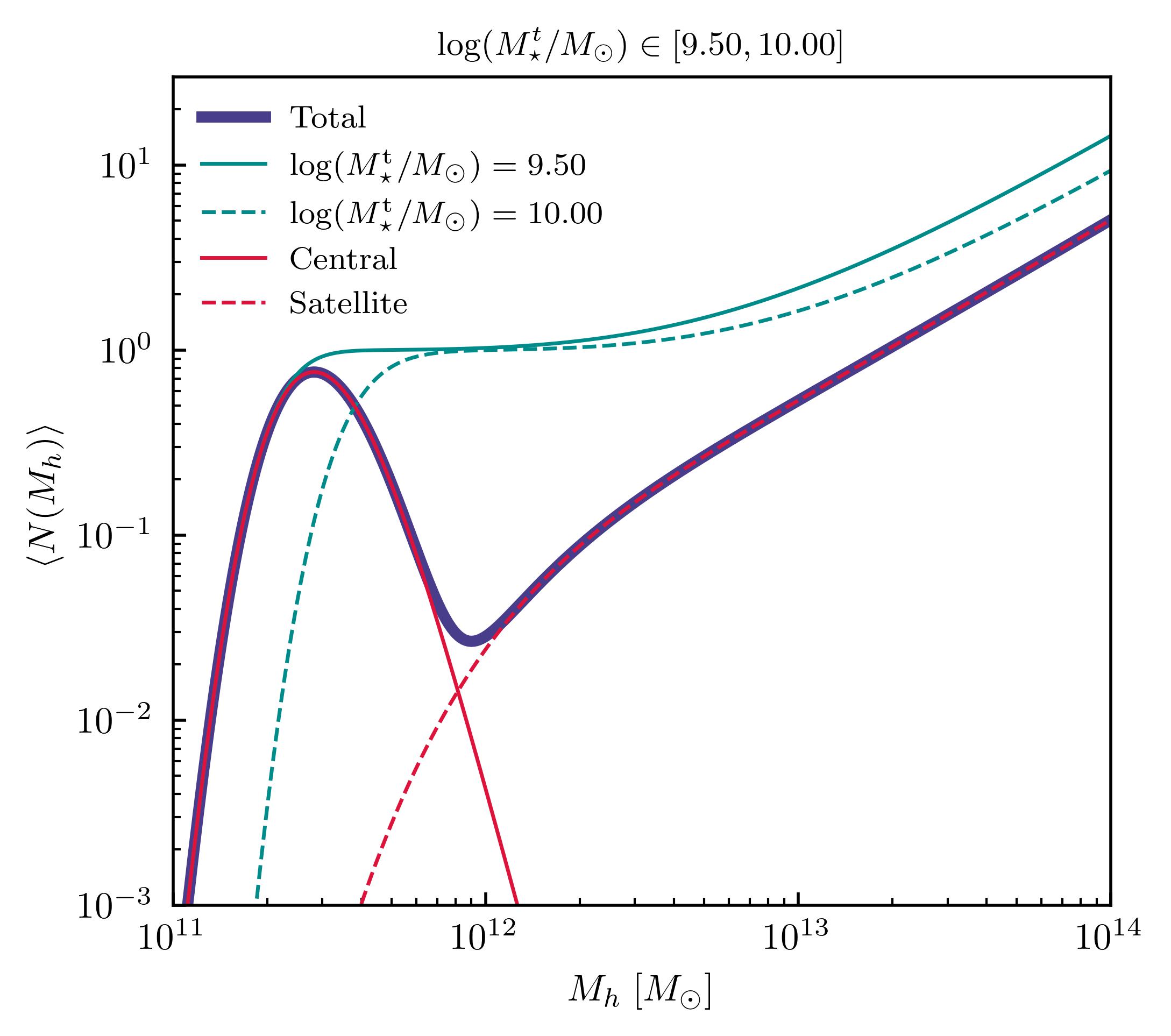}
\caption{\label{fig:SHOD_tophat} 
The total HOD model prediction (thick solid blue) and the two thresholded HOD's (solid and dashed cyan) that produce it via their difference. The central (solid red) and satellite (dashed red) components of the total HOD (thick solid blue) are also presented. The HOD parameters we used in this plot are: $\sigma_{\log M_\star} = 0.3$, $\alpha_{\rm sat} = 1$, $B_{\rm sat}=25$, $\beta_{\rm sat}=0.8$, $B_{\rm cut}=10$, $\beta_{\rm cut}=1.5$, $f_{\rm cen}=1=f_{\rm sat}$, $\kappa_{\rm sat}=1.5$; the SHMR parameters are: $\log M_1=11.45$, $\log \epsilon=-1.702$, $\alpha = -1.736$, $\gamma = 0.613$, $\delta=4.273$.}
\end{figure}

In Figure~\ref{fig:SHOD_tophat} we show the total (centrals and satellites combined) HOD from our fiducial model (blue), together with the two total thresholded HODs (cyan) that created it via their difference. 
We go into further details of the full model which assumes a more complex stellar-mass distribution and how accurate our approximation where we take the difference of two thresholded HOD's is in Appendix~\ref{app:HODexact}. 

The above computations only involve the first moment of the HOD PDF. We, however, need to supplement it with the second moment which enters the modeling of galaxy clustering in Section~\ref{sec:GGmodel}. To that end, we need to make an assumption for the form of the  satellite PDF, which we model as a \textit{Poisson distribution}. Under this assumption, we write the second moment for the HOD as follows:
\begin{align}\label{eq:HOD2ndmoment}
    &\langle N(M_h | M_\star^t) (N(M_h | M_\star^t) - 1) \rangle = \lambda_s(M_h) \nonumber \\ 
    &\qquad \times \left[ 2 \langle N_c(M_h | M_\star^t) \rangle + \lambda_s(M_h) \rangle \right] \; .
\end{align}
Having all that, we can continue with modeling the observables of interest, before we fit our model to the data.

\section{Modeling the observables}\label{sec:modelobservables}

Equipped with the new HOD model, the main goal in this section is to combine the HOD model with the Halo Model \citep{Seljak2000,Ma2000,PeacockSmith2000} to predict our two observables: galaxy-galaxy lensing (Section~\ref{sec:GGLmodel}) and galaxy clustering (Section~\ref{sec:GGmodel}).  

Following the Halo Model, the first step is to write down the 3D galaxy-matter power spectrum $P_{\rm gm}$ and the 3D galaxy power spectrum $P_{\rm gg}$, each consisting of a ``one-halo'' (1h) and a ``two-halo'' (2h) term and combine the two. Notation wise, we use $P_{\rm gt}^{\rm 1h/2h}(k,z)$ to denote the power spectrum between galaxies (${\rm g}$) and a tracer (${\rm t}$, either ${\rm g}$ or ${\rm m}$), for the 1-halo/2-halo (${\rm 1h/2h}$) components. We combine the two in real space using the 3D correlation function $\xi_{\rm gt}^{\rm 1h/2h}(r,z)$ as:
\begin{equation}\label{eq:PgmTotal}
    \xi_{\rm gm}(r,z) = \xi_{\rm gm}^{\rm 1h}(r,z) + \xi_{\rm gm}^{\rm 2h}(r,z)
\end{equation}
\begin{equation}\label{eq:PggTotal}
    \xi_{\rm gg}(r,z) = \xi_{\rm gg}^{\rm 1h}(r,z) + \xi_{\rm gg}^{\rm 2h}(r,z)
\end{equation}
as a function of distance $r$. Following \cite{hayashi2008, zu2015, clampitt2017}, we combine the 1-halo and 2-halo term in the 3-dimensional real space, as it is found to more accurately describe observations at the quasi-linear regime. In the above, the 1-halo terms can be further decomposed into the \textit{central} and \textit{satellite} terms, as introduced in \cite{KravtsovHOD2004}. We describe how we model each of these terms in $P_{\rm gm}$ and $P_{\rm gg}$ in Sections~\ref{sec:GGLmodel} and \ref{sec:GGmodel}, respectively. 
We also note that each of the terms is a function of redshift ($z$) and scale ($k$). The central and satellite terms refer to the same definitions used in the HOD model in Section~\ref{sec:HODtophat}. The 1- and 2-halo terminology refers to whether the two tracers of interest (galaxy or dark matter) live in the same halo or two separate halos. There is a delicate choice to make when combining the 1- and 2-halo terms, which we discuss in Section~\ref{sec:transition}. Once each term in Equations~\eqref{eq:PgmTotal} and \eqref{eq:PggTotal} are specified, together with the prescription for combining the two terms, it is projected into 2D (the projected spetra $C_{\rm gm}$ and $C_{\rm gg}$, of $P_{\rm gm}$ and $P_{\rm gg}$ respectively), then Fourier transformed into the real-space estimators that we measure ($\gamma_{t}$ and $w$) in Section~\ref{sec:gammat_wtheta}. We describe the approach we take for incorporating the redshift and stellar mass uncertainties in Section~\ref{subsec:nofzSys}, and finally we discuss any additional modeling components in Section~\ref{sec:addition_model}

\subsection{Galaxy-matter cross power spectrum $P_{\rm gm}$}\label{sec:GGLmodel}

Following closely our previous work in \cite{Zacharegkas2022}, which follows \cite{Seljak2000,mandelbaum2004}, we can write down the expressions for the terms in Equation~\eqref{eq:PgmTotal} as follows.

The 1-halo galaxy-cross-matter term can be written as 
\begin{align}\label{eq:smPgm1h}
    &P_{\rm gm}^{\rm 1h}(k,z) 
    = \frac{1}{\rho_m n_g} \int dM_h \; \frac{dn}{dM_h} M_h \nonumber \\
    &\quad \times u_{\rm dm}(k|M_h) \left[ \langle N_c(M_h) \rangle + \langle N_s(M_h) \rangle u_{\rm sat}(k|M_h) \right] \; , 
\end{align}
where $u_{\rm dm/sat}(k|M_h)$ is the Fourier transform of the dark matter/satellite spatial distribution, for which we assume the NFW profile \citep{navarro1996}, and the average galaxy number density
\begin{align}\label{eq:ngaldef}
    n_g = \int dM_h \; \frac{dn}{dM_h} \langle N(M_h) \rangle \; .
\end{align}
The first and second terms in parenthesis in Equation~\eqref{eq:smPgm1h} create the \textit{central 1-halo}, $P_{\rm gm}^{\rm c1h}$, and \textit{satellite 1-halo}, $P_{\rm gm}^{\rm s1h}$, terms, respectively.

For completeness, we also note here that the average fraction of galaxies that are satellites, henceforth the \textit{satellite fraction}, as predicted from our HOD model, is computed as
\begin{equation}\label{eq:SatFracDef}
    F_{\rm sat} = 1 - F_{\rm cen} = \frac{1}{n_g} \int dM_h \; \frac{dn}{dM_h} \langle N_s(M_h) \rangle \; ,
\end{equation}
where we also define the average fraction of central galaxies as $F_{\rm cen} = 1-F_{\rm sat}$.

Finally, the 2-halo term is computed as
\begin{align}\label{eq:Pgm2h}
    P_{\rm gm}^{\rm 2h}(k,z) =& P_{\rm m}^{\rm nl}(k,z) \int dM_h \; \frac{dn}{dM_h} \frac{M_h}{\rho_m} b_h(M_h) u_{\rm dm}(k|M_h) \nonumber \\
    &\times \int dM_h' \; \frac{dn}{dM_h'} \frac{b_h(M_h')}{n_g} \nonumber \\
    &\times \left[ \langle N_c(M_h') \rangle + \langle N_s(M_h') \rangle u_{\rm sat}(k|M_h') \right] \; ,
\end{align}
where, usually, the first and second term in the brackets are identified as the \textit{2-halo central} and \textit{2-halo satellite} terms, respectively.
In this work we adopt the functions from \citet{tinker2010} for the halo bias, $b_h(M_h)$, and the Halo Mass Function (HMF), $dn / d M_h$, from \cite{Tinker2008}. 
We note here that on the very large scales, the above reduces to
\begin{equation}\label{eq:Pgm2hsimple}
    P_{\rm gm}^{\rm 2h}(k,z) \approx b_g P_{\rm m}^{\rm lin}(k,z) \; ,
\end{equation}
where
\begin{align}\label{eq:galbiasdef}
    b_g \equiv \frac{1}{n_g} \int dM_h \; \frac{dn}{dM_h} b_h(M_h) \langle N(M_h) \rangle
\end{align}
is the \textit{large-scale galaxy bias}, and for the linear matter power spectrum $P_{\rm m}^{\rm lin}(k,z)$ we adopt the fitting functions from \citet{eisenstein1998} which we modify by adopting the Halofit \citep{takahashi2012} approximation to compute the nonlinear matter power spectrum, $P_{\rm m}^{\rm nl}(k,z)$. We furthermore emphasize here that the first integral in Equation~\eqref{eq:Pgm2h} must approach unity on those linear scales. In practice, we need to choose a galaxy bias model that when combined with the halo-mass function satisfies this requirement, i.e. that the dark matter is unbiased with respect to itself, and thus:
\begin{align}\label{eq:galbiasconstrain}
    \lim_{k\rightarrow 0} \int dM_h \; \frac{dn}{dM_h} \frac{M_h}{\rho_m} b_h(M_h) = 1 \; .
\end{align}

\begin{figure}
\centering
\begin{minipage}{.5\textwidth}
  \centering
  \includegraphics[width=1.0\columnwidth]{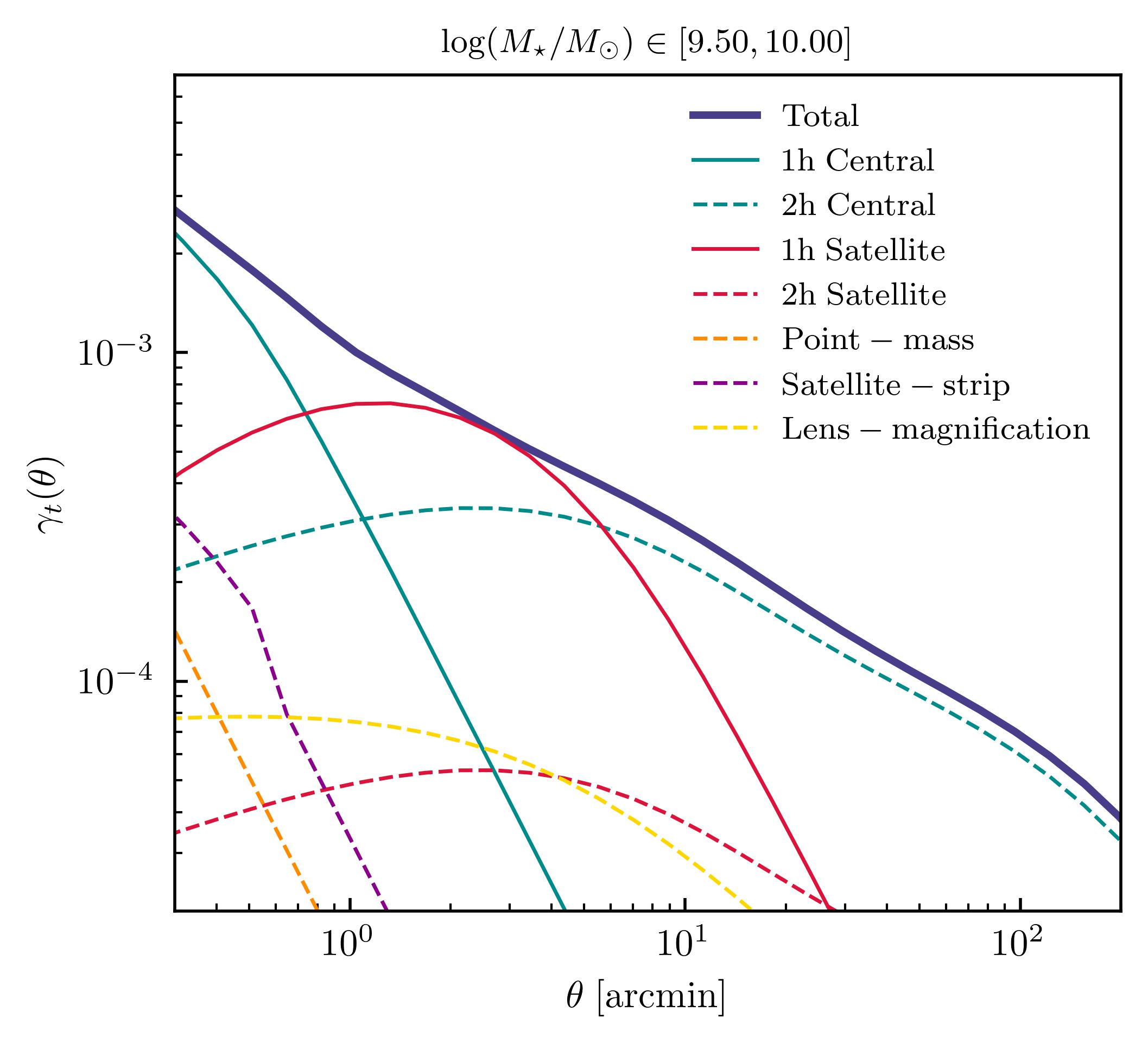}
\end{minipage}%
\hfill
\begin{minipage}{.5\textwidth}
  \centering
  \includegraphics[width=1.0\columnwidth]{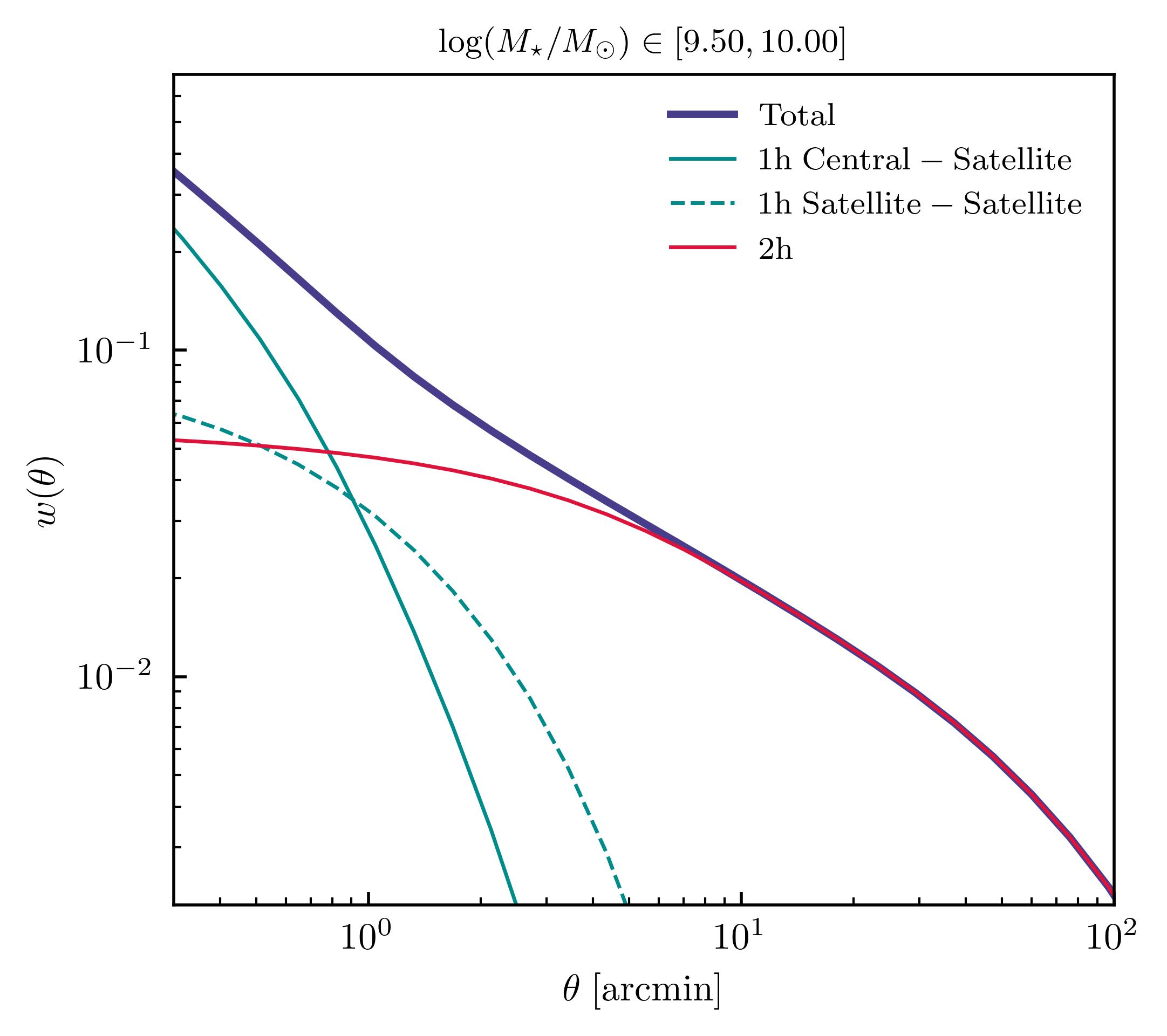}
\end{minipage}
\caption{\label{fig:obs_tophat_vs_1D} \textit{(Upper)} The total $\gamma_t$ model prediction (solid dark blue) and its components, as described in the legend, using the HOD model from Figure~\ref{fig:SHOD_tophat}. In addition to the HOD and SHMR parameters described in Figure~\ref{fig:SHOD_tophat}, for this plot we set $\Delta \log M_\star=0$, $\Delta z_\ell =0 = \Delta z_s$, $m_s=0$, $\Sigma_\ell = 1$, $\kappa_c = 1$, $\alpha_{\rm lmag}=4.5$. \textit{(Lower)} Similar to the left plot but for $w$.}
\end{figure}

\subsection{Galaxy auto-correlation power spectrum $P_{\rm gg}$}\label{sec:GGmodel}

To model the galaxy clustering signal, we follow \cite{vandenBosch13}.

For the 1-halo components of Equation~\eqref{eq:PggTotal}, we follow \citet{Seljak2000} which leads to the 1-halo galaxy auto-correlation power spectrum
\begin{align}\label{eq:Pgg1hTotal}
    &P_{\rm gg}^{\rm 1h}(k,z) = \frac{1}{n_g^2} \int dM_h \; \frac{dn}{dM_h} \lambda_s(M_h) u_{\rm sat}(k|M_h) \nonumber \\
    &\quad \times \left[ 2 \langle N_c(M_h) \rangle + \lambda_s(M_h) u_{\rm sat}(k|M_h) \right] \; ,
\end{align}
where we have made use of Equation~\eqref{eq:HOD2ndmoment}. The first and second terms of the equation above correspond to the \textit{central-satellite} term, $P_{\rm gg}^{\rm cs1h}$, and the \textit{satellite-satellite} term, $P_{\rm gg}^{\rm ss1h}$, respectively. Note here that we ignore the contribution from the central-central term $P_{\rm gg}^{\rm cc1h}=1/n_g$ as it is negligible compared to the other two 1-halo components. We finally calculate the 2-halo component as
\begin{align}\label{eq:Pgg2hTotal}
    &P_{\rm gg}^{\rm 2h}(k,z) = P_{\rm m}^{\rm nl}(k,z) \nonumber \\
    &\; \times \int dM_h \; \frac{dn}{dM_h} \frac{b_h(M_h)}{n_g} \left[ \langle N_c(M_h) \rangle + \lambda_s (M_h) u_{\rm sat}(k|M_h) \right] \nonumber \\
    &\; \times \int dM_h' \; \frac{dn}{dM_h'} \frac{b_h(M_h')}{n_g} \rangle \left[ \langle N_c(M_h') \rangle + \lambda_s(M_h') u_{\rm sat}(k|M_h') \right] \nonumber \\
    &\approx b_g^2 P_{\rm m}^{\rm nl}(k,z) \; .
\end{align}

\subsection{Transition regime}
\label{sec:transition}
Following \cite{mead2021}, we implement a "beyond-linear" halo bias model when combining the 1 and 2-halo terms in Sections~\ref{sec:GGLmodel} and \ref{sec:GGmodel}. This changes the shape of the 2-halo component as a function of $k$, especially around the scales corresponding to the size of individual dark matter halos. This \textit{scale-dependence} of the halo bias becomes important on scales $r \lesssim 3 \; {\rm Mpc}/h$ and we have found that it is especially important in modeling $\xi_{gg}$, as without it the 2-halo term can dominate even in the quasi-to-non-linear regime for low-mass galaxies. Moreover, this modification is especially useful for fitting our model for $\xi_{\rm gm}$ to the low-stellar mass bins, as without it our fits fail in the small-scale regime. In this paper, we implement the scale-dependent halo bias by utilizing the fitting functions from \cite{tinker2005}, which are calibrated on numerical simulations and were found to be accurate enough to model galaxy clustering for galaxies in the Sloan Digital Sky Survey \cite[SDSS;][]{York2000_SDSS} using the HOD framework. In particular, we multiply the 2-halo correlation function by $\zeta(r,z)$, which is defined as
\begin{equation}\label{eq:zetaDef}
    \zeta(r,z) \equiv \sqrt{\frac{[1+1.17 \xi_{\rm m}^{\rm nl}(r,z)]^{1.49}}{[1+0.69\xi_{\rm m}^{\rm nl}(r,z)]^{2.09}}} \; ,
\end{equation}
where $\xi_{\rm m}^{\rm nl}(r,z)$ is the non-linear real-space correlation function of $P_{\rm m}^{\rm nl}(k,z)$. This is the same approach also used in \cite{zu2015} and results in the 2-halo galaxy-matter cross- and galaxy auto- correlation function terms 
\begin{equation}\label{eq:2haloxigm}
    \xi_{\rm gm}^{\rm 2h}(r,z) \approx \zeta(r,z) b_g \xi_{\rm m}^{\rm nl}(r,z)
\end{equation}
and 
\begin{equation}\label{eq:2haloxigg}
    \xi_{\rm gg}^{\rm 2h}(r,z) \approx \zeta^2(r,z) b_g^2 \xi_{\rm m}^{\rm nl}(r,z),
\end{equation}
respectively.
We find that using this model we are able to obtain a better fit to our data on all scales and improve the fits significantly in the 1-halo to 2-halo regime compared to what we assumed previously in \cite{Zacharegkas2022}, which was also followed by \cite{hayashi2008,zu2014,clampitt2017}. As we discuss in Section~\ref{sec:Fits}, this model is especially useful in fitting the low-redshift, low-mass galaxy bins. 

\subsection{The observables: $\gamma_{t}$ and $w$}\label{sec:gammat_wtheta}

Combining Sections~\ref{sec:GGLmodel}, \ref{sec:GGmodel} and \ref{sec:transition}, we can now write down the real-space model prediction from the HOD framework for galaxy-galaxy lensing and galaxy clustering. In Figure~\ref{fig:obs_tophat_vs_1D} we show our model prediction of $\gamma_{t}$ and $w$, as well as their different components. 

For a single lens-source redshift pair $(z_\ell,z_s)$ and for a stellar mass value $M_\star$, the projected 2D galaxy-matter power spectrum is computed, under the Limber approximation, by:
\begin{equation}\label{eq:Limber2DSpectrum}
    C_{\rm gm}(\ell|z_\ell,z_s) = \frac{\rho_m \Sigma_c^{-1}(z_\ell,z_s)}{\chi^2(z_\ell)} P_{\rm gm}\left( \frac{\ell+1/2}{\chi(z_\ell)},z_\ell \right) \; ,
\end{equation}
where $\rho_m$ is the average mass density of the universe today, and we define the critical surface density as
\begin{equation}\label{eq:SigmaCrit}
    \Sigma_c (z_\ell,z_s) = \frac{c^2}{4\pi G} \frac{\chi(z_s) (1+z_\ell)^{-1}}{\chi(z_\ell) \chi(z_\ell,z_s)} \; ,
\end{equation}
with $c$ and $G$ being the speed of light and Newton's gravitational constant, respectively, and $\chi(z_1,z_2)$ is the comoving distance between redshifts $z_1$ and $z_2$. We then calculate the average spectrum in the following way: 
\begin{align}\label{eq:ClpAverage}
    C_{\rm gm}(\ell) =& \int_{z_{s,1}}^{z_{s,2}} dz_s \; n_s(z_s) \int_{z_{\ell,1}}^{z_{\ell,2}} dz_\ell \; C_{\rm gm} (\ell|z_\ell,z_s) \;,
\end{align}
and the real-space galaxy-galaxy lensing model is then
\begin{equation}\label{eq:gammatHankel}
    \gamma_t(\theta) = \int \frac{\ell d\ell}{2\pi} J_2(\ell \theta) C_{\rm gm}(\ell) \;.
\end{equation}
The above transformation uses the Hankel transform with the Bessel function $J_2$.

Similarly, for galaxy clustering, we first project $P_{\rm gg}$ via the Limber approximation:
\begin{equation}\label{eq:CggDef}
    C_{\rm gg}(\ell) = \int dz_\ell \; \frac{n_\ell^2(z_\ell)}{\chi^2(z_\ell)} \left( \left. \frac{d\chi}{dz} \right|_{z_\ell} \right)^{-1} P_{\rm gg}\left( \frac{\ell+1/2}{\chi(z_\ell)},z_\ell \right) \; ,
\end{equation}
where $(d\chi/dz)|_{z_\ell}$ is the derivative of the comoving distance with respect to redshift at $z=z_\ell$, and we have used the galaxy selection function, which is essentially the normalized redshift distribution of our lenses, twice and hence the $n_\ell^2$ factor. Via a Hankel transform, using $J_0$, we can then model the final galaxy clustering observable as
\begin{equation}\label{eq:wthetaHankel}
    w(\theta) = \int \frac{\ell d\ell}{2\pi} J_0(\ell \theta) C_{\rm gg}(\ell).
\end{equation}

\subsection{Uncertainty in shear calibration, redshift and stellar-mass}\label{subsec:nofzSys}

To account for uncertainties in the shear calibration, redshift and stellar mass distributions, we make some substitutions to the equations of the previous sections, as we describe in the remaining of this section. Before we proceed, we note that aside from the marginalization of the stellar mass distribution, our approach is the same as that used in \cite{Prat2022}.\\

\noindent \textit{Shear calibration:}
\begin{equation}\label{eq:ms}
    \gamma_t^{ij}(\theta) \rightarrow (1+m_s^j) \gamma_t^{ij}(\theta),
\end{equation}
for the $i$-lens and $j-$source bin combination, where $m_s$ is the multiplicative shear calibration bias. \\

\noindent \textit{Redshift distribution (lens):}
\begin{equation}\label{eq:LensNofzFinal}
    n_\ell^i (z_\ell) \rightarrow \frac{1}{\Sigma_\ell^i} n_\ell \left( \langle z_\ell \rangle^i + \frac{z_\ell-\langle z_\ell \rangle^i}{\Sigma_\ell^i} - \Delta z_\ell^i \right) \; ,
\end{equation}
for the $i$-lens bin, where $\Delta z_\ell$ and $\Sigma_\ell$ shifts and stretches the lens redshift distributions. \\

\noindent \textit{Redshift distribution (source):}
\begin{equation}\label{eq:SourceNofzFinal}
    n_s^j(z_s) \rightarrow n_s^j \left( z_s - \Delta z_s^j \right) \;,
\end{equation}
for the $j$-source bin, where $\Delta z_s$ shifts the source redshift distributions. \\

\noindent \textit{Stellar mass distribution (lens):}
\begin{equation}\label{eq:LensNofMstarFinal}
    n_\star^i(\log M_\star) \rightarrow n_\star^i (\log M_\star - \Delta \log M_\star^i) \;,
\end{equation}
for the $i$-lens bin, where $\Delta \log M_\star$ shifts the stellar mass distributions of the lenses. 

Each of these components introduce an additional nuisance parameter per bin for us to marginalize over. In particular, when we consider jointly fitting $\gamma_{t}$ and $w$ for each combination of lens bin $i$ and source bin $j$, we marginalize over the following five parameters ($m_{s}^{j}$, $\Sigma_{\ell}^{i}$, $\Delta z_\ell^i$, $\Delta z_s^j$, $\Delta \log M_\star^i$).

\subsection{Additional modeling components}
\label{sec:addition_model}

Similarly to \cite{Zacharegkas2022}, we consider a number of additional modeling components to our data vectors $\gamma_t$ and $w$. We briefly summarize each of these components and show the full model in Figure~\ref{fig:obs_tophat_vs_1D}. \\

\noindent \textit{Lens magnification:} In addition to the distortion (shear) of galaxy shapes, weak lensing also changes the observed flux and number density of galaxies -- this effect is referred to as magnification. We only consider the magnification in flux for the lens galaxies, as that is the dominant effect in our work. In our fiducial model we ignore the contribution to galaxy clustering from magnification, as it is negligible compared to all other terms, while accounting for its contribution in galaxy-galaxy lensing, 
    \begin{equation}\label{eq:gammatLensMag}
        \gamma_t^{\rm lmag}(\theta) = 2(\alpha_{\rm lmag}-1) \int \frac{\ell d\ell}{2\pi} J_2(\ell \theta) C_{\rm gm}^{\rm lmag}(\ell) \; ,
    \end{equation}
    where we model the 2D projected power spectrum $C_{\rm gm}^{\rm lmag}(\ell)$ following \cite{Unruh2019,Zacharegkas2022} and where $\alpha_{\rm lmag}$ is a free parameter. More specifically, in our notation, for lens bin $i$ and source bin $j$, we vary one additional parameter due to lens magnification that we call $\alpha_{\rm lmag}^{i,j}$. \\
    
\noindent \textit{Tidal stripping of satellites:} As tidal forces in the outskirts of the host halo strips off the dark matter content of the satellite subhalo, the density profile of the subhalos drops off at large scales. Therefore, this term is usually modeled as a truncated NFW profile \citep[see e.g.][]{mandelbaum2004,velander2013} which is similar to that of the central 1-halo, $\gamma_t^{\rm c1h}$, out to the truncation radius $R$ and falls off as $\propto r^{-2}$ at larger radii $r$. For this work, we fix the truncation radius to $R = 0.4 R_{200c}$, which is roughly equivalent to having $50\%$ of the mass stripped, consistent with the average mass loss for subhaloes observed in cosmological
simulations \citep{Gao_etal_2004b,Nagai_Kravtsov_2005}. Additionally, since this is a satellite term, it needs to be multiplied by the satellite fraction $F_{\rm sat}$, and we thus write: 
    \begin{align}\label{eq:smSatStrip}
    	\gamma_t^{\rm strip}(\theta) = F_{\rm sat} \times \left\{ 
    	\begin{tabular}{l l}
    		$\gamma_t^{\rm c1h}(\theta)$ & if $r \leq R$ \\
    		$\gamma_t^{\rm c1h}(R) \left( \dfrac{R}{r} \right)^2$ & if $r > R$
    	\end{tabular}
    	\right. \; ,
    \end{align}
    where $r=r(\theta;z_\ell)$ is the radius from the center of the sub-halo at redshift $z_\ell$ that corresponds to angular scale $\theta$. There is no free parameters in this component. \\
    
\noindent \textit{Point-mass contribution:} This term quantifies the contribution to lensing by the baryonic content of the central galaxy \citep[e.g.][]{velander2013}. This term is simply modelled as a point-source term:
    \begin{align}
    	\gamma_t^{\rm PM}(\theta) =& \int dz_\ell \; n_\ell (z_\ell) \frac{\langle M_\star \rangle}{\pi r^2(\theta;z_\ell)} \int dz_s n_s(z_s) \Sigma_c^{-1}(z_\ell, z_s)  \; .
    \end{align}
    In this work, $\langle M_\star \rangle$ is the average stellar mass content in a given lens bin from Table~\ref{tab:MstarBinInfo}. The final contribution of this term to $\gamma_t$ is computed as in \cite{Zacharegkas2022}. There is no free parameters in this component. \\

\noindent \textit{Density profiles for satellite galaxies:} For the spatial distribution of dark matter and satellite galaxies within a dark matter halo we have assumed the NFW proflie \citep{navarro1996}. Thus, their Fourier transforms have analytic forms with concentration parameters $c_{\rm dm}$ and $c_{\rm sat}$, respectively for dark matter and satellite galaxies. However, the distribution of satellite galaxies is typically less concentrated than that of the dark matter \citep[e.g.][]{Carlberg1997,Lin2004,Nagai2004,Hansen2004}. To account for this, in this work we allow $c_{\rm sat}$ to be different than $c_{\rm dm}$ by introducing the free parameter $\kappa_c^{i}=c_{\rm sat}/c_{\rm dm}$ per lens bin $i$. \\
    
\noindent \textit{Intrinsic alignment:} Galaxies are not randomly oriented even in the absence of lensing. On large scales, galaxies can be stretched in a preferable direction by the tidal field of the large scale structure. On small scales, other effects such as the radial orbit of a galaxy in a cluster can affect their orientation. This phenomenon, where the shape of the galaxies is correlated with the density field, is known as {\it intrinsic alignment} \citep[IA,][]{troxel2015}. The contamination of shear by IA can become important when the source galaxies are physically close to the lenses and gravitational interactions can modify the shape of the galaxies. A good proxy for whether IA's contribution to $\gamma_t$ is significant is the measured boost factors (see Section~\ref{sec:BinSecectionAndScaleCuts} and Appendix~\ref{app:systematics} for the relevant discussions and definitions). Therefore, given our bin selection and scale cuts, which are described in Section~\ref{sec:BinSecectionAndScaleCuts}, we have ignored the contribution from IA in this work.

\section{Data}\label{sec:Data}

In this section we introduce the data used in this work. This work is based on the first three years (Y3) of DES data \citep{DESY3GOLD} and inherits all the infrastructure and value-added products associated with that dataset \citep[e.g.][]{DESY3shear,Myles2021,DESY3DeepFields}. The most novel aspect in the dataset here is the new stellar mass-selected lens sample we construct and validate in this work. The source sample follows the same selection as described in \cite{Zacharegkas2022,Prat2022}. As such, we describe in more detail the lens sample in Section~\ref{sec:LensSample} and briefly introduce the source sample in Section~\ref{sec:SourceSample}. 

\subsection{The stellar mass-selected lens sample}\label{sec:LensSample}

To generate our stellar mass-selected sample, we start from the DES Y3 \texttt{GOLD} catalog \citep{DESY3GOLD}. We first define an approximately complete galaxy sample by performing the following steps:
\begin{itemize}
\item Select all objects with i-band magnitude \texttt{sof\_cm\_mag\_corrected\_i$<23$}
\item Define footprint to only contain regions with limiting magnitude deeper than  \texttt{sof\_cm\_mag\_corrected\_i$<23$}
\item Select objects with \texttt{EXTENDED\_CLASS$\geq$2} to ensure the objects are high-confidence galaxies
\end{itemize}
This catalog is then fed into the \texttt{GalPro} \citep{Mucesh2020} package to generate photometric redshift and stellar mass for each galaxy. We briefly introduce \texttt{GalPro} and our training set in Section~\ref{sec:galpro}, then define the actual sample used for this work in Section~\ref{sec:sample}. In Section~\ref{sec:LSSweights} we describe the procedure in which we obtain weights for this lens sample to ensure it is not significantly impacted by the spatial variation of observing conditions.

\subsubsection{\texttt{GalPro} and the training set}\label{sec:galpro}

\texttt{GalPro} \citep{Mucesh2020} is a python-based machine learning code based on the Random Forest algorithm for estimating multivariate posterior probability distributions of galaxy properties. It can operate generically on a wide range of galaxy properties (e.g. redshift, stellar mass, star formation rate, metallicity), while we use it to extract redshift and stellar mass information.

The input quantities that \texttt{GalPro} accepts are \textit{luptitudes} \citep{Lupton1999}, $\mu$, their uncertainties $\sigma_\mu$, as well as \textit{lupti-colors} and their uncertainties. Luptitudes behave like magnitudes for bright photometry and
like fluxes for faint photometry, with the turning point in the behaviour determined by the softening parameter. Thus, by converting to luptitudes, we avoid introducing an additional selection effect by not discarding galaxies with negative fluxes. 

Starting from the \textit{griz} fluxes $f_\rho$, with $\rho \in \{g,r,i,z\}$, and their corresponding uncertainties $\sigma_\rho$, we can calculate the corresponding luptitudes as:
\begin{equation}\label{eq:luptitudes}
    \mu_\rho = \mu_{0,\rho} - a \sinh^{-1} \left( \frac{f_\rho}{2b_\rho} \right) \; ,
\end{equation}
where $\mu_{0,\rho} = m_{0,\rho} - 2.5 \log b_\rho$, $a = 2.5 \log e$ and $b_\rho = \sigma_{\mu,\rho} \sqrt{a}$. In these definitions, $e$ is the base of natural logarithms, $m_0 = 2.5 \log f_0$, and $f_0$ is a normalization that we set to $f_0 = 10^{12}$. The flux errors, $\sigma_\rho$, are also converted into errors in luptitude:
\begin{equation}\label{eq:LuptitudeError}
    \sigma_{\mu,\rho} = \sqrt{\frac{a^2 \sigma_\rho^2}{4b_\rho^2+f_\rho^2}}\; .
\end{equation}

For the colors, we calculate three lupti-colors $g-r$, $r-i$ and $i-z$ and their corresponding errors. These are simply calculated as $\mu_\rho-\mu_\nu$ for two bands $\rho,\nu \in \{g,r,i,z\}$ and $\sqrt{\sigma_{\mu,\rho}^2 + \sigma_{\mu,\nu}^2}$, respectively. 

Before we can apply \texttt{GalPro} to the DES data, we need to train the random forest with a training set where we have ``truth'' information for the mapping between the photometry and the desired quantities (redshift and stellar mass in our case). We choose to train our model on galaxies in the DES Y3 \texttt{Deep Fields} \citep[DF,][]{DESY3DeepFields}, matched to an overlapping subset of galaxies in the Cosmic Evolution Survey \texttt{COSMOS2020} \citep[COSMOS,][]{COSMOS2020}. In matching the two samples, we adjust the stellar mass slightly according to the difference in amplitude between the DF photometry and the COSMOS photometry. We note that this may not address the underlying cause of the mismatch, which could be coming from e.g. background subtraction.
However, as we see below, the output from \texttt{GalPro} is posteriors of stellar mass and redshift that are most likely broad enough to subsume the bias due to the photometry offset.

With this training set compiled, we can now train the \texttt{GalPro} network, keeping the default basic setting of the algorithm as described in \cite{Mucesh2020}. 
The trained random forest is then applied to all galaxies that survive the basic cuts described in the beginning of this section. The output of each \texttt{GalPro} run on a single galaxy is a 2D posterior distribution of redshift and stellar mass. 
For each galaxy we save both the mean redshift and stellar mass values, $z_{\rm m}$ and $M_{\star, \rm m}$, which is used to define the bins as we describe in Section~\ref{sec:sample}, as well as a random sample of the 2D posterior distribution, $z_{\rm samp}$ and $M_{\star,\rm samp}$. The latter, when combining a sample of galaxies, allows us to construct an estimate of the 2D distribution for the sample, as is also discussed in Section~\ref{sec:sample}.

In Appendix~\ref{app:GALPRO validation} we show a set of validation tests with our \texttt{GalPro} model -- we find that our derived redshift and stellar mass distributions shown in Figure~\ref{fig:1D_distributions_all} are unbiased in these tests. However, we also point out that our training set has an upper limit in stellar mass. This incompleteness in the training set cannot be diagnosed in the tests in Appendix~\ref{app:GALPRO validation}, but it propagates through \texttt{GalPro} and produces a sharp cutoff in the predicted stellar mass posteriors in the highest redshift bins. Due to this, we will exclude the results from fitting our model to the highest-mass bins in our fiducial analysis. In addition, we introduce nuisance parameters that shift the stellar-mass and redshift distributions (see Table~\ref{tab:smParamPriors}). These nuisance parameters could absorb any residual biases coming from the photometry mismatch between DF and COSMOS, and the upper limit of the stellar mass.

\subsubsection{The stellar mass sample}\label{sec:sample}

With the \texttt{GalPro} output from our approximately complete galaxy sample, we now define the stellar mass sample for this work. In particular, we like to define bins in both redshift and stellar mass, as we are interested in the redshift-dependence of the SHMR. For the redshift binning, we define the binning similar to that used in DES Y3 \citep[e.g.][]{Prat2022,y3-3x2pt}. That is, we bin the galaxies into four tomographic redshift bins based on their mean redshift $z_{\rm m}$, as shown in Table~\ref{tab:MstarBinInfo}. For each redshift bin, we first impose a stellar mass cut on the mean stellar mass $M_{\star, \rm m}$ so that the remaining galaxies are complete in stellar mass at that redshift\footnote{This is done using a similar procedure as that is described in Section~5.2 in \cite{Pozzetti2010}. For the four redshift bins we get $\log M_{\rm lim} = (9.5617, 9.9996, 10.3549, 10.5923)$.}. We then split the remaining galaxies into stellar mass bins again according to their mean stellar mass $M_{\star, \rm m}$ where aside from the highest mass bins, each bin contains $3 \times 10^6$ objects. 
The number of objects, number density, as well as the limits in stellar-mass per bin are shown in Table~\ref{tab:MstarBinInfo}. 
From now on, we will denote a single bin with the indices $(\ell,m)$, the first of which refers to the reshift bin and the latter to the stellar-mass bins in that $\ell$. The approximate redshift bin edges are defined, per $\ell$ index, as $\{ (0.20,0.40), (0.40,0.55), (0.55,0.70), (0.70,0.85) \}$ for $\ell = \{1,2,3,4\}$ respectively. 

\begin{figure}
  \centering
    \includegraphics[width=1.0\columnwidth]{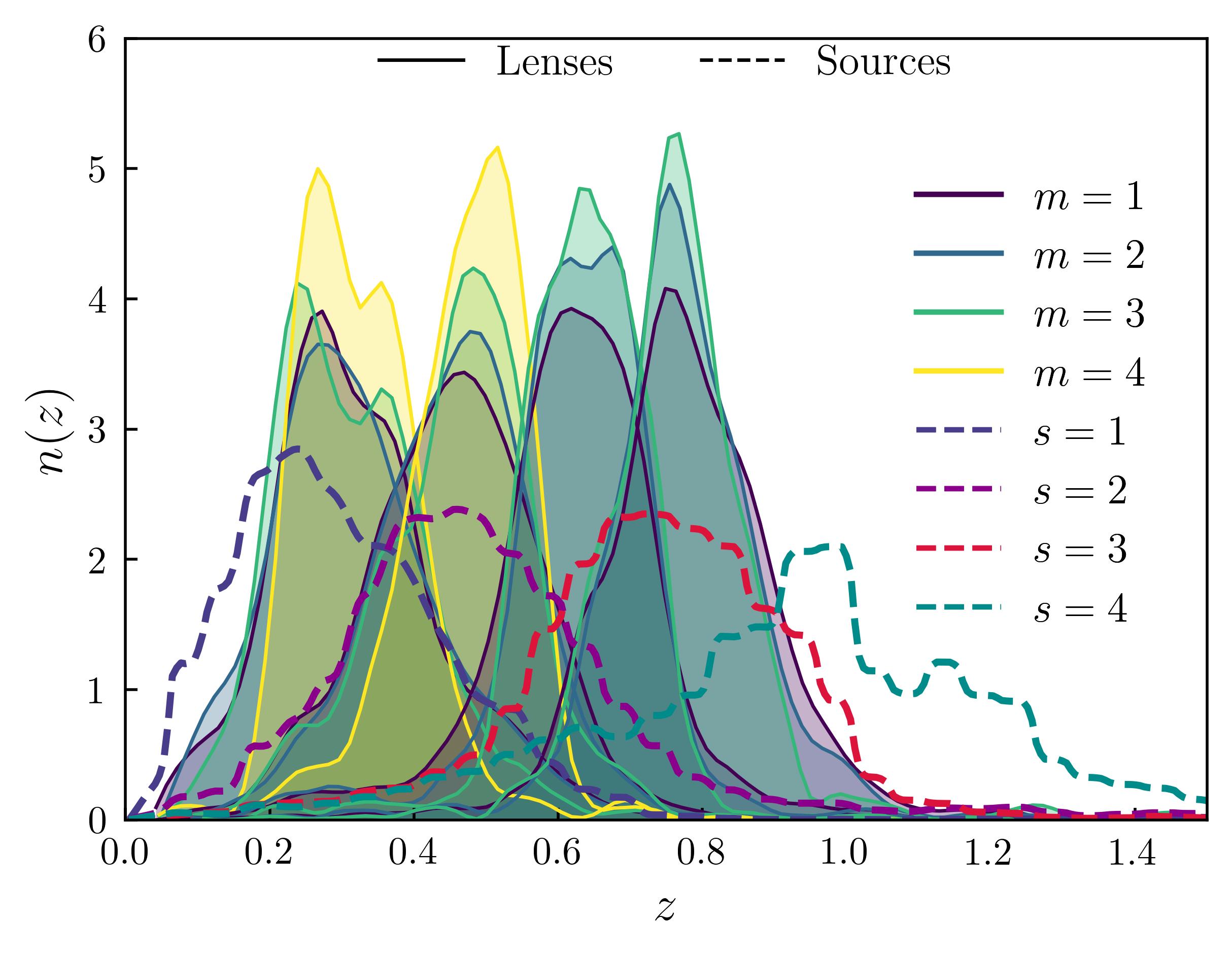}
  \includegraphics[width=1.0\columnwidth]{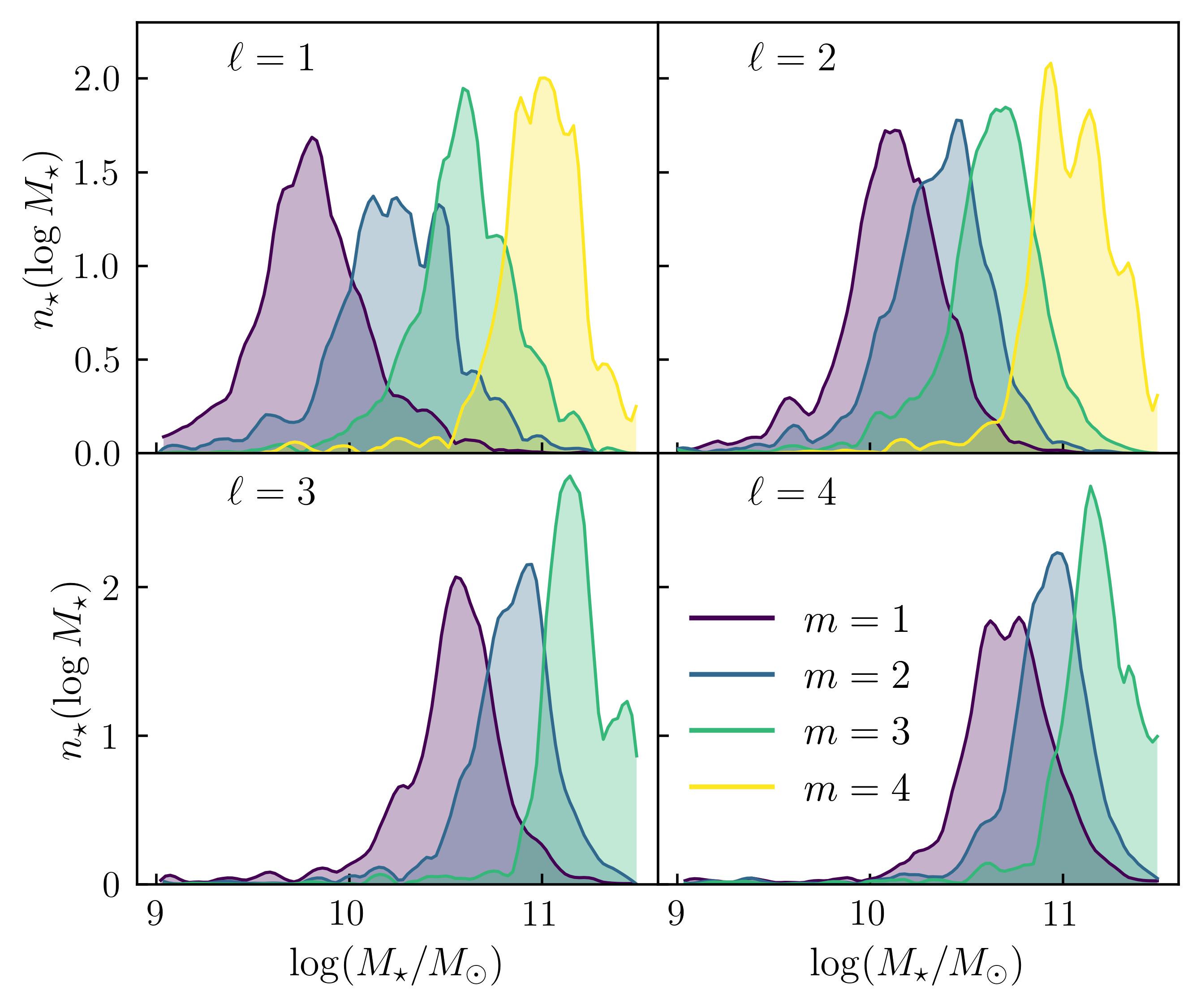}
  \caption{\label{fig:1D_distributions_all} (\textit{Upper})  Redshift distributions, $n(z)$, of the lenses (solid filled; different mass bins are plotted with different color per redshift bin) and of the source (dashed) galaxies. (\textit{Lower}) Stellar-mass distributions, $n_\star(\log M_\star)$, of our lenses. Each panel corresponds to a different redshift bin $\ell$, and in each panel all stellar-mass bins $m$ are shown.}
\end{figure}

With the bins defined, we can now stack the $z_{\rm samp}$ and $M_{\star, \rm samp}$ values to obtain the redshift and stellar mass distributions for each bin as shown in Figure~\ref{fig:1D_distributions_all}. We note that the sharp cutoff in the stellar mass distributions at the high stellar mass end is a result of the limit in the training set. This could partially be accounted for in the uncertainty in the stellar mass distribution as described in Section~\ref{subsec:nofzSys}, but still a major limitation for these bins. 
In our main results we exclude the highest-mass bins in each redshift range. 


\begingroup

\setlength{\tabcolsep}{10pt}
\renewcommand{\arraystretch}{1.5}

\begin{table*}
\centering
	\begin{tabular}{l l l l l l l l}
	    \hline
		\textrm{Redshift bin} & \textrm{Mass bin} & $N_{\rm gal}$& $n_{\rm gal}$ & $\log \langle M_h^{\rm cen} \rangle$ & $F_{\rm sat}$ & $b_{\rm gal}$ & $\chi^2_r$ \\
         & ($\log M_\star/M_\odot$) & $[\times 10^6]$  & $[{\rm arcmin^{-2}}]$ & $[M_\odot]$
        \\ \hline \hline
		$\ell=1$ $(0.20,0.40)$ & 
		$m=1$ $(9.56,9.98)$ & $3.00$ & $0.218$ & $11.430^{+0.065}_{-0.064}$ & $0.053^{+0.005}_{-0.005}$ & $0.926^{+0.010}_{-0.006}$ & $3.92$\\ 
		& 
		$m=2$ $(9.98,10.40)$ & $3.00$ & $0.218$ & $11.776^{+0.055}_{-0.054}$ & $0.219^{+0.013}_{-0.024}$ & $1.100^{+0.020}_{-0.015}$ & $2.65$ \\ 
		& 
		$m=3$ $(10.40,10.78)$ & $3.00$ & $0.218$ & $12.125^{+0.048}_{-0.045}$ & $0.233^{+0.013}_{-0.024}$ & $1.250^{+0.010}_{-0.040}$ & $3.55$ \\ 
		& 
		$m=4$ $(10.78,11.50)$ & $1.59$ & $0.115$ & $12.599^{+0.036}_{-0.042}$ & $0.249^{+0.033}_{-0.039}$ & $1.460^{+0.040}_{-0.041}$ & $2.76$ \\ 
		\hline
		$\ell=2$ $(0.40,0.55)$ & 
		$m=1$ $(10.00,10.21)$ & $3.00$ & $0.218$ & $11.699^{+0.068}_{-0.064}$ & $0.218^{+0.015}_{-0.016}$ & $1.280^{+0.030}_{-0.020}$ & $3.35$ \\ 
		& 
		$m=2$ $(10.21,10.46)$ & $3.00$ & $0.218$ & $11.853^{+0.067}_{-0.059}$ & $0.254^{+0.047}_{-0.031}$ & $1.460^{+0.100}_{-0.060}$ & $2.89$ \\ 
		& 
		$m=3$ $(10.46,10.83)$ & $3.00$ & $0.218$ & $12.122^{+0.066}_{-0.067}$ & $0.245^{+0.035}_{-0.032}$ & $1.531^{+0.020}_{-0.020}$ & $2.28$ \\ 
		& 
		$m=4$ $(10.83,11.56)$ & $1.92$ & $0.139$ & $12.689^{+0.048}_{-0.054}$ & $0.281^{+0.013}_{-0.013}$ & $1.705^{+0.020}_{-0.020}$ & $2.14$ \\ 
        \hline
		$\ell=3$ $(0.55,0.70)$ & 
		$m=1$ $(10.35,10.66)$ & $3.00$ & $0.218$ & $12.014^{+0.097}_{-0.090}$ & $0.258^{+0.048}_{-0.040}$ & $1.632^{+0.020}_{-0.030}$ & $2.18$ \\ 
		& 
		$m=2$ $(10.66,11.04)$ & $3.00$ & $0.218$ & $12.343^{+0.081}_{-0.084}$ & $0.305^{+0.033}_{-0.039}$ & $1.783^{+0.040}_{-0.030}$ & $3.25$ \\ 
		& 
		$m=3$ $(11.04,11.53)$ & $0.88$ & $0.065$ & $13.075^{+0.047}_{-0.040}$ & $0.195^{+0.062}_{-0.022}$ & $2.240^{+0.150}_{-0.110}$ & $1.95$ \\ 
		\hline
		$\ell=4$ $(0.70,0.85)$ & 
		$m=1$ $(10.59,10.79)$ & $3.00$ & $0.218$ & $12.186^{+0.089}_{-0.074}$ & $0.085^{+0.017}_{-0.019}$ & $1.651^{+0.041}_{-0.040}$ & $1.74$ \\ 
		& 
		$m=2$ $(10.79,11.05)$ & $3.00$ & $0.218$ & $12.398^{+0.050}_{-0.059}$ & $0.068^{+0.016}_{-0.018}$ & $1.782^{+0.047}_{-0.030}$ & $3.13$ \\ 
		& 
		$m=3$ $(10.05,11.63)$ & $1.26$ & $0.091$ & $12.837^{+0.079}_{-0.078}$ & $0.092^{+0.013}_{-0.025}$ & $2.486^{+0.238}_{-0.241}$ & $2.71$ \\ 
		\hline \hline
		$s=1$ & 
		 & $24.94$ & $1.476$ \\ \hline
		$s=2$ & 
		 & $25.28$ & $1.479$ \\ \hline
		$s=3$ & 
		 & $24.89$ & $1.484$ \\ \hline
		$s=4$ & 
		 & $25.09$ & $1.461$ \\ 
		\hline
	\end{tabular}
	\caption{\label{tab:MstarBinInfo}%
		Lens (indexed by $(\ell,m)$) and source (indexed by $s$) galaxy bin information. The first column corresponds to the redshift bins, while the second one corresponds to the stellar-mass bins of our lens galaxies within each redshift bin. 
		The third and fourth columns show, respectively, the number of galaxies in each of the bins and their number density in units of ${\rm gal}/{\rm arcmin}^2$. The next three columns present out constraints on the average host halo mass of the central galaxies, $\langle M_h^{\rm cen} \rangle$, the average satellite fraction, $F_{\rm sat}$, and the average galaxy bias, $b_{\rm gal}$, which are discussed in Section~\ref{sec:smResults}. In the final column we list the reduced $\chi^2$, which is a measure of the goodness-of-fit. These constraints are derived from jointly fitting our model to $\gamma_t$ and $w$, as described in Section~\ref{sec:Fits}.}
\end{table*}

\endgroup


\subsubsection{Weights} 
\label{sec:LSSweights}

Thus far, we have constructed a sample of galaxies that is complete in stellar mass and binned by stellar mass and redshift. However, in most cases, the rather simplistic selection will still leave low-level spatial patterns in the galaxy distributions that are imprints of observing conditions such as depth or seeing. A common way to further remove these effects is to assign each galaxy a \textit{weight} to compensate the over- or under-estimation of the galaxy densities -- we will refer these weights as large-scale structure (LSS) weights \citep[see also][]{Rodriguez-Monroy2022,HighzSample}.

\begin{figure*}
\centering
\includegraphics[width=1.5\columnwidth]{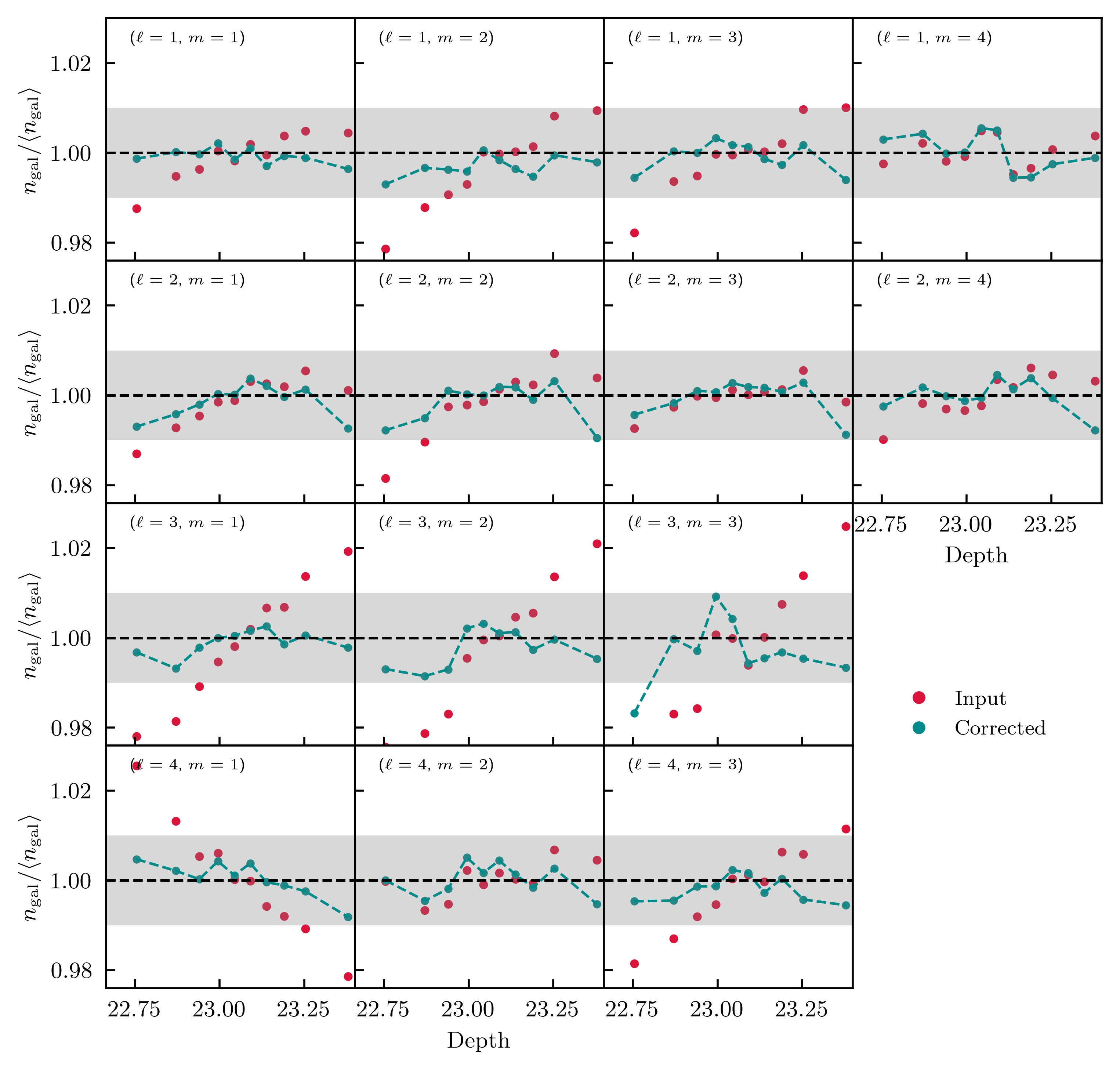}
\caption{\label{fig:NNcorrection} 
Correlations between the $i$-band depth survey property and observed galaxy density relative to the mean density over the full footprint, before (red) and after (blue) the correction using the Neural-Network weights from Section~\ref{sec:LSSweights}. We show this relationship for all $(\ell,m)$ lens bins as noted in each panel. The gray shaded region in every panel corresponds to 1\% deviation from unity.}
\end{figure*}

The methodology we implement in this work is very similar to what is done in \cite{HighzSample}. It relies on using a neural network to find the mapping between the galaxy number density and a number of survey conditions, and then correcting for them, with the underlying assumption that the cosmological LSS signal should not be correlated with observing conditions. In particular, we examine the following observational parameters via a set of survey property (SP) maps, each constructed in all four $griz$ bands:
\begin{itemize}
    \item {\bf Depth:} Mean survey depth, computed as the mean magnitude for which galaxies are detected at S/N = 10.
    
    \item {\bf Sky brightness:} Estimated sky brightness, or more precisely, the standard deviation of sky pixels due to shot noise and read noise, measured in units of electrons/second/pixel.
    
    \item {\bf Exposure time:} Total exposure time at a given point in the survey footprint, measured in seconds.
    
    \item {\bf Airmass:} Mean airmass, computed as the optical path length for light from a celestial object through Earth’s atmosphere (in the secant approximation), relative to that at the zenith for the altitude of the telescope site.
    
    \item {\bf Seeing:} Mean seeing, measured in arcseconds, computed as the full width at half maximum of the flux profile.
    
    \item {\bf FWHM:} Full width at half maximum of seeing.

    \item {\bf Galactic extinction:} We use the SFD dust extinction map from \cite{schlegel98}, which measures the $E(B-V)$ reddening, in magnitudes.
    
    \item {\bf Stellar density:} We use a map of stellar density, in ${\rm deg}^{-2}$, using stellar sources from Gaia EDR3 \citep{gaia21}.
\end{itemize}
 
We describe in more detail how the LSS weights are calculated in Appendix~\ref{sec:NNweights}. As an illustrative example, we show in Figure~\ref{fig:NNcorrection} the correction from applying the LSS weights to the correlation between the galaxy number density $n_{\rm gal}$, scaled to the average density over the full footprint, $\langle n_{\rm gal} \rangle$, and the $i$-band depth SP in Figure~\ref{fig:NNcorrection}. We can see how the weights are able to capture the non-linear relation between $n_{\rm gal}/\langle n_{\rm gal} \rangle$ and that SP and produce weights that correct for it: The input points deviate from unity significantly, especially in some bins, but the corrected ones are within the 1\% error band. 

We have examined these diagnostics for all the SP maps, and find that the LSS weights are able to capture the non-linear relation between $n_{\rm gal}/\langle n_{\rm gal} \rangle$ and that SP and produce weights that correct for it. 

\subsection{Source galaxies}\label{sec:SourceSample}

We use the DES Y3 shear catalog presented in \citet*{y3-shapecatalog}. 
The galaxy shapes are estimated using the \textsc{Metacalibration} \citep{huff2017,sheldon2017} algorithm. The shear catalog has been thoroughly tested in \citet*{y3-shapecatalog}, and tests specifically tailored for tangential shear have been presented in \citet{Prat2022}. 
In this paper, we perform additional tests on this shear catalog for tangential shear measurement on small scales (Section~\ref{subsec:smSystematics}). 

We further note that, it was recently discovered that the Year 3 DES cosmological inference results in \cite{DESY3_cosmological_inference} used a tomographic binning
of the source catalog that was inconsistent with the redshift distribution of the galaxies. The first iteration of our results also suffered from this inconsistency but we have since updated all measurements and our Bayesian analysis to use the correct version of the source catalog, also used in \cite{McCullough2024}. Comparing the two analyses, we found mostly negligible effects from updating the catalog, although some of the trends in our host halo mass, galaxy bias and satellite fraction constraints changed as a function of stellar mass and redshift. We only present results using the correct version of the catalog in this current paper.

Following \citet{Prat2022} we bin the source galaxies into four redshift bins, where details of the redshift binning and calibration is described in \citet*{y3-sompz}. The redshift distributions of the sources is shown in Figure~\ref{fig:1D_distributions_all}. The number of galaxies per source bin and the galaxy number density per square arcmin are both listed in Table~\ref{tab:MstarBinInfo}. The approximate redshift bin edges, per $s$ index, defined as $\{ (0.20,0.43), (0.43,0.63), (0.63,0.90), (0.90,1.30) \}$ for $s=\{1,2,3,4\}$ respectively.

\section{Measurements}\label{sec:smMeasurements}

\begin{figure*}
\centering
\includegraphics[width=1.8\columnwidth]{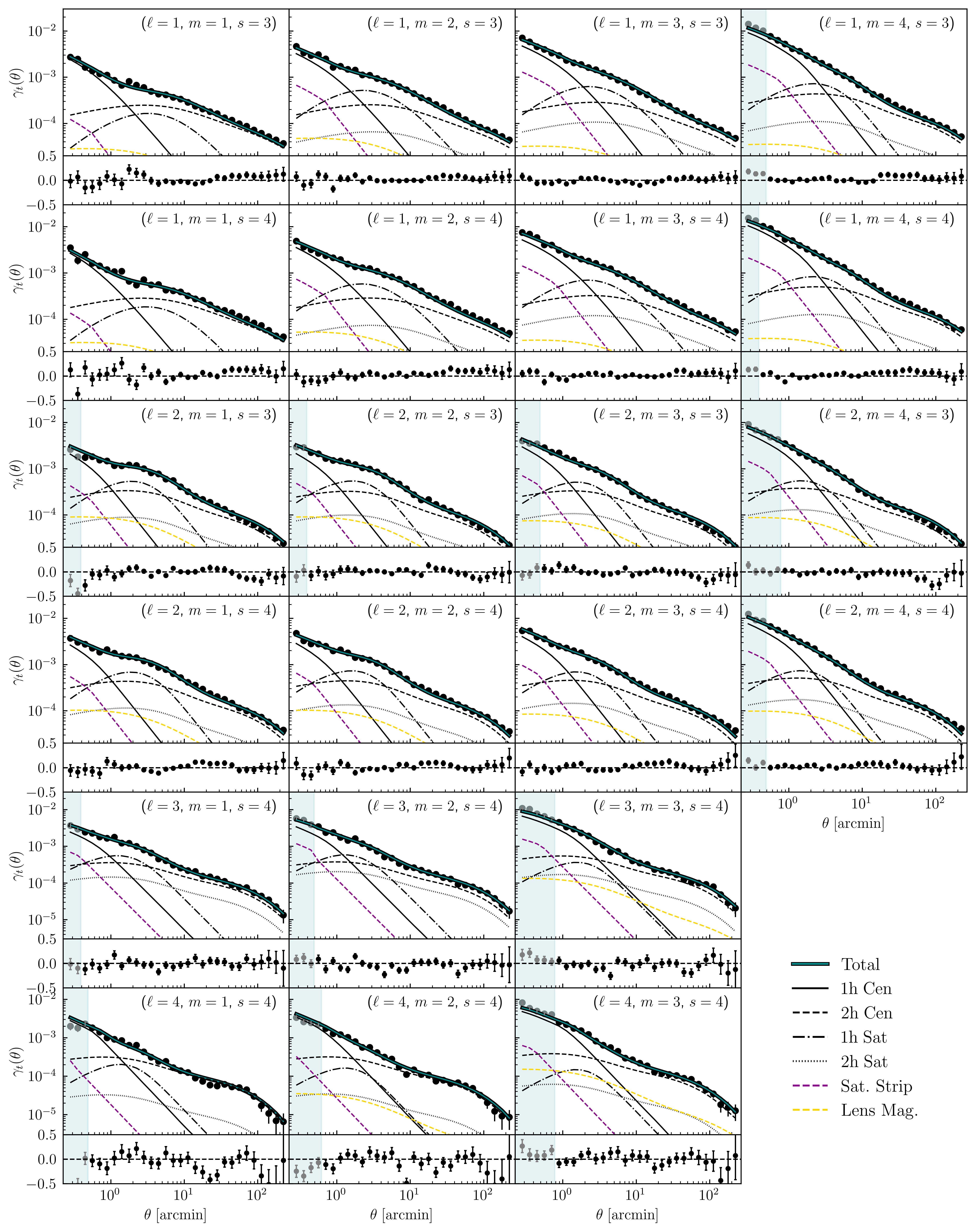}
\caption{\label{fig:GGLfits} 
Measurements (points) and best-fit model (thick solid outlined) predictions for galaxy-galaxy lensing as a function of angular scale in ${\rm arcmin}$ from jointly fitting $\gamma_t$ and $w$ of each individual panel. Each panel corresponds to a lens-source bin combination from the ones that we choose to model in this work, as indicated in parenthesis $(\ell,m,s)$. Each model component that makes up the total (outlined solid cyan) line is also shown: 1-halo central (${\rm 1h \; Cen}$, solid black), 2-halo central (${\rm 2h \; Cen}$, dashed black), 1-halo satellite (${\rm 1h \; Sat}$, dash-dotted black), 2-halo satellite (${\rm 2h \; Sat}$, dotted black), satellite stripping (${\rm Sat\; Strip}$, dashed magenta) and lens magnification (${\rm Lens Mag.}$, dashed gold). Gray points under the shaded areas are removed from the fits by the scales cuts. We also plot the residuals between the data points and the best-fit model below each panel.}
\end{figure*}

\begin{figure*}
\centering
\includegraphics[width=1.8\columnwidth]{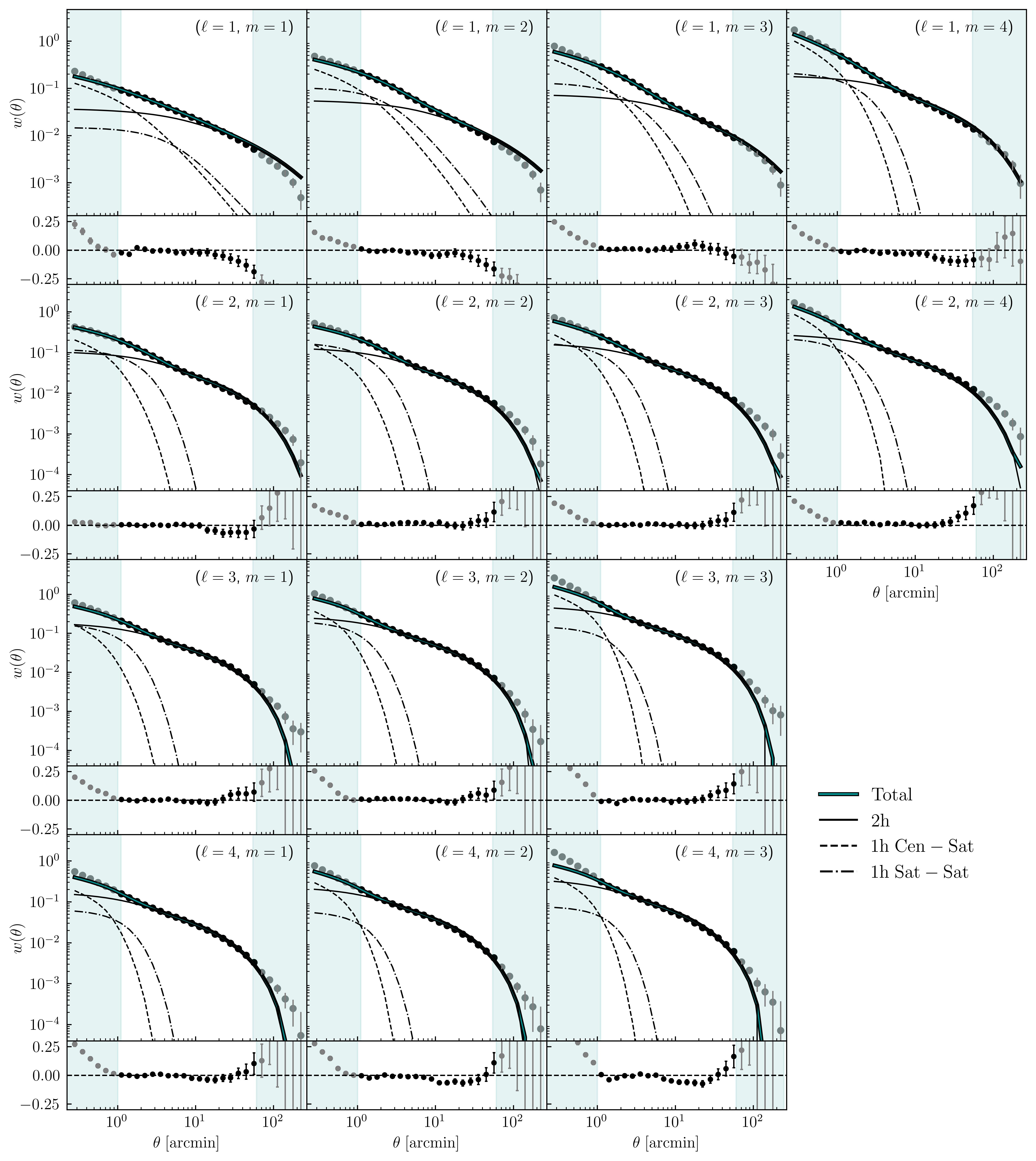}
\caption{\label{fig:GGfits} 
Measurements (points) and best-fit model (thick solid outlined) predictions for galaxy clustering as a function of angular scale in ${\rm arcmin}$ from jointly fitting $\gamma_t$ and $w$ of each individual panel. Each panel corresponds to a lens bin from the ones that we choose to model in this work, as indicated in parenthesis $(\ell,m)$. Each model component that makes up the total (outlined solid cyan) line is also shown: 2-halo (${\rm 2h}$, solid), 1-halo central-satellite (${\rm 1h \; Cen-Sat}$, dashed) and 1-halo satellite-satellite (${\rm 1h \; Sat-Sat}$, dash-dotted). Gray points under the shaded areas are removed from the fits by the scales cuts. We also plot the residuals between the data points and the best-fit model below each panel.}
\end{figure*}

In this section we discuss in detail how we measure the observables of interest, namely galaxy-galaxy lensing and galaxy clustering (Section~\ref{sec:dv}), the covariance matrices for both (Section~\ref{sec:covariance}), and tests we perform to ensure there is no significant systematic effects in our measurements (Section~\ref{subsec:smSystematics}). We then discuss our choices in bins and scales according to the results of the tests in Section~\ref{sec:BinSecectionAndScaleCuts}. 
The final measurements that will be used in this work are shown in Figure~\ref{fig:GGLfits} for galaxy-galaxy lensing and in Figure~\ref{fig:GGfits} for galaxy clustering. 

\subsection{Data vectors}
\label{sec:dv}

The galaxy-galaxy lensing estimator is the same as in \cite{Prat2022} and written as
\begin{align}\label{eq:gammat_full_estimator}
    \gamma_t(\theta) =& \frac{1}{\langle R \rangle} \left[ \frac{\sum_k w_{r_k}}{\sum_i w_{\ell_i}} \frac{\sum_{ij} w_{\ell_i} w_{s_j} e_{t,ij}^{\rm LS}(\theta)}{\sum_{kj} w_{r_k} w_{s_j}} \right. \nonumber \\
    &\qquad \left. - \frac{\sum_{kj} w_{r_k} w_{s_j} e_{t,kj}^{\rm RS}(\theta)}{\sum_{kj} w_{r_k} w_{s_j}} \right] \; ,
\end{align}
where $w_{\ell_i}$, $w_{r_k}$ and $w_{s_j}$ are the weights associated with the lens galaxy $i$, random point $k$ and source galaxy $j$, respectively. $\langle R \rangle = \sum_j w_{s_j} R_{s_j} / \sum_j w_{s_j}$ is the weighted average \textsc{Metacalibration} response. 
$e_{t,ij}^{\rm LS}$ and $e_{t,kj}^{\rm RS}$ are the measured tangential ellipticity of the source galaxy $j$ around the lens galaxy $i$ and random point $k$, respectively. Finally, the factor $\sum_k w_{r_k} / \sum_i w_{\ell_i}$ in front of the first term in brackets is called the \textit{boost factor} and is discussed in more detail in Appeddix~\ref{app:systematics}.

For galaxy clustering we adopt the Landy-Szalay \citep{landy93} estimator, written as:
\begin{equation}\label{eq:wtheta estimator}
    w(\theta) = \frac{{\rm DD} - 2{\rm DR} + {\rm RR}}{{\rm RR}},
\end{equation}
where ${\rm DD}$, ${\rm DR}$ and ${\rm RR}$ are normalized weighted counts of galaxy-galaxy, galaxy-random point and random point-random point pairs within the angular bin $\theta$. 

For both $\gamma_t(\theta)$ and $w(\theta)$ we use $N_\theta = 30$ angular bins in the range of $[0.25,250] \; {\rm arcmin}$. 
Each galaxy $i$ in our sample is weighted by $w_{g,i}$ as described in Section~\ref{sec:LSSweights}. Our catalog of random points contains $30$ times the number of galaxies in our lens galaxy catalog. 

\subsection{Covariance matrix}\label{sec:covariance}

In this work we use the Jackknife (JK) covariance matrix, which we calculate as
\begin{align}\label{eq:JKCovariance}
\mathcal{C}_{ij} \equiv \mathcal{C}(\gamma_t(\theta_i),\gamma_t(\theta_j)) = \frac{N_{\rm JK}-1}{N_{\rm JK}} \sum \limits_{k=1}^{N_{\rm JK}} \Delta \gamma^k_i \Delta \gamma^k_j \; ,
\end{align}
where $\gamma_t^k(\theta_i)$ is the shear in the $i$'th angular bin for the $k$'th JK resampling, $\langle \gamma_t(\theta_i) \rangle_k$ is the average over all $N_{\rm JK}$ realizations of the shear for the $i$'th angular bin and we have defined $\Delta \gamma^k_i \equiv \gamma_t^k(\theta_i) - \langle \gamma_t(\theta_i) \rangle_k$. We use $N_{\rm JK}=200$ JK patches defined via the \textsc{kmeans}\footnote{\texttt{https://github.com/esheldon/kmeans$\_$radec}} algorithm. $N_{\rm JK}$ is chosen so that the individual JK regions are at least as large as the maximum angular scale we need for our measurements.

When inverting the covariance matrix in the likelihood analysis, a correction factor is needed to account for the bias introduced from the noisy covariance \citep{Friedrich2016}. This correction is often referred to as the Hartlap \citep{hartlap2007} correction. When inverting the JK covariance matrix $\mathcal{C}$ we multiply it by a factor $H$ to get the unbiased covariance \citep{kaufmann1967}
\begin{align}\label{eq:HartlapCovariance}
	\mathcal{C}_H^{-1} = H \mathcal{C}^{-1} = \left(\frac{N_{\rm JK} - N_\theta - 2}{N_{\rm JK}-1}\right) \mathcal{C}^{-1} \; ,
\end{align}
where $N_\theta$ is the number of angular bins we use in each model fit. 
As shown in \cite{hartlap2007}, for $N_{\theta}/N_{\rm JK}<0.8$ the correction produces an unbiased estimate of the inverse covariance matrix; in our case we find that, depending on the bin we consider, $N_{\theta}/N_{\rm JK}$ can vary between $0.215$ and $0.240$. 
In Appendix~\ref{sec:cov_test} we also compare our Jackknife covariance with an analytical Gaussian covariance calculated via the software package \texttt{CosmoCov} \citep{Krause2017_cosmolike,fang20_cosmocov2} to show that they are fairly consistent and that our main results should not be sensitive to our choice of how we model the covariance matrix.

\subsection{Systematics tests}\label{subsec:smSystematics}

We preform a series of systematics tests detailed in Appendix~\ref{app:systematics} to ensure that our data vectors are minimally affected by observational or instrumental systematic effects. 
Our findings are summarized below:
\begin{itemize}
    \item We measure $\gamma_{\times}$, the cross-component of galaxy-galaxy lensing, which is expected to be a red-flag for systematic effects if we detect a nonzero signal. For all the relevant lens-source bin pairs, we find our $\gamma_{\times}$ measurements consistent with zero with a $\chi^2/{\rm dof} \sim 1-2$ for all bins. 
    \item We test that the noise in our data vector coming from the finite number of randoms is subdominant to other sources of uncertainties. This can be seen by the fact that the $\chi^2/{\rm dof}$ between the measurements with and without the random-point subtraction is always $\sim 1-2$.
    \item We examine the effect of the LSS weights on our data vector. In particular, we show that, as expected, the LSS weights have a large impact on the large-scale $w$ data vector, but minimally affect the $\gamma_{t}$ data vectors, as the $\chi^2/{\rm dof}<1$ for all bins.
    \item We measure the boost factor and we show that, as expected, on the linear scales it is around unity and has minimal effects on $\gamma_t$. On small scales we use the fact that the boost factors deviate from unity and make a significant difference in $\gamma_t$ to determine our scale cuts on lensing, as discussed in Section~\ref{sec:BinSecectionAndScaleCuts}. In particular, we use a tolerance threshold of $10\%$, thus removing from our fits all scales with values of boost factor greater than that.
\end{itemize}

\subsection{Bin selection and scale cuts}\label{sec:BinSecectionAndScaleCuts}

The final step before we can fit our model to the measurements is to determine the data points that will go into the fits. In particular, we like to know: 1) What are the lens-source bin combinations we can most reliably model and use? 2) What are the angular scales we can most reliably model and use? We base our decisions on two factors that dominate the modeling and measurement uncertainties: the boost factors (for galaxy-galaxy lensing) and the LSS weights (primarily for galaxy clustering).

When considering boost factors, large values suggest significant overlap between the source and the lens sample in redshift. This introduces uncertainties in the modeling of IA and magnification. We choose to work only with those bins that have boost factors mostly below $\sim 10\%$ from unity. This results in lens and source redshift bin combinations that are largely separated in redshift. That is, we select redshift bin combinations in $\gamma_t$ for the following pairs: $(\ell,s) = \{ (1,3), (1,4), (2,3), (2,4), (3,4), (4,4) \}$ based on the results from Figure~\ref{fig:GGLsystematics}. In those bins we further remove the a small number of small-scale data points with boost factor values 10\% above unity.

On the galaxy clustering part, we do not remove bins, but we determine the scale cuts based on the LSS weights that are discussed in Section~\ref{sec:LSSweights} and Appendix~\ref{sec:NNweights}. 
On the large-scale part, $\theta \gtrsim 60 \; {\rm arcmin}$ in $w$ are removed. 
This is due to how the Neural Network that produces the LSS weights is trained: when the maps are split into sub-areas for training and validation, the largest scale of those sub-maps corresponds to roughly $60\; {\rm arcmin}$ and thus we do not trust that the training will be successful at scales larger than that to correct for the systematics. 
In addition, we remove the small scales in $w$ below the angular scale that corresponds to the resolution of \texttt{NSIDE=4096} of our \texttt{healpix} maps. Thus, we also remove angular scales $\theta \lesssim 0.86 \; {\rm arcmin}$ in $w$. 

The final scale and bin selection for our $\gamma_{t}$ and $w$ data vectors are shown in Figures~\ref{fig:GGLfits} and \ref{fig:GGfits}.


\begingroup

\setlength{\tabcolsep}{10pt}
\renewcommand{\arraystretch}{1.2}

\begin{table*}
    \centering
	\begin{tabular}{ l c c}
	    \hline
		\textbf{Parameter} & \textbf{Prior} & \textbf{Description} \\ \hline
        \hline
		$\sigma_{\log M_\star}$ & $\mathcal{U}[0.01,0.5]$ & {Central galaxy's stellar-mass scatter; eq.~\eqref{eqHOD:CentralThresholded}} \\
		$\alpha_{\rm sat}$ & $\mathcal{U}[0.5,3]$ & {High-mass satellite power-law index; eq.~\eqref{eqHOD:SatelliteThresholdedInit}} \\ 
        $\kappa$ & $\mathcal{U}[0,5]$ & {Halo-mass cutoff parameter in satellite term; eq.~\eqref{eqHOD:SatelliteThresholdedInit}} \\ 
        $f_{\rm cen}$ & $\delta_K(1)$ & {Central galaxy number incompleteness; eq.~\eqref{eqHOD:CentralThresholded}} \\
        $f_{\rm sat}$ & $\delta_K(1)$ & {Satellite galaxy number incompleteness; eq.~\eqref{eqHOD:SatelliteThresholdedInit}} \\
        \hline
        $B_{\rm sat}$ & $\mathcal{U}[0,20]$ & {Satellite SHMR amplitude; eq.~\eqref{eq:SatelliteSHMRinM1}} \\
		$\beta_{\rm sat}$ & $\mathcal{U}[0.1,2]$ & {Satellite SHMR power-law index; eq.~\eqref{eq:SatelliteSHMRinM1}} 
        \\
        $B_{\rm cut}$ & $\mathcal{U}[0,15]$ & {Satellite exp. cutoff SHMR amplitude; eq.~\eqref{eq:SatelliteSHMRinMcut}} \\
		$\beta_{\rm cut}$ & $\mathcal{U}[-0.5,2]$ & {Satellite exp. cutoff SHMR power-law index; eq.~\eqref{eq:SatelliteSHMRinMcut}}
        \\
        \hline
        $\log M_1$ & $\mathcal{U}[11,13]$ & {Central SHMR characteristic halo mass; eq.~\eqref{eq:SHMRdef}} \\
        $\log \epsilon$ & $\mathcal{U}[-2,-1]$ & {Central SHMR characteristic value; eq.~\eqref{eq:SHMRdef}} \\
        $\alpha$ & $\mathcal{U}[-3,-1]$ & {Central SHMR low-mass power-law index; eq.~\eqref{eq:SHMRdef}} \\
        $\gamma$ & $\mathcal{U}[0,1]$ & {Central SHMR high-mass power-law index; eq.~\eqref{eq:SHMRdef}} \\
        $\delta$ & $\mathcal{U}[3,6]$ & {Central SHMR subpower-law strength at massive end; eq.~\eqref{eq:SHMRdef}} \\
        \hline
        $\kappa_c$ & $\mathcal{U}[0.1,1.3]$ & {Satellite concentration parameter; sec.~\ref{sec:addition_model}} \\
        \hline 
        $\alpha_{\rm lmag}$ & $\mathcal{U}[1,3]$ & {Lens magnification coefficient; sec.~\ref{sec:addition_model}} \\
        \hline 
        $\Delta \log M_\star$ & $\mathcal{U}[-0.15, 0.15]$ & {Lens stellar-mass uncertainty; eq.~\eqref{eq:LensNofMstarFinal}} \\
		$\Delta z_\ell$ & $\mathcal{N}(0, 0.1)$ & {Lens photo-$z$ uncertainty; eq.~\eqref{eq:LensNofzFinal}} \\
        $\Sigma_\ell$ & $\mathcal{U}[0.8, 1.3]$ & {Stretch parameter of lens $n(z)$; eq.~\eqref{eq:LensNofzFinal}} \\
		\hline
        $\Delta z_s^3$ & $\mathcal{N}(0, 0.01)$ & {Source photo-$z$ uncertainty in $3$rd bin; eq.~\eqref{eq:SourceNofzFinal}} \\
        $\Delta z_s^4$ & $\mathcal{N}(0, 0.006)$ & {Source photo-$z$ uncertainty in $4$th bin; eq.~\eqref{eq:SourceNofzFinal}} \\
		$m_s^3$ & $\mathcal{N}(-0.0255, 0.0013)$ & {Multiplicative bias in $3$rd source bin; eq.~\eqref{eq:gammatHankel}} \\
		$m_s^4$ & $\mathcal{N}(-0.0322, 0.0118)$ & {Multiplicative bias in $4$th source bin; eq.~\eqref{eq:gammatHankel}} \\
		\hline 
        \hline
	\end{tabular}
	\caption{\label{tab:smParamPriors}%
		Priors of model parameters. If the prior is flat we present its range in the brackets of the \textit{Uniform distribution}, $\mathcal{U}$, while for Gaussian priors we list the mean and variance in the parenthesis of the \textit{Normal distribution}, $\mathcal{N}$. For fixed parameters we use the \textit{Kronecker delta} notation, $\delta_K$.}
\end{table*}

\endgroup


\section{Parameter inference}\label{sec:Fits}

We continue to perform parameter inference by fitting our model described in Sections~\ref{sec:HODmodel} and \ref{sec:modelobservables} to the data vector described in Section~\ref{sec:smMeasurements}. In our fiducial model, we vary the seven SHMR parameter vector $\boldsymbol{\theta}_{\rm SHMR} = ( \log M_1, \log \epsilon, \alpha, \gamma, \delta, B_{\rm sat}, B_{\rm cut}, \beta_{\rm sat}, \beta_{\rm cut} )$. We also vary three HOD parameters $\boldsymbol{\theta}_{\rm HOD}=(\sigma_{\log M_\star}, \alpha_{\rm sat},\kappa)$, as well as the additional model-complexity parameters $\boldsymbol{\theta}_{\rm complex}=(\kappa_c, \alpha_{\rm lmag})$. Since the distribution of satellite galaxies is typically less concentrated than that of the dark matter \citep[e.g.][]{Carlberg1997,Lin2004,Nagai2004,Hansen2004}, we vary $\kappa_c$ within a range mostly below unity. For the magnification parameter $\alpha_{\rm lamg}$ we use values between $1$ (for no magnification) and $3$ in order to be broadly consistent with the constraints for the \texttt{MagLim} sample from DES Y3 \citep{DESY3_magnification}. Given that our samples are expected to be complete in our redshift-stellar mass bins, we decided to set the central and satellite parameters $f_{\rm cen}$ and $f_{\rm sat}$, respectively, to unity. 

For observational systematic effects, we further vary a set of systematic parameters for each lens-source combination $\boldsymbol{\theta}_{\rm sys}=(m_{s}^{j}, \Sigma_{\ell}^{i}, \Delta z_\ell^i, \Delta z_s^j, \Delta \log M_\star^i)$ for lens bin $i$ and source bin $j$. 
The functions for all these uncertainty parameters are defined in Section~\ref{subsec:nofzSys}. 
To the source-bin systematics parameters $\Delta z_s$ and $m_s$ we apply the same Gaussian priors as in the  DES Y3 main analyses \citep[see][]{Myles2021,MacCrann2022}. 
The priors on the lens-bin redshift uncertainty parameters $\Delta z_\ell$ and $\Sigma_\ell$ are set so that they are broadly consistent with the \texttt{MagLim} sample from DES Y3. 
More specifically, we apply a Gaussian prior on $\Delta z_\ell$ centered at $0$ with a variance of $0.1$ and a uniform prior from $0.8$ to $1.3$ to the stretch parameter $\Sigma_\ell$. Finally, we allow the stellar-mass distribution of our lenses to have a small shift, $\Delta \log M_\star$, to which we apply uniform priors from $-0.15$ to $0.15$, to account for about $0.3$ dex uncertainty on the mean stellar mass which is typical for inference of this parameter from SED fitting algorithms. This is a fairly conservative choice given our validation tests on the stellar mass catalog (Appendix~\ref{app:GALPRO validation}).

The priors on the lens magnification parameter $\alpha_{\rm lmag}$ is allowed to vary from $1$, which translates to no magnification, to $3$, which gives a lot of freedom to this parameter component. 
The parameter $\kappa_c$, described in Section~\ref{sec:addition_model}, is allowed to mostly get values less than unity since the satellite galaxies are expected to be less concentrated than central. 
The SHMR parameter priors are informed by \cite{Kravtsov2018} and allow for considerable freedom during our model fits. 
As for the HOD parameters, we set broad priors similar to relevant studies, such as \cite{leauthaud2011,zu2015}. 

A short description of all the model parameters as well as the priors on them are listed in Table~\ref{tab:smParamPriors}.

To sample the posterior of each data set we utilize the \textsc{Multinest}\footnote{\texttt{https://github.com/JohannesBuchner/MultiNest}} sampler, which implements a nested sampling algorithm \citep[see for example][]{feroz2009}, and assume a Gaussian likelihood. 
For data vector $\mathbf{d}$ of length $N_d$ and model prediction vector $\mathbf{m}$ of the same length we express the log-likelihood as
\begin{equation}\label{eq:Likelihood}
	\ln \mathcal{L}(\boldsymbol{\theta}) = -\frac{1}{2} (\mathbf{d}-\mathbf{m})^T \mathcal{C}_H^{-1} (\mathbf{d}-\mathbf{m}) \equiv -\frac{1}{2} \chi^2 \; ,
\end{equation}
where $\boldsymbol{\theta}$ is the parameter vector of our model $\mathcal{M}$ and $\mathcal{C}_H^{-1}$ is the Hartlap-corrected data covariance matrix. Notice that we have neglected the constant factors which are not useful while sampling the likelihood.

For our model fits, we report the total $\chi^2$ of our best-fit model to the data, as a measure of the goodness of fit. Alongside this we report the number of degrees of freedom (dof), which we calculate as the effective number of parameters that are constrained by the data, $N_{\rm eff} = N_{\rm par} - {\rm tr} \left[ \mathcal{C}_\Pi^{-1} \mathcal{C}_p \right]$, where $N_{\rm par}$ is the total number of parameters we vary, subtracted from the number of data points, $N_d$:
\begin{equation}\label{eq:Neff}
	N_{\rm dof} = N_{\rm d} - N_{\rm eff} = N_{\rm d} - N_{\rm par} + {\rm tr} \left[ \mathcal{C}_\Pi^{-1} \mathcal{C}_p \right] \; ,
\end{equation}
where $\mathcal{C}_\Pi$ and $\mathcal{C}_p$ are the prior and posterior covariance, respectively. We should note here that a goodness-of-fit estimation based on finding an effective number of parameters is not always straightforward when the parameters do not enter the model linearly, as discussed in Section~6.3 of \cite{Joachimi2021}. 

\section{Results}\label{sec:smResults}

We now present the main results from this work -- the model fits to our data as well as constraints on the physical quantities of our model (galaxy bias, satellite fraction, and the SHMR). We first discuss our fiducial results in Section~\ref{sec:fit} and then explore some of the dependencies of our model fit to some analysis choices in Section~\ref{sec:sensitivity}. We use our pipeline to jointly fit lensing and galaxy clustering. As we have previously shown in \cite{Zacharegkas2022}, our HOD model is capable of describing galaxy-galaxy lensing well. The two main advances compared to that work are: 1) we are using the stellar mass-selected generic galaxy sample described in Section~\ref{sec:smMeasurements}, and 2) we include galaxy clustering as part of the model fit.

\subsection{Constraints from DES Y3 galaxy clustering and galaxy-galaxy lensing}
\label{sec:fit}

Figures~\ref{fig:GGLfits} and ~\ref{fig:GGfits} show the predictions of the best-fit model to all our galaxy-galaxy lensing and galaxy clustering data vectors. All of these fits are performed on a combined data vector of $\gamma_{t}+w$. In the case of $\ell=1$ and $\ell=2$, we combine each lens redshift bin with two source redshift bins. 
This is not the case for $\ell=3$ and $\ell=4$ for which we have only one source redshift bin to combine with each lens redshift bin for lensing. The best-fit parameters are listed in Appendix~\ref{app:parameter_constraints}. 

For all the fits, we show the model broken down into the 1- and 2-halo, central and satellite terms. We find that our model generally provides a good enough description of the data points, in all lens and source pairs and on all scales, to capture the main features of the dataset. However, our model is not a perfect fit to the data, especially for $w(\theta)$. In particular, we find that to fit the 1-halo to 2-halo transition regime for low-mass low-redshift bins, it is necessary to use the prescription described in Section~\ref{sec:transition} instead of the model used in \cite{Zacharegkas2022}.

We have used the reduced $\chi^2$ values, which we list in Table~\ref{tab:MstarBinInfo}, from fitting our model to the data as a goodness-of-fit guide when determining how well the fits perform using different iterations of our fiducial model. The goodness-of-fit numbers we obtain are much larger than unity per degree of freedom, which, at first glance, implies a poor fit of the model to the data. In principle, we could add additional complexities (see Section~\ref{sec:addition_model}) to the model to make the $\chi^2/N_{\rm dof}$ closer to 1, but that would create additional degeneracies that can be difficult to interpret. Rather, we keep the most relevant and physically motivated model components here to capture the qualitative behaviors of the data. In future work, the basis HOD model would need to improve in order to obtain a better description of the data, especially for clustering. We have found that fitting $\gamma_t$ only results in much better $\chi^2$ values than jointly fitting with $w$. However, clustering controls the satellite fraction better, especially for high-mass galaxies. In more detail, at high-mass, high-redshift bins where the contribution from the 1-halo central term is smaller, the 2-halo central and satellite terms can become degenerate, and thus the fit cannot distinguish between the two components. We have seen something similar in \cite{Zacharegkas2022} where we imposed a prior on the satellite fraction. In this work, clustering helps to control the fraction of satellites due to how sensitive $w$ is to the satellite terms.

\begin{figure}
\centering
\includegraphics[width=1.0\columnwidth]{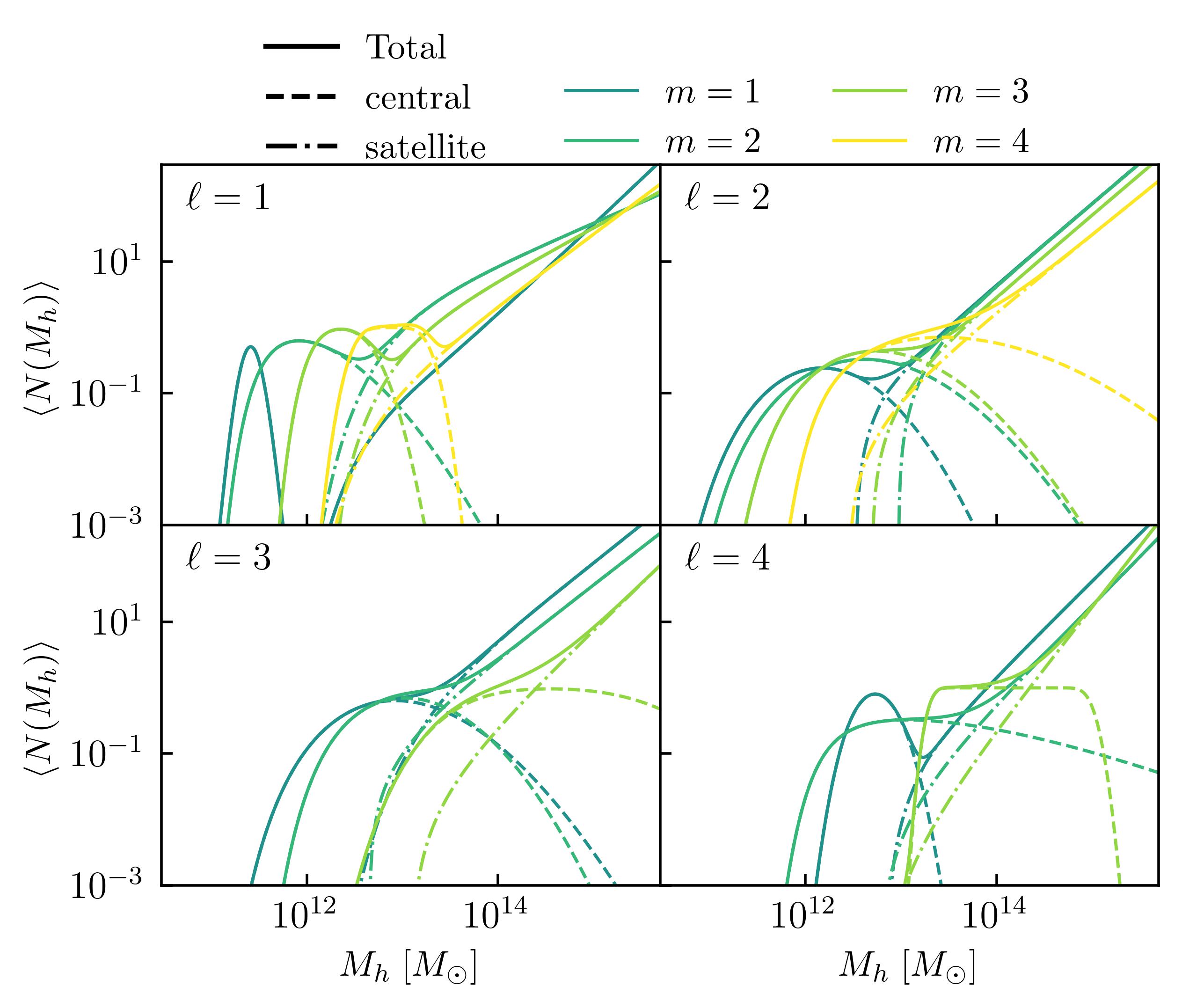}
\caption{\label{fig:HODconstraints} 
The best-fit HOD model from fitting $\gamma_t$ and $w$, as shown in Figures~\ref{fig:GGLfits} and \ref{fig:GGfits}. The total HOD is shown in solid, while the central and satellite terms are, respectively, shown in dashed and dash-dotted lines. Different colors correspond to different stellar-mass bins, as indicated in the legend, for each lens redshift bin labeled by the index $\ell$ in each panel.}
\end{figure}

In each row of Figure~\ref{fig:GGLfits}, when going from left to right we notice an increase in the 1-halo term as we would expect since the mass increases from left to right. Furthermore, the only non-negligible extra components are the satellite stripping and lens magnification. The satellite stripping becomes more important the higher the satellite fraction is, which is why it generally contributes more in low-mass bins. The lens magnification term becomes more important at higher redshifts, increasing in relative contribution at high $\ell$, as we would expect.

In Figure~\ref{fig:GGfits}, we see the increase in the amplitude of the 1-halo central-satellite term from left to right of each row, showing consistency with lensing in Figure~\ref{fig:GGLfits}. The model seems to have difficulties modeling the lowest mass bins, in particular $(\ell,m)=\{(1,1),(2,1)\}$. In those bins the 1-to-2 halo transition regime is particularly hard to model and necessitated the use of the modified model described in Section~\ref{sec:transition} in the first place.

The constraints on the mean HOD itself from the fits discussed above are shown in Figure~\ref{fig:HODconstraints}. In each panel we present the mean occupation for centrals and satellites for all mass bins $m$ at a given redshift bin $\ell$, as a function of host halo mass. The overall trend we observe is that higher stellar mass bins $m$ shift the central occupation to higher halo masses. This implies that higher stellar-mass galaxies occupy a higher-mass halo, on average. As the halo mass increases the satellite occupation also increases which means that high-mass halos host more satellite galaxies. Moreover, for cluster masses, $M_h \gtrsim 10^{14}\; M_\odot$, the halo occupation is clearly dominated by satellite galaxies. 

In addition, we notice that bins $\ell=\{1,2\}$ look qualitatively different from the last two, the former having a narrower central occupation component, and a steeper satellite term. The narrower central HOD in the lower-redshift bins can be understood by the fact that at lower redshift the presence of the 1-halo term is stronger and we can thus constrain the HOD a lot better. This is both due to the fact that as the distance to the lenses increases with redshift, a given physical scale corresponds to a larger scale, as well as due to the scale cuts that remove more of the small scales at larger redshift. The steeper satellite fraction at higher halo masses corresponds to the fact that higher-mass halos host more satellite galaxies compared to the less massive halos.

\begin{figure}
\centering
\includegraphics[width=\columnwidth]{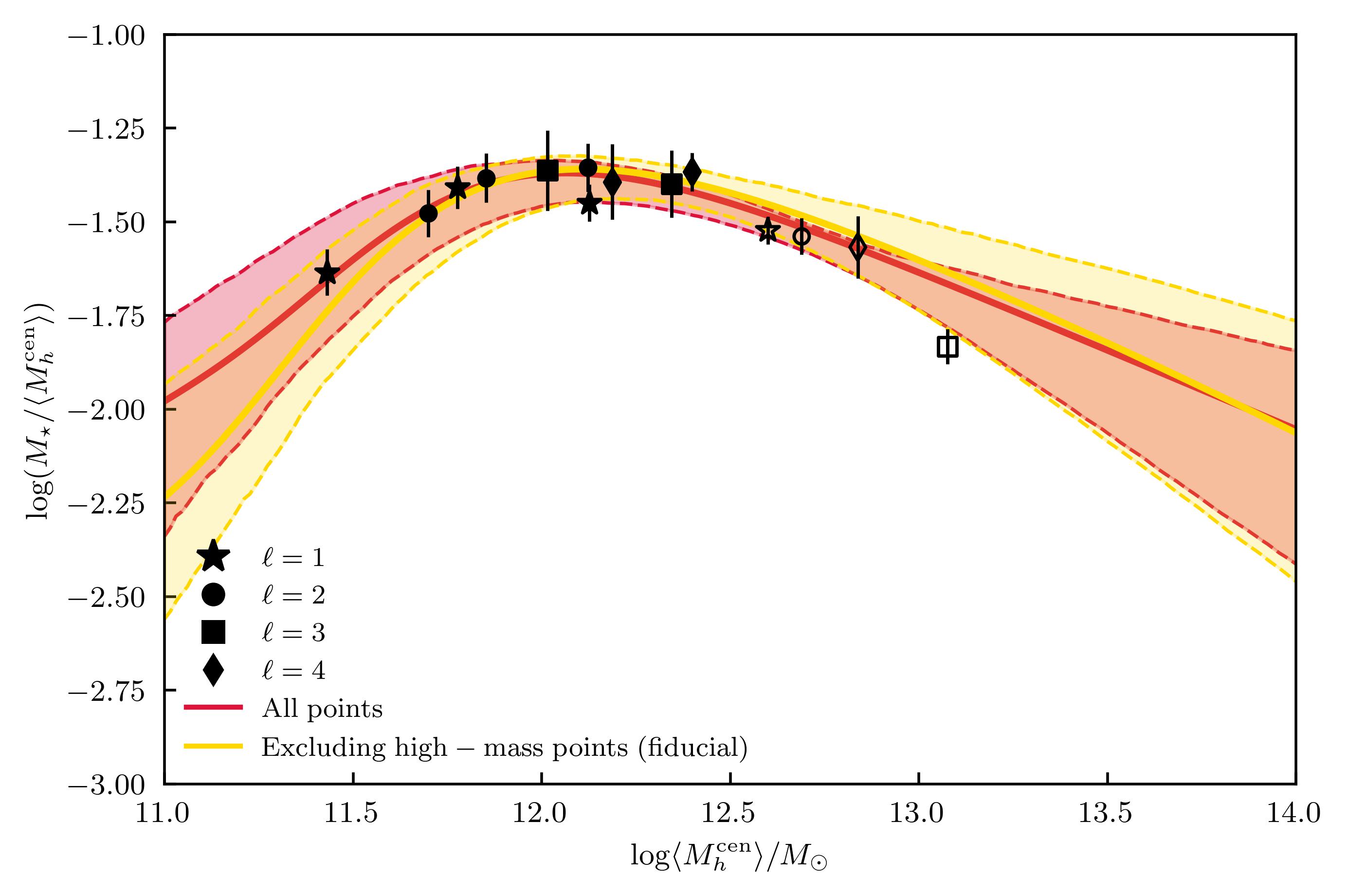}
\caption{\label{fig:shmrFittingFunctions} 
Constraints on the stellar-to-halo mass relation from our analysis (bold outlined points) from jointly fitting $\gamma_t$ and $w$. The lines correspond the best-fit lines from fitting our mode in Equation~\eqref{eq:SHMRdef} to our points. The bands around the lines show the uncertainty in these constraints. We fit separately two times, one including all data points, and one where the highest-mass points per redshift bin are removed. In the latter case we lose the constraining power on high masses under the shaded area. Hollow points mark the highest-mass bins which we exclude in our fiducial analysis.}
\end{figure}

\begingroup

\setlength{\tabcolsep}{10pt}
\renewcommand{\arraystretch}{1.5}

\begin{table*}
\centering
	\begin{tabular}{c c c c c c}
	    \hline
		& $\log M_1$ & $\log \epsilon$ & $\alpha$ & $\gamma$ & $\delta$ \\
        \hline \hline
        \textrm{Fiducial} & $11.506^{+0.325}_{-0.404}$ & $-1.632^{+0.306}_{-0.181}$ & $-1.638^{+0.108}_{-0.099}$ & $0.596^{+0.251}_{-0.210}$ & $3.810^{+2.045}_{-1.811}$ \\
        \hline
        \textrm{Including highest-mass bins } & $11.485^{+0.287}_{-0.366}$ & $-1.722^{+0.298}_{-0.211}$ & $-1.645^{+0.106}_{-0.099}$ & $0.571^{+0.222}_{-0.220}$ & $4.326^{+1.703}_{-1.512}$ \\
        \hline

	\end{tabular}
	\caption{\label{tab:SHMRfitfunc}%
		Constraints on the SHMR parameters in Equation~\ref{eq:SHMRdef}, with the priors and descriptions in Table~\ref{tab:smParamPriors}, from directly fitting the parametric form to our point-mass constraints on the average central host halo mass shown in Figure~\ref{fig:shmrFittingFunctions} and the numbers presented in Table~\ref{tab:MstarBinInfo}. The first row shows our fiducial result, while the second row shows the best-fit parameters when including the highest-mass bin per each $\ell$ redshift bin.}
\end{table*}

\begin{figure*}
\centering
\includegraphics[width=0.7\paperwidth]{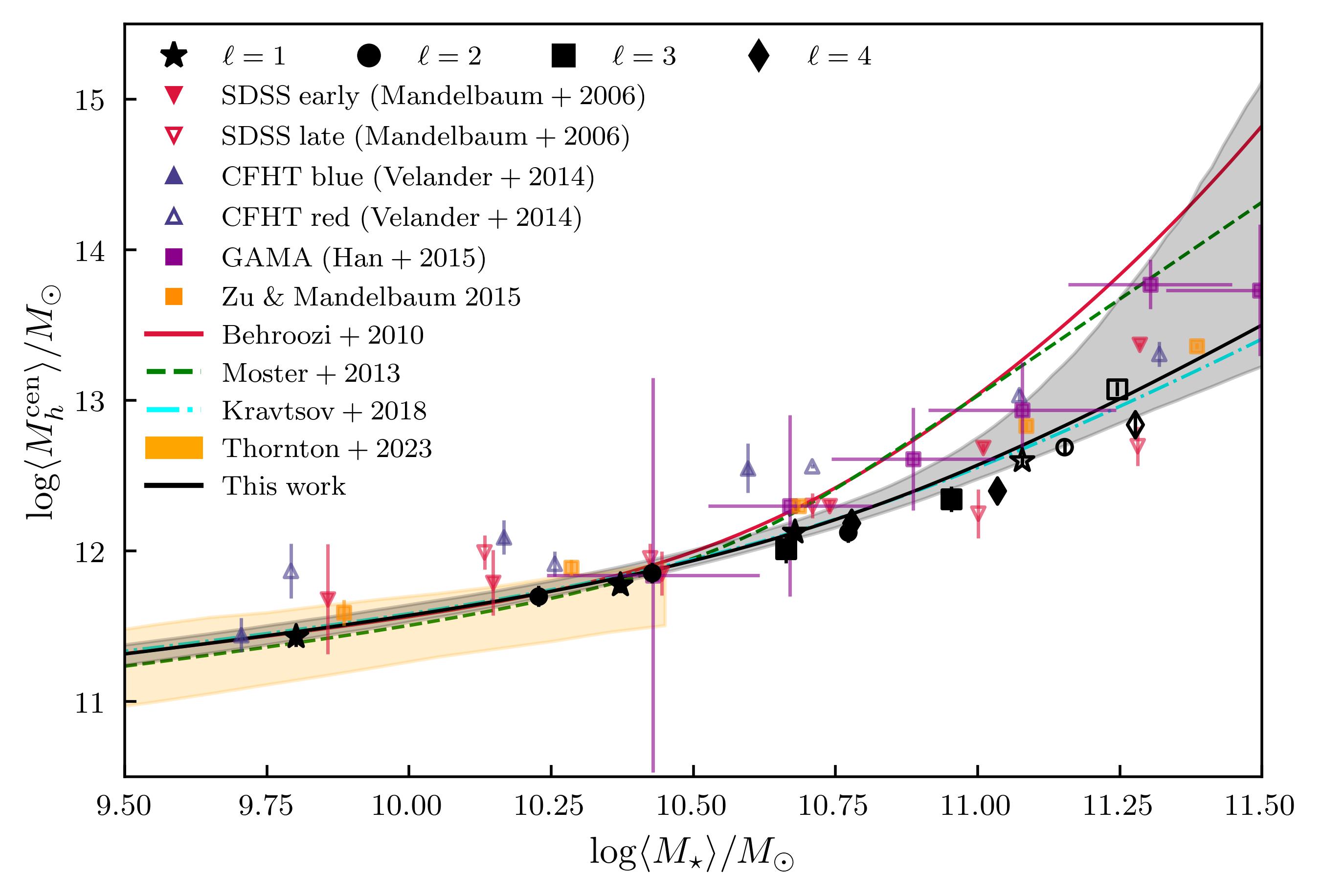}
\caption{\label{fig:shmrResults} 
Constraints on the stellar-to-halo mass relation from our analysis (black markers, split by redshift bins for different marker type, as indicated in the legend) from jointly fitting $\gamma_t$ and $w$, compared to constraints from the literature, as shown in the legend. The error bars on halo mass shows the uncertainty on our inference from the data fits. The hollow point in each redshift bin $\ell$ corresponds to the highest-mass bin $m$.}
\end{figure*}

\endgroup

\subsubsection{SHMR}
\label{subsec:results_SHMR}

We present constraints on the stellar-to-halo mass relation as inferred from our data fits in Figure~\ref{fig:shmrFittingFunctions}, while the constraints on the mean halo mass are also presented in Table~\ref{tab:MstarBinInfo}. Our best-fit results when they are inferred from our joint $\gamma_t$ and $w$ fits to the data over all scales within our scale cuts are shown in Figure~\ref{fig:GGLfits} for $\gamma_t$ and in Figure~\ref{fig:GGfits} for $w$. In each plot, the model components are shown together with the total one as outlined in the legend.

We fit the parametric form of Equation~\eqref{eq:SHMRdef} to the data points in Figure~\ref{fig:shmrFittingFunctions}, and the best-fit parameters are listed in Table~\ref{tab:SHMRfitfunc}.  The best-fit model and its 1$\sigma$ bounds on the model are shown in Figure~\ref{fig:shmrFittingFunctions}. We present two sets of parameter constraints and the corresponding best-fit lines: one where we use only the bins with reliable stellar mass estimates (Fiducial), and one where we include all bins. We find that, as expected, the two SHMR constraints differ mostly at the high-mass end, at $\log M_h/M_\odot \gtrsim 12.5$, where the constraints which include all points are more constraining. In terms of the parameter constraints, presented in Table~\ref{tab:SHMRfitfunc}, when removing the high-mass bins, we find that the strongest effect is an increase in the constraining power of $\delta$, which controls the subpower-law strength at the massive end. The rest of the parameters associated with the low-mass end are practically unaffected.

We compare our results with independent constraints from the literature, as well as with the prediction of \cite{Behroozi2013}, \cite{Moster2013}, \cite{Kravtsov2018}, and \cite{Thornton2023} in Figure~\ref{fig:shmrResults}. The filled and open blue triangles present the results from CFHTLenS \citep{velander2013}, where the samples they considered were flux-limited and were split into blue and red galaxies, respectively. The filled and open red triangles in the same plot present the constraints from SDSS \citep{Mandelbaum2006} and galaxies are, respectively, split into early and late types. The magenta squares denote the average halo masses inferred from individual galaxy stellar mass samples using the weak lensing measurements in GAMA \citep{Han2015} where the central galaxies of groups and clusters were almost volume-limited. We finally also compare our results with those from \cite{zu2015}, where a HOD model similar to this work is used to jointly fit galaxy-galaxy lensing and galaxy clustering in SDSS. We note that all those studies were carried out in similar redshift ranges where the SHMR is only weakly dependent on redshift. In addition, we find that we do not have the constraining power to independently fit to our redshift bins, whereas by combining all data points we can obtain meaningful constraints on the SHMR model parameters. Overall, these constraints are broadly in agreement. We emphasize, however, that all of these analyses make different assumptions in their models, priors and covariances, use different cosmological models, halo mass definitions, and they even differ in how they measured and reported stellar masses; thus, it is not straightforward to compare the datasets directly to make statements beyond the high-level comparisons we comment on here. To correct and account for all these differences in the reported results would be complicated and outside the scope of this work. We have, however, made sure that we convert all halo masses to the same mass definition, in particular $M_{\rm 200c}$. Below we briefly comment further on these comparisons.

Our model constraints are in excellent agreement with \cite{Kravtsov2018}, whose formulation of the SHMR is very similar to this work. In that paper, the authors used abundance matching and re-analyzed data of stellar masses of clusters in SDSS to constrain the stellar mass-halo mass relation. They argue that previous studies that reported similar constraints using abundance matching underestimated the stellar mass in clusters. That could explain why compared to \cite{Behroozi2013} and \cite{Moster2013} we find overall agreement with our constraints on the small masses, while there is tension at higher masses $\log (M_{\rm h}^{\rm cen}/M_\odot) \gtrsim 12$. This is an interesting observation, further investigation of which we leave for future work. Our results are also in good agreement with \cite{zu2015} who used a similar HOD framework to derive their constraints with galaxy-galaxy lensing and galaxy clustering. On the small-mass regime we find good agreement with \cite{Thornton2023} within 1-$\sigma$, where the SHMR was constrained using galaxy-galaxy lensing around photometrically selected dwarf galaxies. Our constraints demonstrate a different shape for the SHMR compared to \cite{velander2013}, who utilized weak lensing to constrain the galaxy-halo connection. A key difference between that study and this work is that our model is more flexible and we have combined lensing with galaxy clustering which adds significantly more constraining power. In \cite{Mandelbaum2006} the authors used weak gravitational lensing to infer the galaxy-dark matter relation; our results are broadly in agreement within 1-$\sigma$, however we note that in this work we have made various different assumptions, including the assumed cosmological parameter values.

\subsubsection{Satellite fraction}
\label{sec:fsat}

\begin{figure}
\centering
\includegraphics[width=\columnwidth]{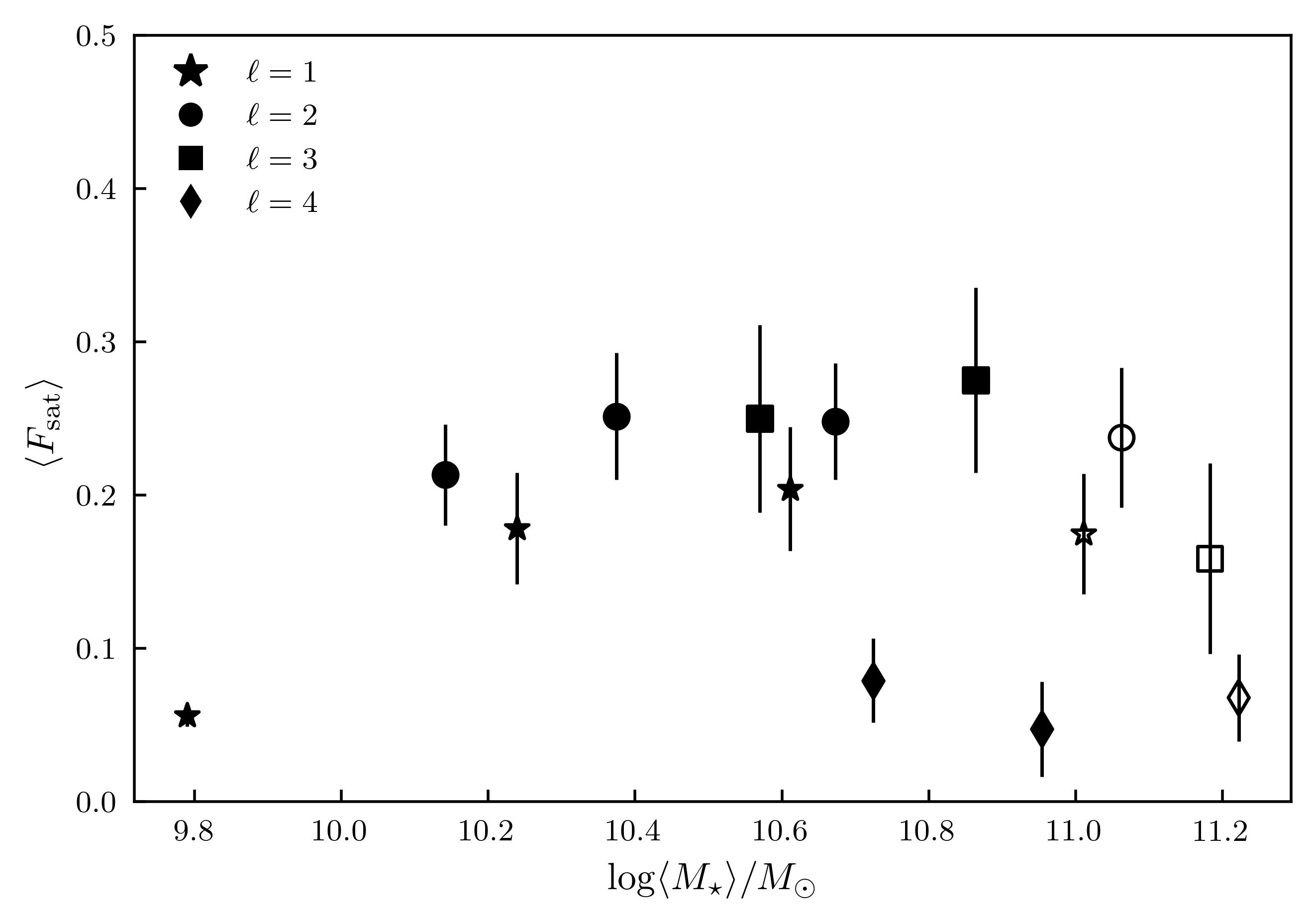}
\caption{\label{fig:fsatResults} 
Constraints on satellite fractions from the same fits as the results in Figure~\ref{fig:shmrResults}. Different markers correspond to different redshift $\ell$ bins.}
\end{figure}

Figure~\ref{fig:fsatResults} shows our constraints on the satellite fraction, as also quoted in Table~\ref{tab:MstarBinInfo}. 
The constraints on satellite fraction are generally in the range $\sim 5-35\%$ which is typical for a generic galaxy sample in the redshift range of $\sim 0-1$, excluding the open markers which we do not take into account in our fiducial results. We can see that the $\ell=4$ bins have some of the lowest satellite fractions, which is expected since very massive galaxies tend to be centrals. Besides those bins, the lowest stellar-mass bin we consider, namely $(\ell,m)=(1,1)$, has the lowest value of $\langle F_{\rm sat} \rangle$. This could potentially be a result of the difficulty of the model to fit the low-mass bins, especially with clustering, as Figure~\ref{fig:GGfits} demonstrates.

\subsubsection{Galaxy bias}
\label{sec:bgal}

\begin{figure}
\centering
\includegraphics[width=\columnwidth]{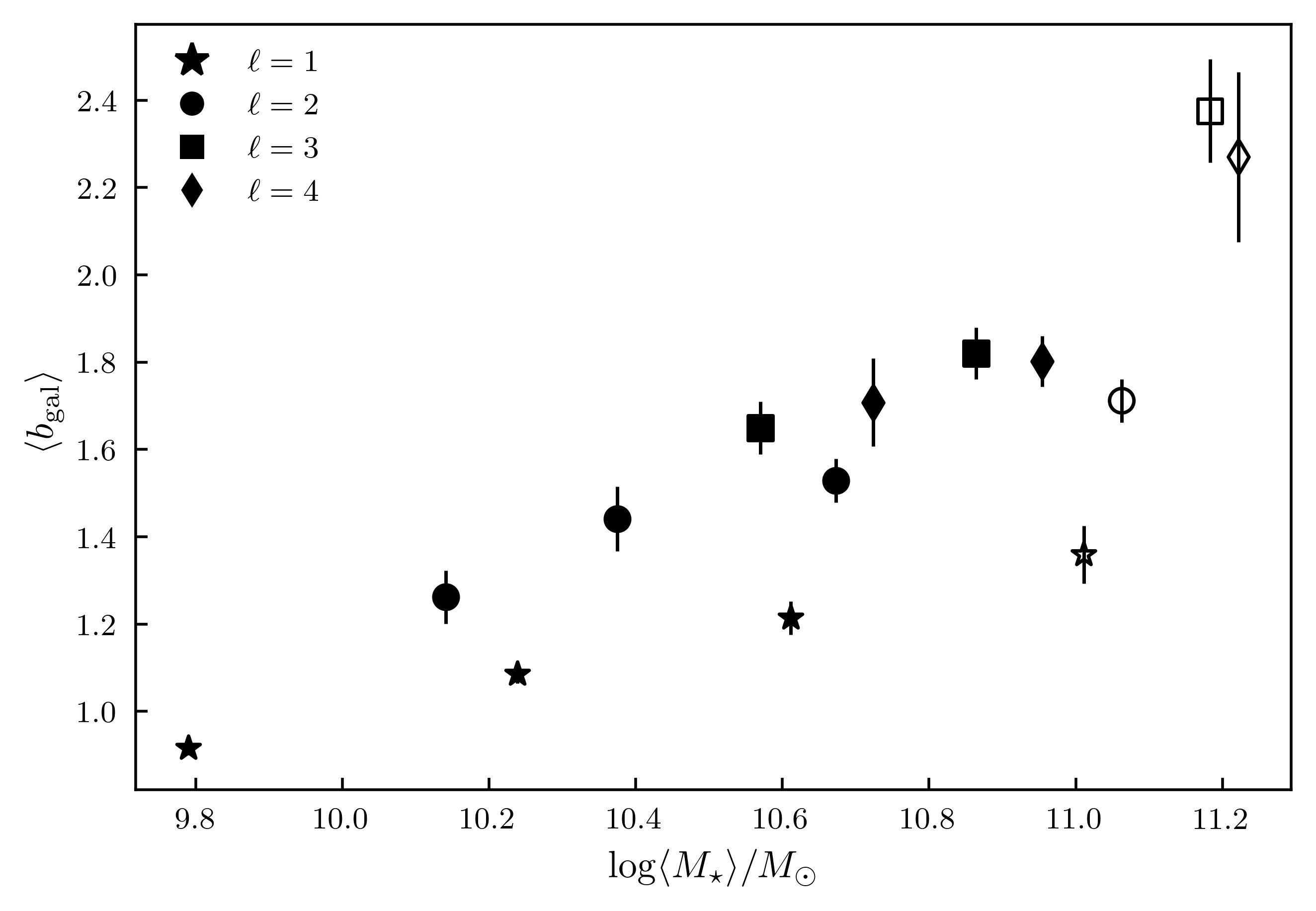}
\caption{\label{fig:bgalResults} 
Constraints on galaxy bias from the same fits as the results in Figure~\ref{fig:shmrResults}. Different markers correspond to different redshift $\ell$ bins.}
\end{figure}

We present our constraints on the galaxy bias in Figure~\ref{fig:bgalResults} and the value constraints are also shown in Table~\ref{tab:MstarBinInfo}. 
We find that the galaxy bias is generally increasing with mass, as we would expect. In addition, higher redshift bins tend to generally have higher galaxy bias values, as we would also expect.

\subsection{Sensitivity to analysis choices}
\label{sec:sensitivity}

We examine three analysis choices in our pipeline and study the sensitivity of our main results to modifying them. These three variations are: 1) exclude galaxy clustering from the fit and only use galaxy-galaxy lensing, similar to \cite{Zacharegkas2022}; 
2) assume a different set of fiducial cosmological parameters; 3) use a different SHMR functional form Equation~\ref{eq:SHMRdef} in our modeling. 
All three tests are summarized in Figures~\ref{fig:shmr_robustness} and~\ref{fig:shmr_robustness_bands}.

\subsubsection{Fitting $\gamma_t$ only}\label{sec:GAMMATrobustness}

The model in this paper is developed to fit simultaneously $w$ and $\gamma_t$. When we exclude galaxy clustering from our model fitting, the resulting halo mass constraints do not change significantly for the lowest mass bins but become underestimated for the medium-to-high mass ranges, as shown in Figure~\ref{fig:shmr_robustness} as the black points. This is related to the fact that galaxy clustering is significantly more sensitive to the satellite terms in the model than galaxy-galaxy lensing. Our results imply that with a complex galaxy-halo connection model like the one we have utilized in this paper, which has a lot of flexibility, careful choices of parameter priors are necessary. For instance, in the absence of $w$, as we have also seen in \cite{Zacharegkas2022}, the central and satellite 2-halo terms in $\gamma_t$ can be indistinguishable by the fitting algorithm and priors might need to be applied to the satellite fraction to break this behavior. When clustering is included, this is mostly avoided without additional priors because of how sensitive $w$ is to the satellite galaxies. 

In Figure~\ref{fig:shmr_robustness_bands} we present constraints on the SHMR relation from fitting to our mass point estimates. In contrast to what one might expect based on Figure~\ref{fig:shmr_robustness}, the fit to the $\gamma_t$-only points is in agreement with the fiducial constraints. This demonstrates that the shift to lower masses at $\log (M_\star/M_\odot) \gtrsim 10.5$ is not significant, thus leading to the conclusion of robustness of our fiducial SHMR results when clustering is omitted.

\subsubsection{Assumption of fiducial cosmology}\label{sec:COSMOrobustness}

As described in Section~\ref{sec:Introduction}, throughout the paper we have assumed a hybrid cosmology of DES Y3 and \textit{Planck} 2018 best-fit. We now test whether assuming only the \textit{Planck} 2018 best-fit cosmology \citep{Planck2018VI} changes the main result of the paper. The resulting changes in mean halo mass for each bin as well as the SHMR constraints are shown in Figures~\ref{fig:shmr_robustness} and \ref{fig:shmr_robustness_bands}, respectively. 
We find that the change in cosmology has negligible effect on both.

\subsubsection{Assumption of SHMR model}\label{sec:SHMRrobustness}

\begin{figure}
\centering
\includegraphics[width=\columnwidth]{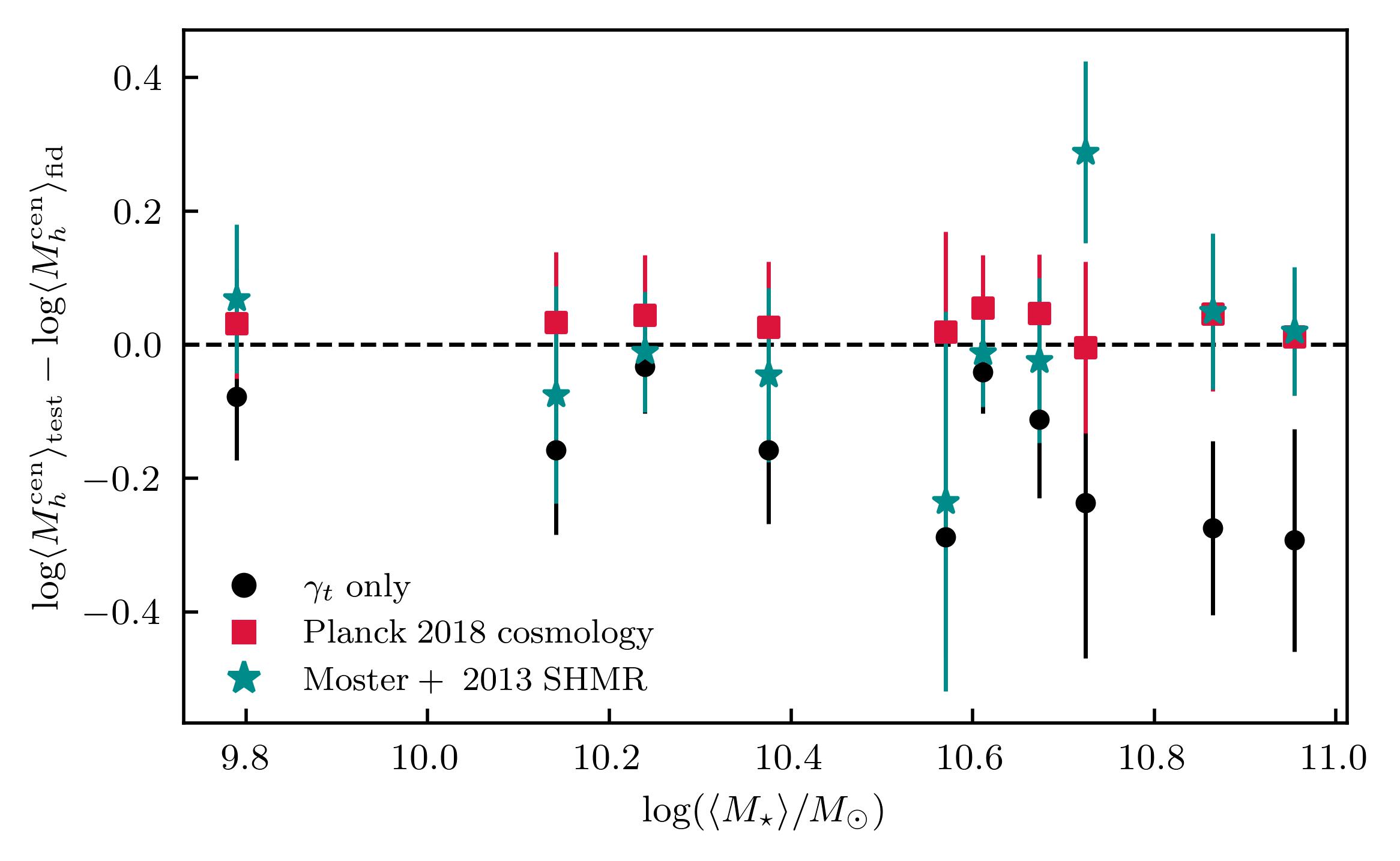}
\caption{\label{fig:shmr_robustness} 
SHMR robustness tests, presented as the difference between the logarithm of our halo mass inference from the various tests and the inference from our fiducial model. The tests shown here are: 1) Fitting only $\gamma_t$, shown in black points; 2) Repeating our fiducial analysis assuming the best-fit \textit{Planck 2018} cosmology, presented in red squares; 3) Repeating the fiducial analysis while assuming the \textit{Moster et al. 2013} SHMR functional form, shown as cyan stars. In all cases, the highest-mass bins are ignored due to the fact that we do not consider them part of our fiducial SHMR fit.}
\end{figure}

\begin{figure}
\centering
\includegraphics[width=\columnwidth]{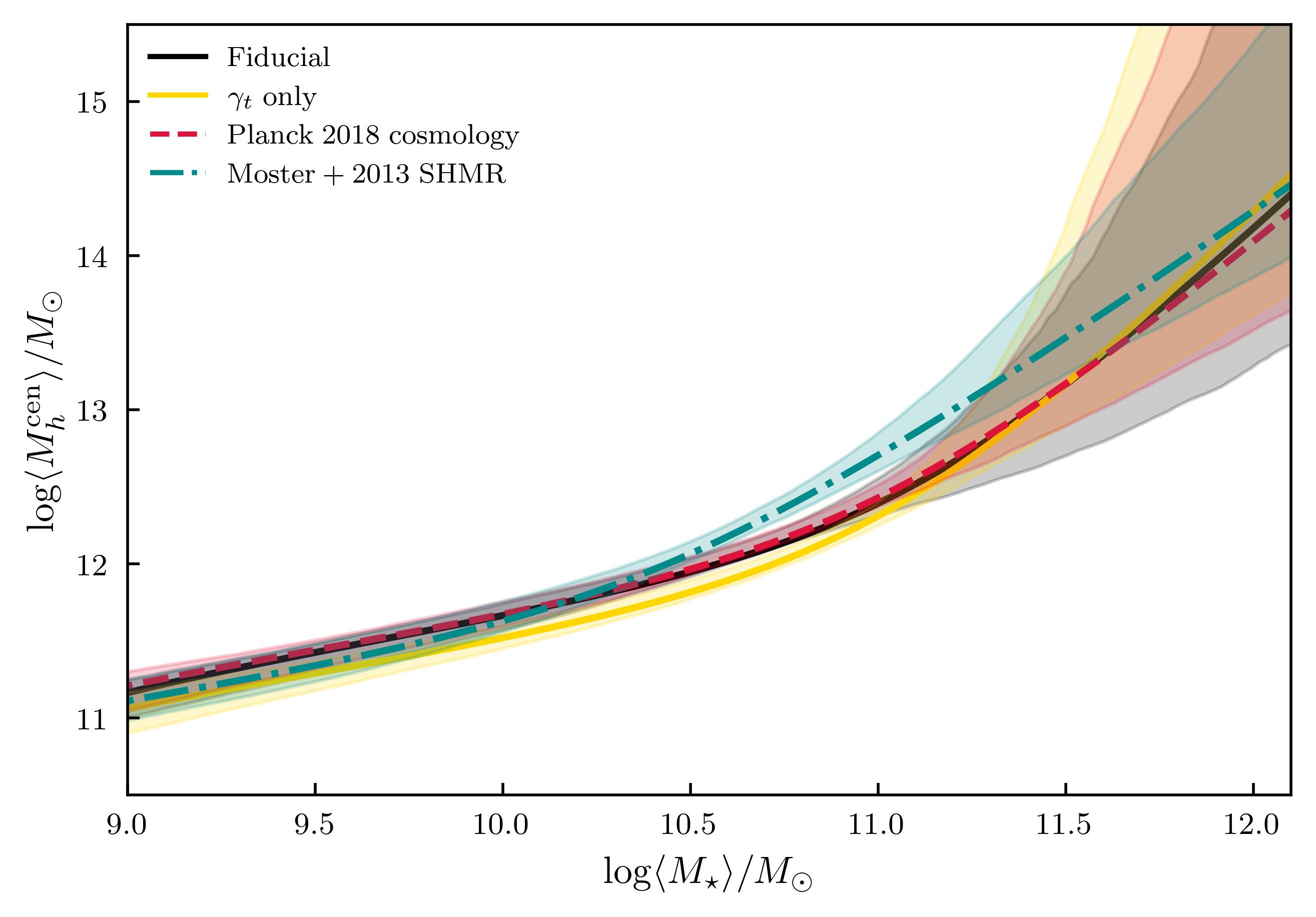}
\caption{\label{fig:shmr_robustness_bands} 
SHMR constraints from fitting the mass point estimates from our various model tests, which are also discussed in Figure~\ref{fig:shmr_robustness}. We are also showing the fit to our fiducial constraints in gold for comparison.}
\end{figure}

Throughout the paper, we assumed the SHMR model of \cite{Kravtsov2018}. There are many other parametrizations of the SHMR, and there can be significant statistical differences between
different forms of the SHMR, as pointed out e.g. in \cite{Romeo2020}, and we would like to test whether using a different model would change the main conclusion of the paper. We choose to test with the \citet{Moster2013} model. This change is both in the HOD model as well as the final fit of the SHMR to all the redshift/stellar mass bins. We show the resulting halo mass constraints as well as the SHMR fit in Figures~\ref{fig:shmr_robustness} and \ref{fig:shmr_robustness_bands}. We find that the adoption of a different SHMR model does not change significantly the constrained mass as well as the actual SHMR for the most part, except for some of the highest stellar mass bins. Given that the model is also the least constraining at high stellar mass, the resulting SHMR are largely in agreement within 1$\sigma$. The difference is about 1$\sigma$ in the intermediate range of $\log (M_\star/M_\odot) \sim 10.8$.

\section{Summary and discussion}\label{sec:Conclusions}

In this study, we jointly fit galaxy clustering and galaxy-galaxy lensing on the small, highly nonlinear scales, for a stellar mass-selected sample in the DES Y3 dataset. We use these measurements and model fits to help us interpret how these galaxies connect to their host halos. 

To do this, we first compiled and validated a new lens galaxy sample from the DES Y3 data that includes the redshift and stellar mass information of each individual galaxy. We utilized a machine learning-based software \texttt{GalPro}, which also provided us with the full 2D joint redshift-stellar mass distributions. We ran a number of tests on this sample to ensure that the stellar mass and redshift estimates are robust and the large-scale systematic effects from spatially varying observational properties are removed. This sample can be of general interest to other science applications that require stellar mass information. We measured and validated the data vectors, which consist of galaxy-galaxy lensing ($\gamma_t$) and galaxy clustering ($w$). 

In order to analyze our stellar-mass sample, we developed a HOD model that is an update from that used in \cite{Zacharegkas2022}. The new model accounts for the stellar mass information of the galaxies and further incorporate galaxy clustering into the framework. The final goal of this work is to fit the data across all redshift and stellar mass range in order to constrain the stellar-to-halo mass relation for this sample. Our main findings are the following:
\begin{itemize}
    \item We can jointly fit $\gamma_t$ and $w$ over all scales we consider after applying the necessary scale cuts.

    \item Our HOD model is able to capture the main features in the data but we found that further improvements are needed in order to improve the goodness-of-fit to the data. In particular, the model for galaxy clustering seems to be less of a good description of our data than the model for galaxy-galaxy lensing, possibly implying an insufficient HOD model and/or unmodeled systematics in $w(\theta)$.

    \item In order to obtain good data fits to all scales, especially in low-stellar mass bins, we needed to update how we model the 1-to-2-halo transition scales compared to what was utilized in \cite{Zacharegkas2022}. Our updated model required a more careful treatment of those intermediate scales, which implies that a much better understanding of the galaxy-halo connection is required in order to describe our high-precision measurements.

    \item Our final SHMR constraints are in good agreement with a large number of independent studies in the literature.

    \item The inclusion of $w$ in the fits changes and better constrains the satellite fraction in the galaxy sample.
    
\end{itemize}

Although our model is flexible enough to fit the data, we emphasize here that it can be further improved. The 1-to-2 halo regime, as mentioned above, had to be modified specifically to fit the low mass bins and we expect that further modification, e.g. by adding halo exclusion, could further add freedom to the model to fit the data better. In addition, galaxy clustering is very sensitive to the modeling choices, especially those concerning the satellite terms. Decisions regarding the halo mass function and matter power spectrum on the nonlinear and semi-linear regime can also make a difference in the data fits. In short, we find that practically each model component and every model complexity ultimately leads to a framework that needs to be calibrated specifically to fit the data at hand in the HOD framework we presented for such high-precision data. This can be seen by how swapping the SHMR, for instance, in our test in Section~\ref{sec:SHMRrobustness}, slightly changes our final results. In order to improve the fits in that case we would need to recalibrate our model, given the new model parameters and reduced flexibility from the new SHMR. In addition, we discuss how in Section~\ref{sec:GAMMATrobustness} our results are also changed when only lensing is considered. When adding new observables we see significant changes to the fits and the model also needs to be recalibrated. We saw a similar, albeit simpler, manifestation of this in \cite{Zacharegkas2022} in the case of the \textsc{redMaGiC} sample where our constraints immediately improved by adding the galaxy number density as part of the data vector. 

We have demonstrated in this paper that the HOD framework is capable of describing our complex dataset while accounting for the stellar mass of the galaxies in our sample, in a computationally efficient way. However, such a simple approach to studying the complex galaxy-halo connection has its drawbacks. For one, modeling the 1-halo-to-2-halo transition regime is usually hard and if not done properly it can lead to biased results given that these scales carry a lot of weight in the model fits. Furthermore, models for additional physics, such as halo exclusion and intrinsic alignment of the galaxies can be challenging in the HOD framework.

The fact that empirical models such as the HOD are being gradually made more complex as the data sets become more constraining could be a manifestation of the need for a more complete and physics-oriented approach to studying the galaxy-halo connection. This paper is an example of this, as we had to improve upon our previous work \citep{Zacharegkas2022} significantly in order to be able to describe our new dataset. Today, alongside the empirical models, such as the HOD, Conditional Luminosity Function (CLF) or abundance matching, techniques such as \textit{forward-modeling} are commonly utilized \citep[e.g.,][]{Hearin2017_halotools,Chittenden2022,ChangHoon2023}. The latter are based on producing synthetic galaxy catalogs to make observations on, which are then compared to real data. The simulations that produce the synthetic galaxy populations can incorporate the physics of interest, albeit limitations exist such as resolution and simulation volume, and they can be a lot more computationally expensive, and hard to fully interpret, than empirical models that depend on a few parameters and are based on a few transparent assumptions. 

Overall, the synergy of these different and complementary approaches to studying the galaxy-halo connection is crucial moving forward. We imagine the more physics-oriented forward models would depend on empirical models to inform the range of physics that are preferred by the data, while the empirical models can incorporate new physics based on any significant effects we see in the forward models. This synergy would be especially important as we usher into the new era with even more data and tighter statistical uncertainties from new surveys such as the Vera C. Rubin Observatory Legacy Survey of Space and Time (LSST), the Nancy Grace Roman Space Telescope and the Euclid mission. 

\appendix

\section{A generalized HOD model given joint redshift-stellar mass distribution}\label{app:HODexact}

In this appendix we present a more general case of our fiducial model, where the full 2D distribution of redshift and stellar mass for each lens bin is considered, similar to \cite{leauthaud2011,zu2015}. Our fiducial model is a special case in which 1) the redshift and stellar mass distributions are uncorrelated and 2) the stellar mass distribution is approximated as a tophat function. 

In this generalized model, we define the joint distribution $p(M_\star,z_\ell)$ which is the probability density function (PDF) to find a galaxy of mass $M_\star$ at redshift $z_\ell$. The normalization of $p(M_\star,z_\ell)$ is such that:
\begin{equation}\label{eq:pMsznorm}
    \int_{M_{\star,1}}^{M_{\star,2}} dM_\star \int_{z_{\ell,1}}^{z_{\ell,2}} dz_\ell \; p(M_\star,z) = 1,
\end{equation}
given a stellar mass range $[M_{\star,1},M_{\star,2}]$ and a redshift range $[z_{\ell,1},z_{\ell,2}]$. Henceforth, we will be denoting joint probability distributions by the letter "$p$", and 1D distributions by "$n$". 

We start from considering the probability $p(M_h,M_\star,z)$ that a galaxy of stellar mass $M_\star$ exists in a halo of mass $M_h$ at redshift $z$, which is normalized to unity:
\begin{equation}\label{eq:pMhMsnorm}
    \int \int \int dM_h dM_\star dz \; p(M_h,M_\star,z) = 1 \; .
\end{equation}
The goal is then to: 1) relate this to the HOD $\langle N(M_h) \rangle$ presented in Section~\ref{sec:HODtophat}, and 2) model this PDF so that we can obtain a parametrized theory prediction. In the usual HOD notation, we define the occupation probability as
\begin{align}\label{eq:HODdef}
    &\langle N(M_h|z_\ell) \rangle = n_g \left( \frac{dn}{dM_h} \right)^{-1} \nonumber \\
    &\qquad \times \int dM_\star \; p(M_h,M_\star) n_\star(\log M_\star | z_\ell) \; ,
\end{align}
where $p(M_h,M_\star)$ is the probability of finding a galaxy with $M_\star$ in a halo of mass $M_h$, $n_\star(M_\star|z_\ell)$ is the distribution of stellar mass at redshift $z_\ell$, and where we have defined $n_\star(\log M_\star) = [M_\star/\log(e)] n_\star(M_\star)$. Also, in the above, $dn/dM_h$ is the Halo Mass Function (HMF) for which we use \cite{Tinker2008} in the paper, and $n_g$ is the average number density of galaxies. In addition, in Equation~\eqref{eq:HODdef} above we can write 
\begin{align}\label{eq:pMhMstar}
    p(M_h,M_\star) &= \frac{\log(e)}{M_\star n_g} \left( \frac{dn}{dM_h} \right) \langle N(M_\star|M_h) \rangle \; ,
\end{align}
using the $\langle N(M_\star|M_h) \rangle$, the expectation number of galaxies that have stellar mass $M_\star$ in a halo of mass $M_h$. Hence, plugging Equation~\eqref{eq:pMhMstar} into \eqref{eq:HODdef} we have: \begin{align}\label{eq:HODformzavg}
    \langle N(M_h) \rangle &= \int _{z_{\ell,1}}^{z_{\ell,2}} dz_\ell\int_{M_{\star,1}}^{M_{\star,2}} d\log M_\star\; \nonumber \\
    &\quad \times \langle N(M_\star|M_h) \rangle n_\star(\log M_\star| z_\ell) \nonumber \\
    &= \int _{z_{\ell,1}}^{z_{\ell,2}} dz_\ell\int_{M_{\star,1}}^{M_{\star,2}} d\log M_\star\; \nonumber \\
    &\quad \times \langle N(M_\star|M_h) \rangle \frac{p(\log M_\star,z_\ell)}{n_\ell(z_\ell)} \; ,
\end{align}
where $n_\ell(z_\ell)$ is the redshift distribution of the galaxies. 
We present our parametrization for this model in what follows in this appendix, assuming that the HOD parameters do not evolve with redshift within a single redshift bin. 

\begin{figure}
\centering
\begin{minipage}{.45\textwidth}
  \centering
  \includegraphics[width=1.\columnwidth]{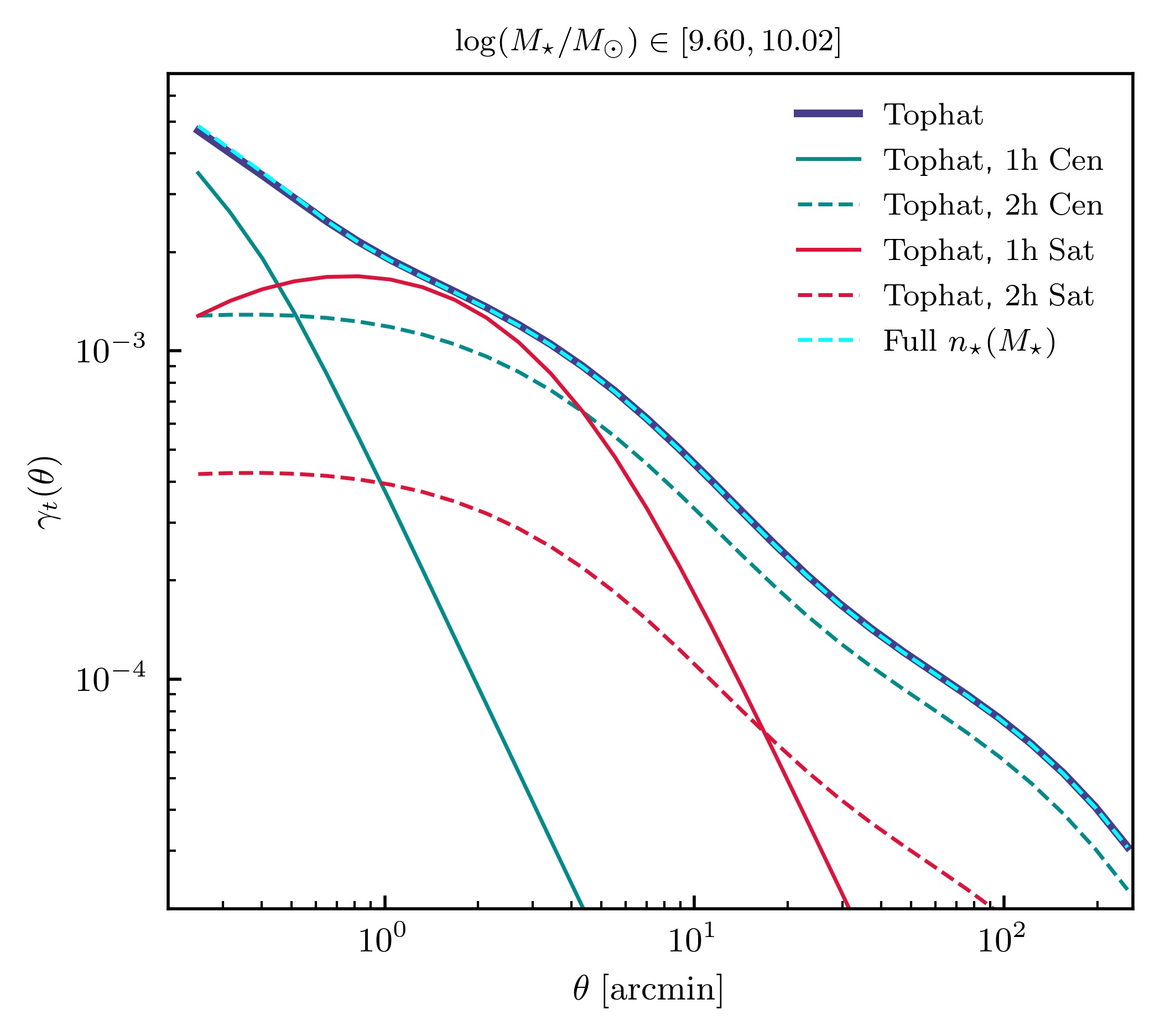}
\end{minipage}%
\hfill
\begin{minipage}{.45\textwidth}
  \centering
  \includegraphics[width=1.\columnwidth]{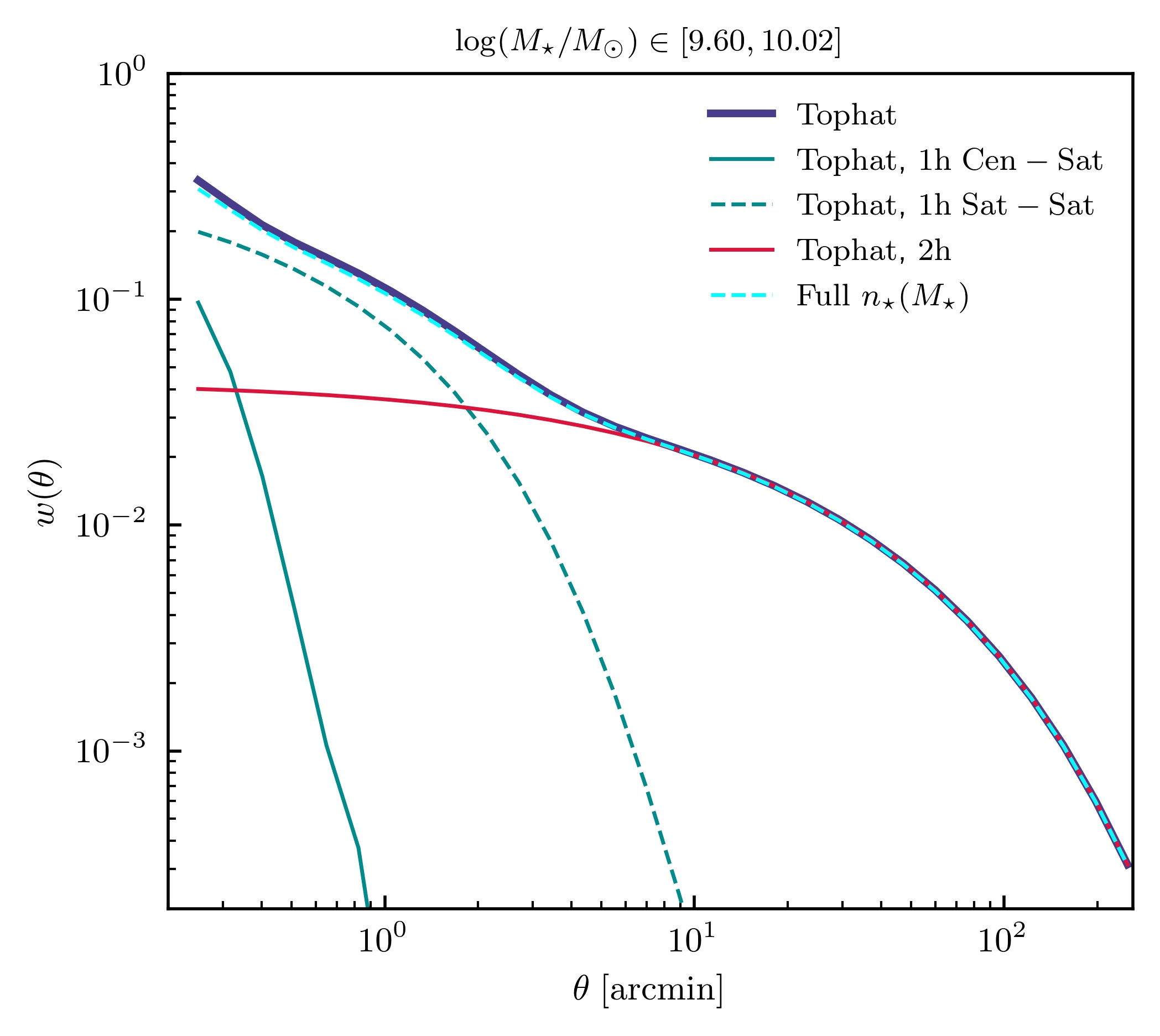}
\end{minipage}
\caption{\label{fig:tophat_vs_full_model} Predicted $\gamma_t$ (upper panel) and $w$ (lower panel) using both the tophat and full joint 2-D stellar mass-redshift distribution.}
\end{figure}

For the central occupation we assume a log-normal distribution in $\log M_\star$ with mean provided by the SHMR (see Section~\ref{sec:SHMR}) as $f_{\rm SHMR} (M_h)$ and a scatter around the mean denoted by $\sigma_{\log M_\star}$. 
We then write
\begin{align}\label{eq:HODcenLogNorm}
    &\langle N_c(M_\star|M_h) \rangle = \frac{f_{\rm cen}}{\sqrt{\pi} \sigma_{\log M_\star}} \nonumber \\
    &\quad \times \exp \left[ - \left( \frac{\log M_\star - \log f_{\rm SHMR} (M_h)}{\sigma_{\log M_\star}} \right)^2 \right] \; .
\end{align}
For the satellites galaxies we have to follow a different approach. Instead of parametrizing $\langle N_s(M_\star | M_h) \rangle$ directly, we model it as the derivative of the thresholded occupation distribution $\langle N_s(M_h | M_\star) \rangle$ (which we have modeled as discussed in Section~\ref{sec:HODtophat}) with respect to $\log M_\star$. Therefore, we have:
\begin{equation}\label{eq:HODsatCumSumDer}
    \langle N_s(M_\star|M_h) \rangle = \left. \frac{d \langle N_s(M_h|\tilde{M}_\star) \rangle}{d\log \tilde{M}_\star} \right|_{\tilde{M}_\star=M_\star} \; .
\end{equation}
We note here that, in the case that we assume a \textit{tophat}, i.e. a flat, distribution in $M_\star \in [M_{\star,1},M_{\star,2}]$, and thus write $p(\log M_\star,z_\ell) = n_\ell(z_\ell) / (\log M_{\star,2} - \log M_{\star,1})$, the above HOD reduces to the one presented in Section~\ref{sec:HODtophat}. In addition to the tophat approximation, in the above we have ignored the weak correlations between $M_\star$ and $z_\ell$, thus treating them as independent variables in $p(\log M_\star,z) \approx n_\star(\log M_\star) n_\ell (z_\ell)$. We have made sure that this does not significantly affect the predicted $\gamma_t$ and $w$, as we demonstrate in Figure~\ref{fig:tophat_vs_full_model} where we plot together the model predictions from using our decoupled tophat model and our full exact model. The differences are negligible for the purposes of this study.

\section{Validating the stellar mass catalog}\label{app:GALPRO validation}

We run two validation tests for our \texttt{GalPro}-generated stellar mass catalog based on \cite{Gneiting2007}, which we summarize below. In order to perform these tests and create the plots we present in this section, we performs a $80/20$ split to create the training and validation sets, respectively, from the matched DES Y3 \texttt{Deep Field} and \texttt{COSMOS2020} catalogs (see Section~\ref{sec:galpro}). \texttt{GalPro} is trained with the training set and the tests below are done with the validation set. See also \cite{Mucesh2020} for more details regarding the validation tests. More specifically, we test for \textit{probabilistic calibration}, using the Probability Integral Transform (PIT), and for \textit{marginal calibration}, by comparing the average predictive CDF and the true empirical CDF. We present the results, while briefly describing the process, for both tests below.

\begin{figure}
\centering
\begin{minipage}{.45\textwidth}
  \centering
  \includegraphics[width=1.\columnwidth]{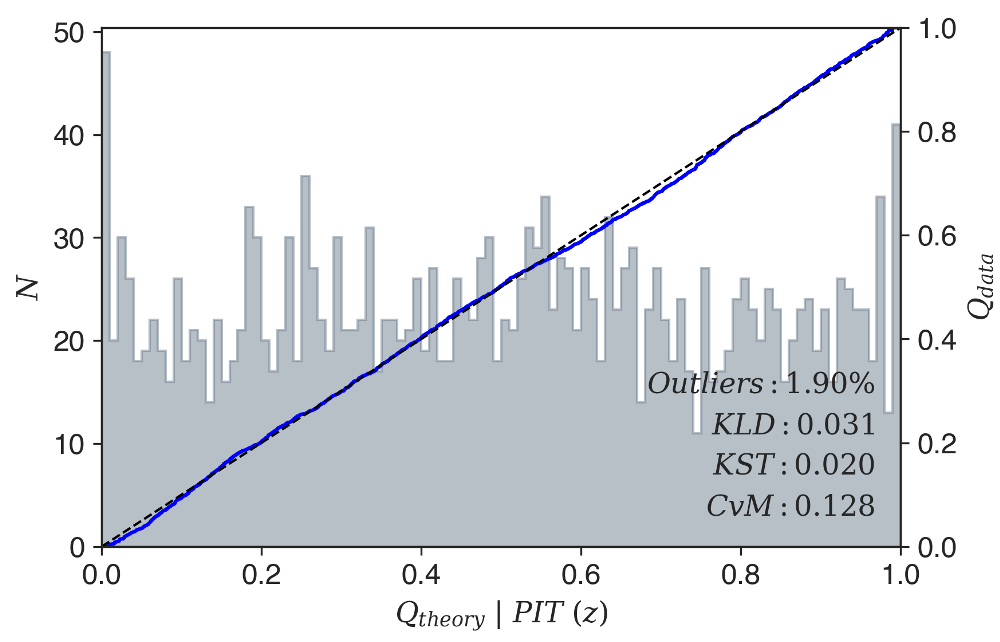}
\end{minipage}%
\hfill
\begin{minipage}{.45\textwidth}
  \centering
  \includegraphics[width=1.\columnwidth]{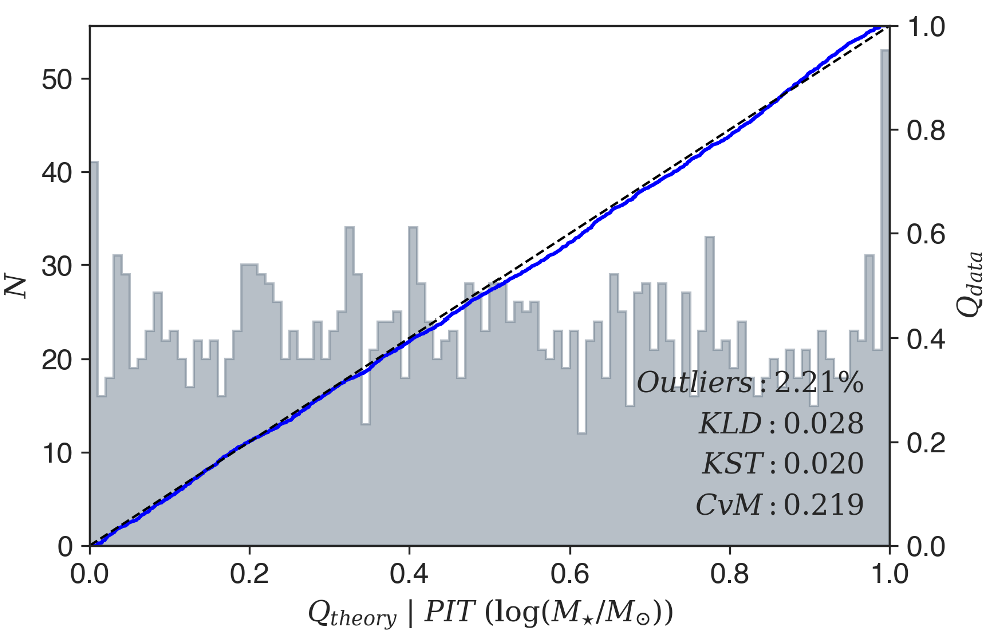}
\end{minipage}
\caption{\label{fig:PIT_validation} Redshift (upper panel) and stellar-mass (lower panel) PIT distribution plots (number of objects $N$ versus $PIT$ values) for the ML model trained to produce the DES Y3 stellar-mass sample. The quantile-quantile ($Q_{\rm theory}$–$Q_{\rm data}$) plots are also shown in each plot to highlight deviations from uniformity. The black-dashed and solid blue lines represent the quantiles of the uniform and PIT distributions, respectively. The percentage of catastrophic outliers along with the values of the Kullback–Leibler (KL) divergence, Kolmogorov–Smirnov (KS) test, and Cram\'er-von Mises (CvM) metrics are also stated to quantify uniformity of the PIT distributions. Values close to zero for each of these statistics would indicate uniformity. We define a catastrophic outlier to be any galaxy with a redshift or stellar mass completely outside the support of its marginal PDF.}
\end{figure}

\begin{figure}
\centering
\begin{minipage}{.45\textwidth}
  \centering
  \includegraphics[width=1.\columnwidth]{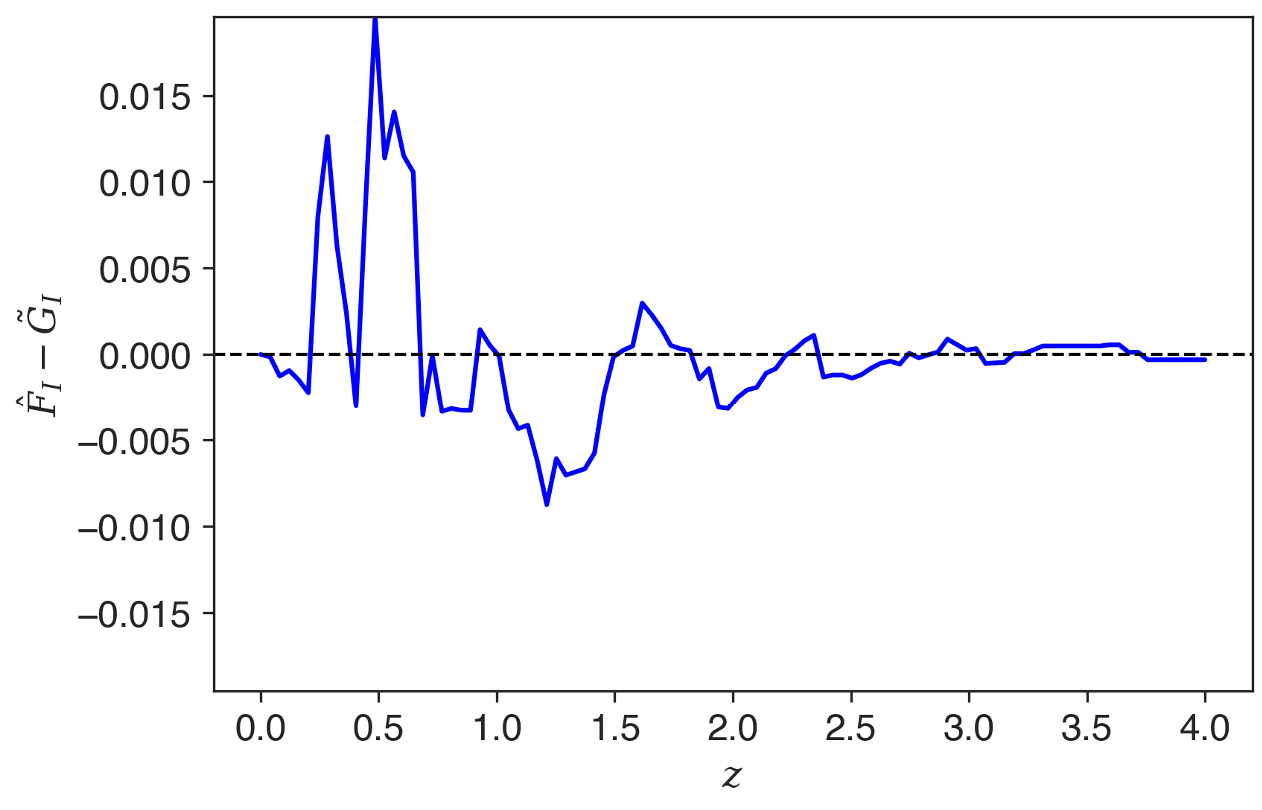}
\end{minipage}%
\hfill
\begin{minipage}{.45\textwidth}
  \centering
  \includegraphics[width=1.\columnwidth]{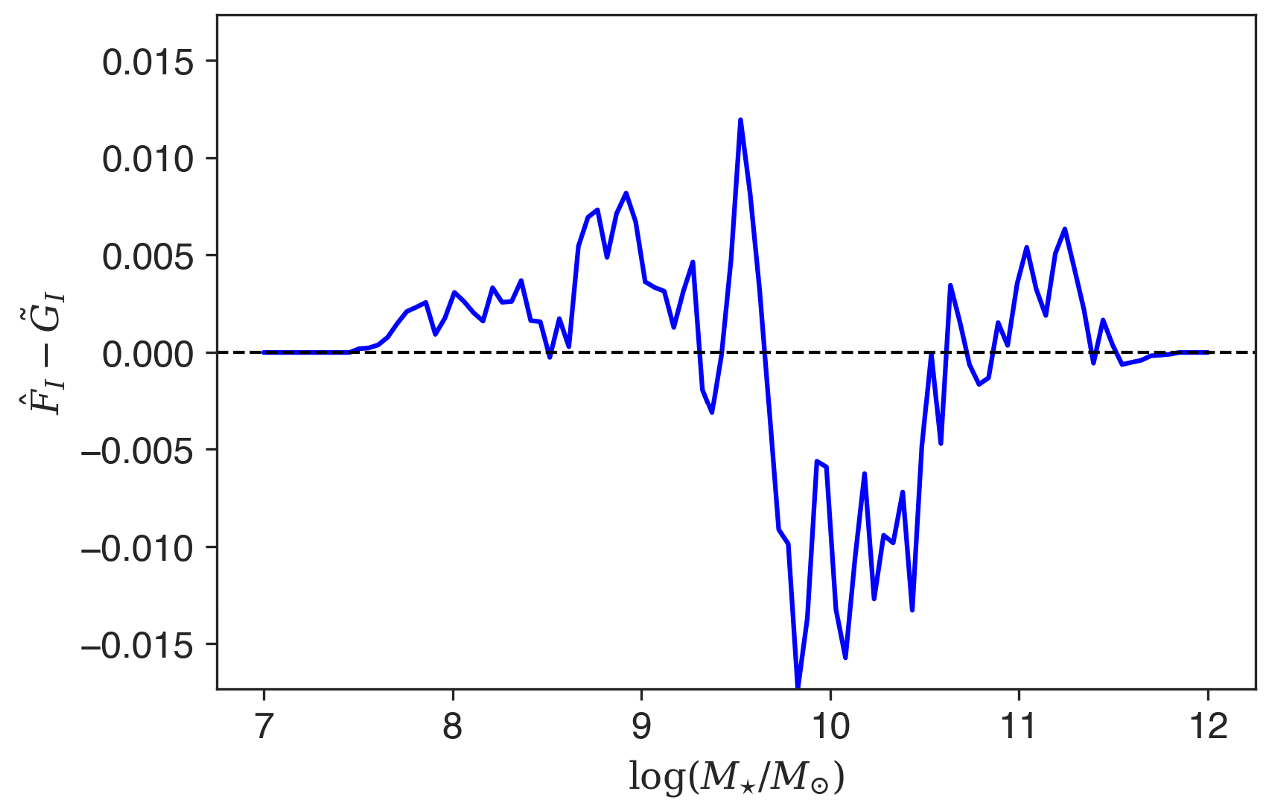}
\end{minipage}
\caption{\label{fig:FminusG_validation} The difference between the average predictive CDF ($\hat{F}_I$) and the true empirical CDF ($\tilde{G}_I$) of redshift (upper panel) and stellar-mass (lower panel) plotted at different intervals in their respective ranges, for the ML models used to create the DES Y3 stellar-mass sample.}
\end{figure}

\begin{itemize}
    \item \textbf{The Probability Integral Transform (PIT).} With this diagnostic we test for the probabilistic calibration of our marginal PDF's. In brief, if our marginal PDF's are probabilistically calibrated, the true values of the redshift and stellar mass for a galaxy should be random draws from each of their PDF. In addition, if each PDF is the true data distribution, the data random variables can be converted to random variables that follow the \textit{Uniform distribution}. 
    
    To test this we take the redshift/stellar mass PDF's generated by \texttt{GalPro} and perform the Probability Integral Transform (PIT) defined as:
    \begin{equation}
        {\rm PIT} \equiv \int_{-\infty}^{\tilde{y}} f(y) dy \; ,
    \end{equation}
    where $\tilde{y}$ is the "true" redshift or stellar mass given by DES Y3 \texttt{Deep Field} and \texttt{COSMOS2020} and $f(y)$ is the marginal PDF output from \texttt{GalPro} for a single galaxy. Then, if $f(y)$ is probabilistically calibrated, at the true values $\tilde{y}$ its cumulative distributions function (CDF) should not have a preferred value. In this case, for our ensemble of galaxies, the PIT would follow the uniform distribution $U(0,1)$. Together with the PIT distribution we present the \textit{Q-Q} plot of the PIT against the Uniform distribution. In more detail, for the Q-Q distribution plots, samples are drawn from the \textit{Quantile functions}, defined as the inverse of the cumulative functions, of the two distributions as a function of quantile and they are plotted against each other. If the two distributions are close to each other their Q-Q plot will fall on the diagonal. Any deviations from the diagonal will give us useful information about how they differ.
    
    The results of the above tests on our model are shown in Figure~\ref{fig:PIT_validation}. The PIT distributions follow closely the uniform. In addition, three statistics listed in the caption of the plot are given to quantitatively assess uniformity: the Kullback–Leibler \citep[KL]{Kullback1951OnIA} divergence, the Kolmogorov–Smirnov \citep[KS]{Kolmogorov1992OnTE} test, and the  Cram\'er-von Mises \citep[CvM]{CramerVonMises1928} test. For each of these tests, values close to zero would indicate that our PIT distribution closely matches the Uniform distribution. We conclude that our model passes this test. 
    
    We notice, however, that despite the remarkable agreement of the PIT to the uniform distribution, our PIT distributions are somewhat concave, as indicated by the small deviations of the Q-Q plots at the extremes, as well as by the significantly larger CvM values. This means that our marginal distributions are a little narrow, i.e. under-dispersed. Our systematics parameters $(\Delta \log M_\star, \Delta z_\ell, \Sigma_\ell)$, listed in Table~\ref{tab:smParamPriors}, should account for this during our model fits.

    Although the PIT test is a necessary condition for the marginal PDF to be valid, it is usually not enough to make concrete statements about how well a PDF is calibrated. In fact, it has been shown \citep{Hamill2001} that biased distributions can lead to uniformly distributed PIT's. This test is thus not sufficient and we also perform the additional test below.

    \item \textbf{Comparing the average predictive CDF and the true empirical CDF.} With this diagnotstic we test for marginal calibration of our PDF's. As mentioned above, uniformity of the PIT distribution is not enough to assess how close our marginal PDF's are to the truth. Thus, with this marginal calibration test we assess the equality of the predicted to the true redshift and stellar-mass distributions for more concrete results when combined with the probabilistic calibration test above. 
    
    For this, we compare the \textit{average predictive CDF}, $\hat{F}_I(y)$, with the \textit{true empirical CDF}, $\tilde{G}_I(y)$, respectively defined as:
    \begin{equation}
        \hat{F}_I(y) \equiv \frac{1}{n} \sum_{i=1}^{n} F_i(y)
    \end{equation}
    and
    \begin{equation}
        \tilde{G}_I(y) \equiv \frac{1}{n} \sum_{i=1}^{n} \mathbbm{1} \{ \tilde{y}_i \leq y \} \; ,
    \end{equation}
    where we have defined the predictive CDF per galaxy as $F_i$ for all $n$ test galaxies $i=1,2,...,n$, and where
    \begin{equation*}
    \mathbbm{1} \{ \tilde{y}_i \leq y \} = 
        \begin{cases}
             1, & \tilde{y}_i \leq y \\
             0, & \tilde{y}_i > y 
        \end{cases}
    \end{equation*}
    is the \textit{indicator function}. If the PDF's are marginally calibrated, the difference between $\hat{F}_I$ and $\tilde{G}_I$ should only slightly deviate around zero. 
    Figure~\ref{fig:FminusG_validation} shows the results from this test for our model, where it can be seen that the maximum deviation between $\hat{F}_I$ and $\tilde{G}_I$ is $\sim 0.015$. Thus, our model passed this test as well.
\end{itemize}

\section{Computing the LSS weights}\label{sec:NNweights}

In this appendix we discuss in more detail how we produce the LSS weights described in Section~\ref{subsec:smSystematics}. There are a total of 26 SP maps per lens redshift-stellar mass bin (4 maps each for depth, sky brightness, exposure time, airmass, seeing and FWHM; 1 map for Galactic extinction; 1 map for stellar density). We take these maps as the {\it features} and we train a neural network to model the correlations between those maps and galaxy count maps, which play the role of the {\it labels}. The main advantage of using a neural network here is that it is able to model nonlinear relations between the SP maps and the galaxy count maps. We note here that in order to avoid the network learning about the cosmological clustering of galaxies in the count maps, each pixel in the latter maps have a label of either 0 (empty pixel) of 1 (occupied by $\geq 1$ galaxy pixel). The \textit{loss function} for the network is the binary \textit{cross-entropy} between the predicted pixel occupancy and the occupancy of the training set. 

The architecture of the network is based on our assumption that the selection function scales primarily as some power law combination of the SP's. To this end, the input SP values are all logarithmically scaled (except those, such as depth, which are already logarithmic quantities), and the output of the network is exponentiated to form the selection probability. The network output is a sum of two branches: the first branch is a simple linear combination of the 26 scaled SP’s, since we expect this to capture most of the functional variation. The second branch is intended to capture departures from a simple power law: it takes the input layer of 22 dimensions through 3 hidden layers of 64, 32 and 4 fully connected neurons, respectively, and a single neuron on the output layer, each with \textit{relu activation}. The output of the network, for each tomographic bin, consists of a single value for each \texttt{Healpix} pixel within our mask, which will be used to weigh the galaxies accordingly.

To prevent the network from overfitting, it is constructed with \textit{k-fold cross-validation}, which works in the following way: The \texttt{NSIDE} = $4096$ maps are re-binned into a coarser grid of \texttt{NSIDE} = $16$ (with a resolution of about 4 degrees). We then randomly divide these cells into $k$ equal-area groups. To derive the weights for a given fold $k$, we train the neural network on the other folds, using fold $k$ as a validation sample (the training halts when the training metric no longer improves on the validation set). This cross-validation scheme will only work to prevent overfitting on scales below the resolution defined by \texttt{NSIDE} split, in this case around 4 degrees. As also mentioned in \cite{HighzSample}, testing the corrected and uncorrected galaxy clustering on log-normal mock catalogs demonstrated no overfitting from this method at scales below 1 degree, and an impact of around 5\% overfitting at scales of 2 degrees. We thus keep the galaxy clustering analysis in this work to angular scales below 1 degree to be conservative. 

\section{Validation of covariance matrix}
\label{sec:cov_test}

\begin{figure}
\centering
  \centering
  \includegraphics[width=\columnwidth]{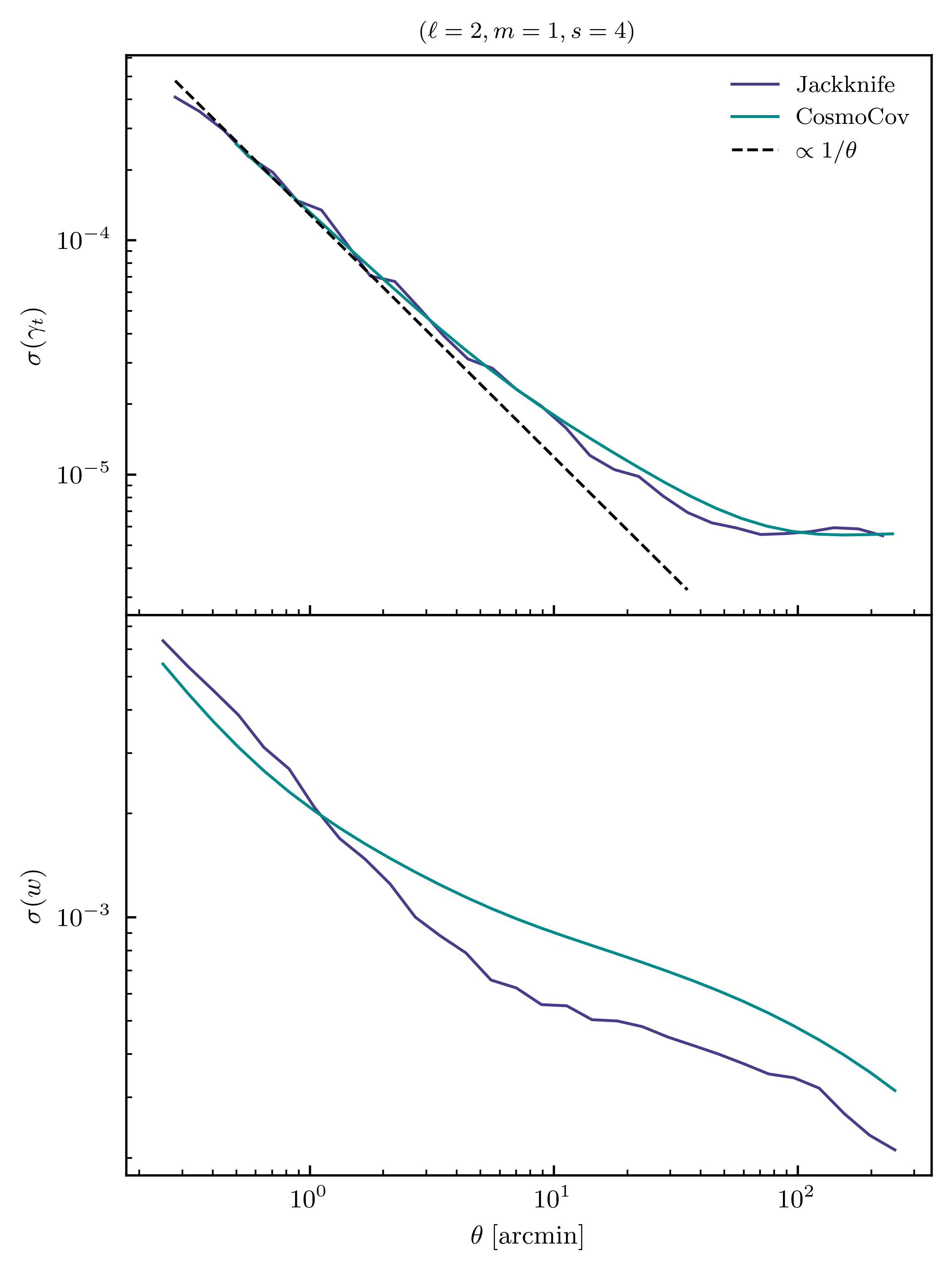}
\caption{\label{fig:covariances} (\textit{Upper}) Square root of the diagonal of the Jackknife covariance matrix that we use in this work (dark blue solid line) and corresponding theory estimation from \texttt{CosmoCov} (light blue solid line). The straight line (dashed black) $\propto 1/\theta$ shows the trend we would expect if the covariance was shape noise dominated. For this plot we used the lens-source bin combination $(\ell,m,s)=(2,1,4)$ for demonstration purposes. (\textit{Lower}) Similar to the left panel but for $w$ for lens bin $(\ell,m)=(2,1)$.}
\end{figure}

As discussed in Section~\ref{sec:covariance}, our main analysis uses a Jackknife covariance, but we also calculate a theoretical covariance matrix via the \texttt{CosmoCov} \citep{Krause2017_cosmolike,fang20_cosmocov2} package. Figure~\ref{fig:covariances} shows the comparison between the Jackknife and \texttt{CosmoCov} error bars (the square root of the diagonal of the covariance matrix) in the case of lens-source bin combination $(\ell,m,s)=(2,1,4)$, for $\gamma_t$ (upper panel) and $w$ (lower panel); we use that bin for demonstration here but we see similar trends in all bins. In the left plot we also show the line $\propto 1/\theta$ which shows the expected behavior of the covariance if it were shape noise dominated. As we see in that plot, $\gamma_t$ is dominated by shape noise on small scales. This also implies that the cross-covariance between $\gamma_t$ and $w$ in a single bin and across different redshift and stellar-mass bins is weak. Therefore, we choose to ignore the cross-covariance in this work. 

The agreement between Jackknife and theory is not perfect. However, this is not very surprising given the differences in the HOD theory used by \texttt{CosmoCov} compared to this study, as the one implemented in the former is based on \cite{zheng2005,zehavi2011}. Nevertheless, the differences we see in the covariance are not expected to have a significant effect on our data fits and we proceed to use the Jackknife covariace matrix in Section~\ref{sec:Fits} while performing the likelihood analysis.

\section{Diagnostic tests}\label{app:systematics}

\begin{figure*}
\centering
\includegraphics[width=1.8\columnwidth]{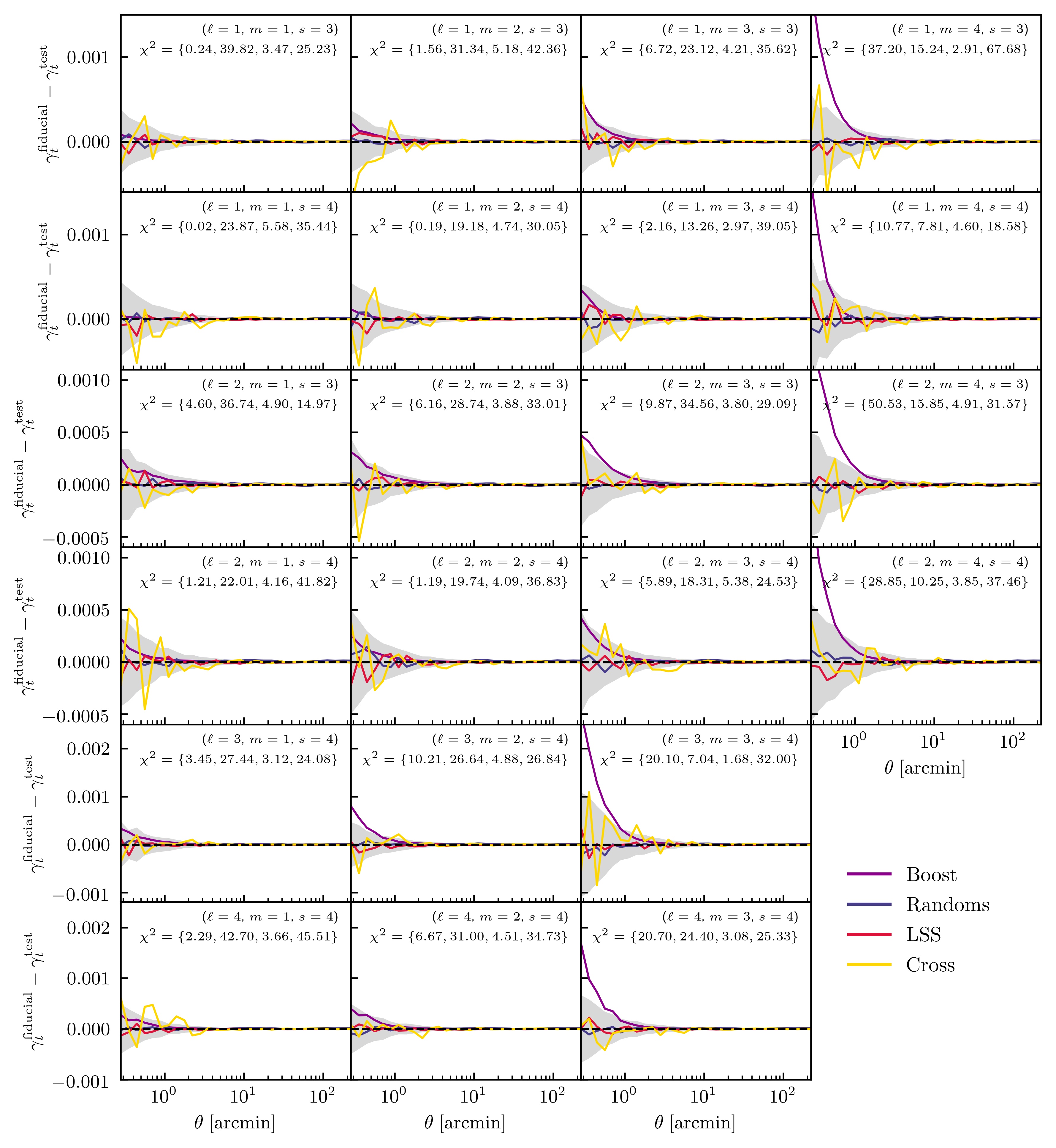}
\caption{\label{fig:GGLsystematics}Systematics tests for the $\gamma_t$ measurements, as discussed in Section~\ref{subsec:smSystematics}, for the stellar-mass sample. {\it Boosts:} Comparison of $\gamma_t$ with and without applying the boost factor correction; {\it Cross component:} The cross-component of shear; {\it Random points}: Effect from applying the random-point subtraction; {\it No LSS weights}: Effect from not applying the LSS weights to correct for observing conditions; {\it Gray area:} The error bars on the shear measurement. In each panel we also list the $\chi^2$ (the order of which follows the order in the legend) between each test and the null, using the covariance of our $\gamma_t$ measurements. The number of points for each of the lines is $30$.}
\end{figure*}

\begin{figure*}
\centering
\includegraphics[width=1.6\columnwidth]{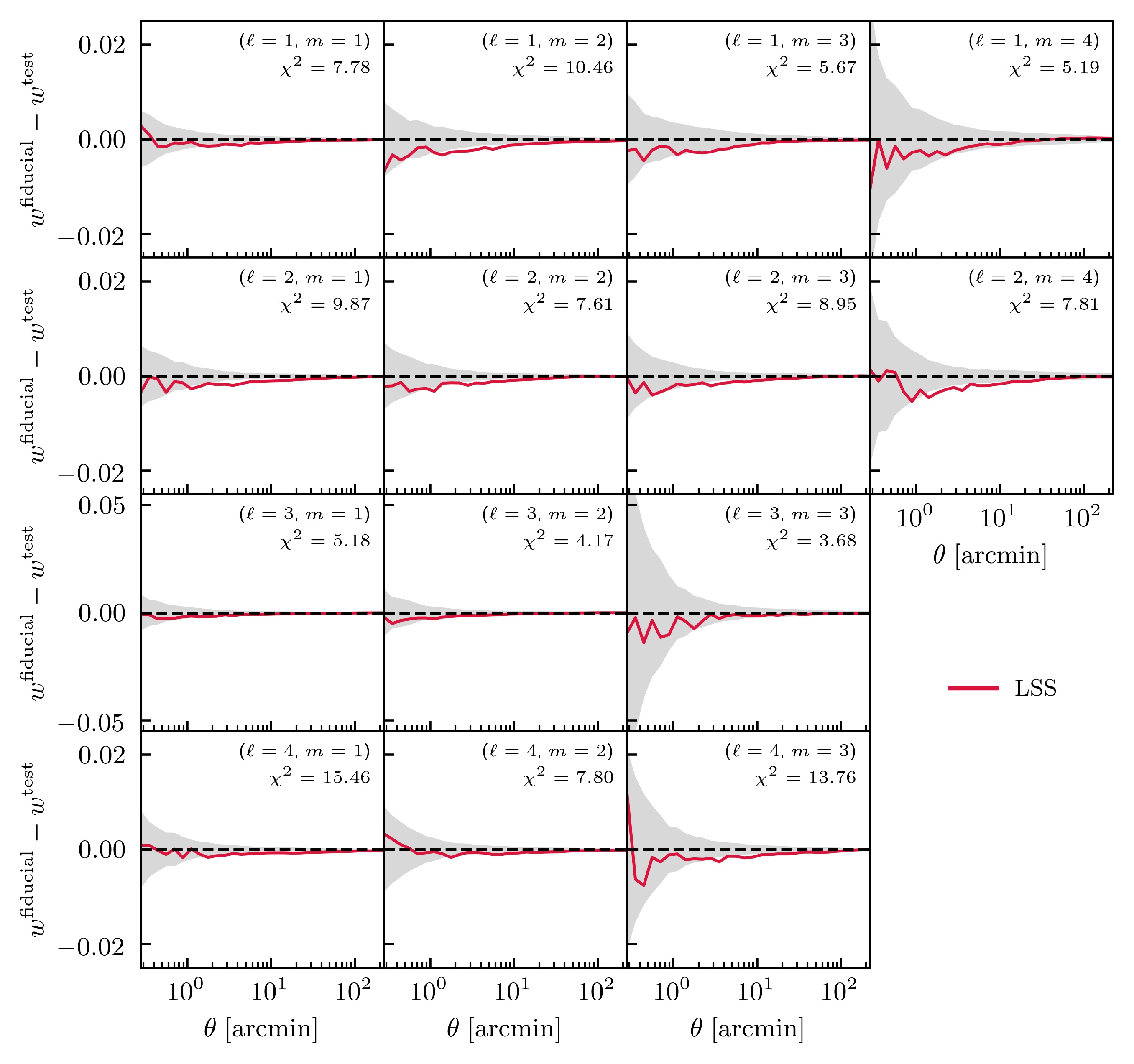}
\caption{\label{fig:GGsystematics}Systematics tests for $w$, as discussed in Section~\ref{subsec:smSystematics}, for the stellar-mass sample. {\it No LSS weights}: Effect from not applying the LSS weights to correct for observing conditions; {\it Gray area:} The error bars on the shear measurement. In each panel we also list the $\chi^2$ between each test and the null, using the covariance of our $w$ measurements. The number of points for each of the lines is $30$.}
\end{figure*}

In this appendix we expand on Section~\ref{subsec:smSystematics} and detail the diagnostic tests performed on the data vectors. The results are summarized in Figures~\ref{fig:GGLsystematics} and \ref{fig:GGsystematics}.
\begin{itemize}
    \item {\bf Cross component}: 
    The tangential shear, $\gamma_t$, is one of the two components when we decompose a spin-2 shear field. 
    The other component is $\gamma_\times$, which  is defined by the projection of the field onto a coordinate system which is rotated by $45^{\circ}$ relative to the tangential frame. 
    For isotropically oriented lenses, the average of $\gamma_\times$ due to gravitational lensing alone should be zero. It is thus a useful test to measure this component in the data and make sure that it is consistent with zero for all angular scales. 
    To be able to decide whether this is the case, we report the total $\chi^2$ calculated for $\gamma_\times$ when compared with the null signal. This test on our shear measurements is presented in Figure~\ref{fig:GGLsystematics}. The measurements of $\gamma_\times$ at large scales are consistent with zero. At smaller scales, below a few arcmin, $\gamma_\times$ fluctuates around zero, roughly within the error bars. Considering that at small scales the level of noise increases, we do not find the behavior of $\gamma_\times$ worrisome. When the reduced $\chi^2$ is examined per bin we see that some bins have a somewhat large value, but the $\chi^2$ distribution is still largely consistent with the theoretical expectation.
    
    \item {\bf Random points}: The mean tangential shear around random points tests the importance of geometrical and mask effects in the signal. Although our estimator of galaxy-galaxy lensing includes the subtraction of tangential shear measurement around random points, it is useful to check that this correction is small, especially for the bins with the highest signal. For this work we use a number of random points that is $30$ times the number of lenses in each bin. We note here, however, that we have tested different numbers of random points, both higher and lower, and we have verified that this number is sufficiently large for our purposes. See also \cite{Prat2022} for a discussion on random-point subtraction and its effects on the measured shear. The effect from including the random-point subtraction is shown on Figure~\ref{fig:GGLsystematics}, where we can see that it is a small effect, and within the error bars in all bins, as also indicated by the reduced $\chi^2$ values we obtain.
    
    \item {\bf LSS weights}: We described in Section~\ref{sec:LSSweights} and Appendix~\ref{sec:NNweights} the LSS weights. 
    Here we compare the measurements with and without the application of the LSS-weights and report the difference between the two in Figures~\ref{fig:GGLsystematics} and \ref{fig:GGsystematics}. 
    We see that the effect is small for $\gamma_{t}$ as indicated by the reduced $\chi^2$. This is expected given that galaxy-galaxy lensing is a cross-correlation measurement and we expect additive systematic effects to drop out during the correlation. 
    For $w$ the effect is larger, especially in small scales, also as expected.

    \item {\bf Boost factors}: Galaxies are clustered, and the number of sources around a lens can be larger than what we would expect from a uniform distribution if the lens and the source redshift distribution overlap. This is usually quantified by the {\it boost factor} \citep{Sheldon_2004}, $B(\theta)$, as estimator quantifying the excess in the number of sources around a lens with respect to randoms. This is a test only relevant for $\gamma_t$ and the results are plotted in Figure~\ref{fig:GGLsystematics}. As can be seen from the plot, the contribution from this effect can be large at small scales, especially when the bins are more overlapped in redshift. In our analysis we take the boost factors into account by correcting for it before carrying out the model fit (Equation~\ref{eq:gammat_full_estimator}). We also use the boost factors to guide our decision in the selection of bins and scales to use in the final analysis (Section~\ref{sec:BinSecectionAndScaleCuts}). 
\end{itemize}

\section{Constraint on parameters}\label{app:parameter_constraints}

In Table~\ref{tab:HODparamConstraints} we list the best-fit parameter values with error bars for our main HOD parameters. We omit parameters that are not well constrained for brevity. These values may be used, e.g., in simulations to populate dark matter halos with galaxies. Due to our complex HOD model, however, it is important that the HOD and other modeling choices, such as the Halo Mass Function, SHMR, etc, are as close to the ones in this current paper as possible.

\setlength{\tabcolsep}{9pt}
\renewcommand{\arraystretch}{1.8}

\begin{table*}

  \rotatebox{90}{
  \centering 
    \begin{minipage}{1.3\textwidth}

    {\setlength{\tabcolsep}{.5em}
    \begin{tabular}{l | c c c c c c c c c c c c c c}
    \hline
    ($\ell$, $m$) & $\sigma_{\rm \log M_{*}}$ & $\alpha_{\rm sat}$ & $B_{\rm sat}$ & $\beta_{\rm sat}$ & $B_{\rm cut}$ & $\beta_{\rm cut}$ & $\log M_{1}$ & $\log \epsilon$ & $\alpha$ & $\gamma$ & $\delta$ \\
    \hline \hline
      (1, 1) & $0.427^{\small +0.144}_{\small -0.165}$ & $1.378^{\small +0.079}_{\small -0.082}$ & $14.077^{\small +2.551}_{\small -3.090}$ & $0.202^{\small +0.073}_{\small -0.061}$ & $9.411^{\small +3.051}_{\small -2.936}$ & $0.419^{\small +0.228}_{\small -0.244}$ & $11.624^{\small +0.101}_{\small -0.104}$ & $-1.068^{\small +0.172}_{\small -0.189}$ & $-2.007^{\small +0.329}_{\small -0.158}$ & $0.208^{\small +0.192}_{\small -0.123}$ & $5.883^{\small +0.473}_{\small -0.743}$ \\
      (1, 2) & $0.329^{\small +0.069}_{\small -0.165}$ & $0.628 ^{\small +0.046}_{\small -0.088}$ & $7.483^{\small +6.117}_{\small -2.162}$ & $1.739^{\small +0.205}_{\small -0.253}$ & $10.133^{\small +3.321}_{\small -2.296}$ & $0.141^{\small +0.380}_{\small -0.351}$ & $11.168^{\small +0.157}_{\small -0.186}$ & $-1.736^{\small +0.095}_{\small -0.101}$ & $-1.940^{\small +0.329}_{\small -0.226}$ & $0.555^{\small +0.198}_{\small -0.116}$ & $3.807^{\small +0.576}_{\small -0.589}$\\
      (1, 3) & $0.142^{\small +0.127}_{\small -0.101}$ & $0.791^{\small +0.046}_{\small -0.059}$ & $7.552^{\small +1.024}_{\small -1.133}$ & $1.999^{\small +0.280}_{\small -0.271}$ & $7.295^{\small +3.541}_{\small -4.336}$ & $0.767^{\small +0.278}_{\small -0.710}$ & $11.287^{\small +0.094}_{\small -0.212}$ & $-1.700^{\small +0.093}_{\small -0.089}$ & $-2.016^{\small +0.262}_{\small -0.251}$ & $0.679^{\small +0.131}_{\small -0.086}$ & $3.874^{\small +0.658}_{\small -0.222}$\\
      (1, 4) & $0.146^{\small +0.015}_{\small -0.055}$ & $1.087^{\small +0.004}_{\small -0.050}$ & $8.193^{\small +0.027}_{\small -2.736}$ & $1.554^{\small +0.290}_{\small -0.320}$ & $5.592^{\small +1.800}_{\small -1.173}$ & $-0.284^{\small +0.485}_{\small -0.269}$ & $11.147^{\small +0.437}_{\small -0.071}$ & $-1.893^{\small +0.082}_{\small -0.035}$ & $-1.977^{\small +0.589}_{\small -0.022}$ & $0.855^{\small +0.089}_{\small -0.103}$ & $5.414^{\small +0.130}_{\small -0.615}$\\
      \hline
      (2, 1) & $0.492^{\small +0.060}_{\small -0.099}$ & $1.232^{\small +0.027}_{\small -0.030}$ & $18.918^{\small +1.351}_{\small -1.817}$ & $1.289^{\small +0.129}_{\small -0.129}$ & $0.963^{\small +3.079}_{\small -2.076}$ & $1.446^{\small +0.443}_{\small -0.654}$ & $10.519^{\small +0.114}_{\small -0.095}$ & $-1.968^{\small +0.088}_{\small -0.065}$ & $-1.220^{\small +0.308}_{\small -0.277}$ & $0.796^{\small +0.062}_{\small -0.093}$ & $4.652^{\small +0.260}_{\small -0.351}$\\
      (2, 2) & $0.414^{\small +0.092}_{\small -0.129}$ & $1.230^{\small +0.053}_{\small -0.052}$ & $9.170^{\small +2.976}_{\small -3.270}$ & $1.147^{\small +0.200}_{\small -0.245}$ & $0.864^{\small +3.844}_{\small -3.268}$ & $1.524^{\small +0.620}_{\small -0.677}$ & $10.609^{\small +0.156}_{\small -0.141}$ & $-1.744^{\small +0.125}_{\small -0.124}$ & $-1.571^{\small +0.288}_{\small -0.302}$ & $0.623^{\small +0.130}_{\small -0.126}$ & $3.948^{\small +0.522}_{\small -0.554}$\\
      (2, 3) & $0.442^{\small +0.109}_{\small -0.125}$ & $1.232^{\small +0.051}_{\small -0.055}$ & $8.058^{\small +2.303}_{\small -2.142}$ & $1.541^{\small +0.182}_{\small -0.245}$ & $1.034^{\small +2.307}_{\small -1.468}$ & $0.348^{\small +0.678}_{\small -0.666}$ & $11.113^{\small +0.180}_{\small -0.180}$ & $-1.864^{\small +0.157}_{\small -0.120}$ & $-1.577^{\small +0.304}_{\small -0.312}$ & $0.635^{\small +0.102}_{\small -0.126}$ & $4.322^{\small +0.527}_{\small -0.587}$\\
      (2, 4) & $0.487^{\small +0.250}_{\small -0.138}$ & $1.173^{\small +0.037}_{\small -0.055}$ & $4.053^{\small +1.528}_{\small -1.217}$ & $1.560^{\small +0.360}_{\small -0.138}$ & $0.300^{\small +1.785}_{\small -0.993}$ & $0.710^{\small +0.371}_{\small -0.524}$ & $11.857^{\small +0.385}_{\small -0.473}$ & $-1.870^{\small +0.228}_{\small -0.095}$ & $-2.056^{\small +0.426}_{\small -0.295}$ & $0.599^{\small +0.170}_{\small -0.052}$ & $3.740^{\small +0.932}_{\small -0.393}$\\
      \hline 
      (3, 1) & $0.236^{\small +0.111}_{\small -0.105}$ & $1.165^{\small +0.063}_{\small -0.066}$ & $7.050^{\small +2.638}_{\small -2.288}$ & $1.004^{\small +0.197}_{\small -0.213}$ & $2.622^{\small +3.662}_{\small -2.779}$ & $1.527^{\small +0.683}_{\small -0.717}$ & $10.746^{\small +0.164}_{\small -0.154}$ & $-1.545^{\small +0.140}_{\small -0.151}$ & $-1.730^{\small +0.312}_{\small -0.279}$ & $0.484^{\small +0.152}_{\small -0.121}$ & $3.441^{\small +0.578}_{\small -0.466}$\\
      (3, 2) & $0.259^{\small +0.128}_{\small -0.091}$ & $1.123^{\small +0.060}_{\small -0.056}$ & $3.098^{\small +0.874}_{\small -0.905}$ & $1.810^{\small +0.141}_{\small -0.209}$ & $1.818^{\small +2.515}_{\small -1.375}$ & $-0.366^{\small +0.681}_{\small -0.583}$ & $11.353^{\small +0.279}_{\small -0.274}$ & $-1.609^{\small +0.204}_{\small -0.189}$ & $-1.582^{\small +0.320}_{\small -0.304}$ & $0.618^{\small +0.154}_{\small -0.144}$ & $3.273^{\small +0.648}_{\small -0.522}$\\
      (3, 3) & $0.162^{\small +0.044}_{\small -0.060}$ & $1.410^{\small +0.098}_{\small -0.113}$ & $2.826^{\small +5.573}_{\small -2.000}$ & $1.210^{\small +0.177}_{\small -0.214}$ & $1.753^{\small +6.226}_{\small -2.328}$ & $0.145^{\small +0.315}_{\small -0.232}$ & $11.375^{\small +0.278}_{\small -0.398}$ & $-1.624^{\small +0.224}_{\small -0.227}$ & $-1.814^{\small +0.301}_{\small -0.261}$ & $0.395^{\small +0.066}_{\small -0.064}$ & $3.463^{\small +0.602}_{\small -0.576}$\\
      \hline
      (4, 1) & $0.102^{\small +0.097}_{\small -0.087}$ & $1.437^{\small +0.087}_{\small -0.084}$ & $19.613^{\small +2.758}_{\small -3.055}$ & $0.853^{\small +0.306}_{\small -0.140}$ & $2.980^{\small +4.147}_{\small -4.074}$ & $0.407^{\small +0.706}_{\small -0.595}$ & $10.656^{\small +0.310}_{\small -0.307}$ & $-1.640^{\small +0.181}_{\small -0.139}$ & $-2.141^{\small +0.296}_{\small -0.283}$ & $0.522^{\small +0.153}_{\small -0.149}$ & $4.467^{\small +0.589}_{\small -0.488}$\\
      (4, 2) & $0.460^{\small +0.102}_{\small -0.115}$ & $1.481^{\small +0.105}_{\small -0.085}$ & $9.834^{\small +3.727}_{\small -4.084}$ & $1.580^{\small +0.292}_{\small -0.260}$ & $3.187^{\small +3.487}_{\small -2.346}$ & $-0.433^{\small +0.520}_{\small -0.393}$ & $11.842^{\small +0.183}_{\small -0.174}$ & $-1.826^{\small +0.077}_{\small -0.073}$ & $-1.848^{\small +0.289}_{\small -0.283}$ & $0.384^{\small +0.188}_{\small -0.181}$ & $3.171^{\small +0.438}_{\small -0.383}$\\
      (4, 3) & $0.039^{\small +0.069}_{\small -0.069}$ & $1.848^{\small +0.131}_{\small -0.163}$ & $5.183^{\small +5.051}_{\small -4.289}$ & $1.255^{\small +0.202}_{\small -0.153}$ & $0.765^{\small +4.549}_{\small -3.520}$ & $0.141^{\small +0.374}_{\small -0.266}$ & $11.603^{\small +0.320}_{\small -0.428}$ & $-1.727^{\small +0.209}_{\small -0.145}$ & $-1.774^{\small +0.325}_{\small -0.282}$ & $0.503^{\small +0.161}_{\small -0.136}$ & $3.478^{\small +0.556}_{\small -0.417}$\\
      \hline 
	\end{tabular}}
    \caption{Best-fit HOD parameter constraints for our model (see Table~\ref{tab:smParamPriors} for the parameter priors). Note that we have omitted parameters that are not constrained.\\}
    \label{tab:HODparamConstraints}
    \end{minipage}}

\end{table*}


\section*{Acknowledgements}

CC and JP were supported by DOE grant DESC0021949.

Funding for the DES Projects has been provided by the U.S. Department of Energy, the U.S. National Science Foundation, the Ministry of Science and Education of Spain, 
the Science and Technology Facilities Council of the United Kingdom, the Higher Education Funding Council for England, the National Center for Supercomputing Applications at the University of Illinois at Urbana-Champaign, the Kavli Institute of Cosmological Physics at the University of Chicago, the Center for Cosmology and Astro-Particle Physics at the Ohio State University, the Mitchell Institute for Fundamental Physics and Astronomy at Texas A\&M University, Financiadora de Estudos e Projetos, Funda{\c c}{\~a}o Carlos Chagas Filho de Amparo {\`a} Pesquisa do Estado do Rio de Janeiro, Conselho Nacional de Desenvolvimento Cient{\'i}fico e Tecnol{\'o}gico and the Minist{\'e}rio da Ci{\^e}ncia, Tecnologia e Inova{\c c}{\~a}o, the Deutsche Forschungsgemeinschaft and the Collaborating Institutions in the Dark Energy Survey. 

The Collaborating Institutions are Argonne National Laboratory, the University of California at Santa Cruz, the University of Cambridge, Centro de Investigaciones Energ{\'e}ticas, Medioambientales y Tecnol{\'o}gicas-Madrid, the University of Chicago, University College London, the DES-Brazil Consortium, the University of Edinburgh, the Eidgen{\"o}ssische Technische Hochschule (ETH) Z{\"u}rich, 
Fermi National Accelerator Laboratory, the University of Illinois at Urbana-Champaign, the Institut de Ci{\`e}ncies de l'Espai (IEEC/CSIC), the Institut de F{\'i}sica d'Altes Energies, Lawrence Berkeley National Laboratory, the Ludwig-Maximilians Universit{\"a}t M{\"u}nchen and the associated Excellence Cluster Universe, 
the University of Michigan, NFS's NOIRLab, the University of Nottingham, The Ohio State University, the University of Pennsylvania, the University of Portsmouth, SLAC National Accelerator Laboratory, Stanford University, the University of Sussex, Texas A\&M University, and the OzDES Membership Consortium.

Based in part on observations at Cerro Tololo Inter-American Observatory at NSF's NOIRLab (NOIRLab Prop. ID 2012B-0001; PI: J. Frieman), which is managed by the Association of Universities for Research in Astronomy (AURA) under a cooperative agreement with the National Science Foundation.

The DES data management system is supported by the National Science Foundation under Grant Numbers AST-1138766 and AST-1536171.
The DES participants from Spanish institutions are partially supported by MICINN under grants ESP2017-89838, PGC2018-094773, PGC2018-102021, SEV-2016-0588, SEV-2016-0597, and MDM-2015-0509, some of which include ERDF funds from the European Union. IFAE is partially funded by the CERCA program of the Generalitat de Catalunya. Research leading to these results has received funding from the European Research Council under the European Union's Seventh Framework Program (FP7/2007-2013) including ERC grant agreements 240672, 291329, and 306478.
We  acknowledge support from the Brazilian Instituto Nacional de Ci\^encia e Tecnologia (INCT) do e-Universo (CNPq grant 465376/2014-2).

This manuscript has been authored by Fermi Research Alliance, LLC under Contract No. DE-AC02-07CH11359 with the U.S. Department of Energy, Office of Science, Office of High Energy Physics.

All analysis in this work was enabled greatly by the following software: \textsc{NumPy} \citep{vanderWalt2011Numpy}, \textsc{SciPy} \citep{Virtanen2020Scipy}, and \textsc{Matplotlib} \citep{Hunter2007Matplotlib}. We have also used the Astrophysics Data Service (\href{https://ui.adsabs.harvard.edu/}{ADS}) and \href{https://arxiv.org/}{\texttt{arXiv}} preprint repository extensively during this project and the writing of the paper.


\section*{Data Availability}

The data underlying this article are available on the Dark Energy Survey website at \url{https://www.darkenergysurvey.org/the-des-project/data-access/}.

The lens galaxy catalog will be made public upon acceptance of the paper at \url{https://des.ncsa.illinois.edu/releases/y3a2/Y3stellarmass}.


\section*{Author contributions}

Georgios Zacharegkas performed the main analysis, produced all the plots in the paper, and led the paper writing. Chihway Chang provided direct supervision for the research and contributed to the paper’s text. Judit Prat provided overall consultation and helped with the measurement and validation of the galaxy-galaxy lensing data vectors. William Hartley and Sunil Mucesh contributed to training the \texttt{GALPRO} machine learning model, and to the generation and calibration of the stellar-mass and redshift properties of the lens galaxy samples. William Hartley, additionally, provided supervision for the research. The authors between A. Alarcon and J. Zuntz, both included, have produced and characterized one or multiple of the following data products used in this paper: the shape catalog, the redshift catalog.


\newpage
\bibliography{HODbib}


\end{document}